\documentstyle[12pt,epsf]{article}

\newlength{\pubnumber} 

\catcode`\@=11
\@addtoreset{equation}{section}

\def\section{\@startsection{section}{1}{\z@}{3.5ex plus 1ex minus .2ex}
 {2.3ex plus .2ex}{\large\bf}}
\def\subsection{\@startsection{subsection}{2}{\z@}{2.3ex plus .2ex}
 {2.3ex plus .2ex}{\bf}}

\def\bh{\bar{h}}

\def\bS{\bar{S}}

\def\beq{\begin{equation}}
\def\eeq{\end{equation}}
\def\beqn{\begin{eqnarray}}
\def\eeqn{\end{eqnarray}}
\def\no{\noindent }
\def\nolabel{\nonumber }

\def\NA{non-Abelian }

\def\gsim{{\buildrel >\over \sim}}
\def\lsim{{\buildrel <\over \sim}}

\def\eq#1{eq.\ (\ref{#1}) }

\def\tenth{{\textstyle{1\over 10}}}

\def\half{{\textstyle{1\over 2}}}
\def\third{{\textstyle {1\over3}}}

\def\sixth{{\textstyle {1\over6}}}
\def\phm{$\phantom{-}$}

\def\Tr{{\rm Tr}\, }
\def\tr{{\rm tr}\, }

\def\MS{M_{str}}

\def\MP{M_{P}}

\def\vev#1{\langle #1\rangle}

\def\lh{\bar{h}}

\def\V{{\bf V}}

\def\eps{\epsilon}

\def\S{{\bf S}}
\def\X{{\bf X}}

\def\mbf{\mathbf}
\def\bone{{\mathbf 1}}
\def\mone{{\mathbf 1}}
\def\bo{{\mathbf 0}}

\def\mzero{{\mathbf 0}}
\def\mF{{\mathbf F}}
\def\mQ{{\mathbf Q}}

\def\bS{{\mathbf S}}
\def\bb{{\mathbf b}}

\def\mk{{\mathbf k}}
\def\mq{{\mathbf q}}
\def\mbp{{\mathbf p}}
\def\mv{{\mathbf v}}
\def\mW{{\mathbf W}}
\def\mN{{\mathbf N}}

\def\balpha{{\mathbf \alpha}}
\def\bbeta{{\mathbf \beta}}
\def\bgamma{{\mathbf \gamma}}

\def\bbeta{ {{\mathbf \beta}}}
\def\bgamma{{\mathbf \gamma}}

\def\NA{non-Abelian }

\def\gsim{{\buildrel >\over \sim}}
\def\lsim{{\buildrel <\over \sim}}

\def\mod#1{{\rm \,\, (mod\, #1)}}

\def\half{{\textstyle{1\over 2}}}
\def\third{{\textstyle {1\over3}}}
\def\fourth{{\textstyle {1\over4}}}

\def\sixth{{\textstyle {1\over6}}}
\def\tenth{{\textstyle {1\over10}}}

\def\twothird{{\textstyle {2\over3}}}

\def\threefourth{{\textstyle {3\over4}}}

\def\MP{M_{P}}

\def\ra{\rightarrow }

\def\MS{M_{str}}

\def\MP{M_{P}}

\def\vev#1{\langle #1\rangle}

\def\UA{U(1)_{\rm A}}

\def\tr{{\rm tr}}

\def\hb#1{\bar{h}_{#1}}

\def\V{{\bf V}}

\def\S{{\bf S}}
\def\X{{\bf X}}

\def\bone{{\mathbf 1}}
\def\bo{{\mathbf 0}}

\def\bS{{\mathbf S}}
\def\bb{{\mathbf b}}

\def\balpha{{\mathbf \alpha}}
\def\bbeta{{\mathbf \beta}}
\def\bgamma{{\mathbf \gamma}}

\def\bbeta{{\mathbf \beta}}
\def\bgamma{{\mathbf \gamma}}

\def\mz{{\mathbf z}}
\def\mZ{{\mathbf Z}}

\def\me{{\mathbf e}}
\def\mv{{\mathbf v}}
\def\mV{{\mathbf V}}
\def\bV{{\mathbf V}}
\def\ml{{\mathbf l}}
\def\mtheta{{\mathbf \theta}}

\def\mchi{{\mathbf \chi}}
\def\mphi{{\mathbf \phi}}

\def\eps{\epsilon}

\def\p#1{{\Phi_{#1}}}
\def\pp#1{{\Phi^{'}_{#1}}}
\def\pb#1{{{\overline{\Phi}}_{#1}}}

\def\ppb#1{{{\overline{\Phi}}^{'}_{#1}}}

\def\h#1{h_{#1}}
\def\hb#1{{\bar{h}}_{#1}}

\def\Q#1{Q_{#1}}
\def\dc#1{d^{c}_{#1}}
\def\uc#1{u^{c}_{#1}}

\def\L#1{L_{#1}}
\def\ec#1{e^{c}_{#1}}
\def\Nc#1{N^{c}_{#1}}

\def\H#1{H_{#1}}
\def\V#1{V_{#1}}
\def\Hs#1{{H^{s}_{#1}}}
\def\Vs#1{{V^{s}_{#1}}}

\def\FD2pv{FD2$^{'}$V }
\def\FD2p{FD2$^{'}$ }

\def\bc{\bar{c}}
\def\lh{\bar{h}}
\def\hb#1{\bar{h}_{#1}}
\def\bh#1{\bar{h}_{#1}}

\def\V{{\bf V}}

\def\mm{\mathbf }

\def\S{{\bf S}}
\def\X{{\bf X}}

\def\bone{{\mathbf 1}}
\def\bo{{\mathbf 0}}

\def\bS{{\mathbf S}}
\def\bb{{\mathbf b}}

\def\balpha{{\mathbf \alpha}}
\def\bbeta{{\mathbf \beta}}
\def\bgamma{{\mathbf \gamma}}

\def\bbeta{{\mathbf \beta}}
\def\bgamma{{\mathbf \gamma}}

\def\Q#1{Q_{#1}}
\def\dc#1{d^{c}_{#1}}
\def\uc#1{u^{c}_{#1}}

\def\L#1{L_{#1}}
\def\ec#1{e^{c}_{#1}}
\def\Nc#1{N^{c}_{#1}}

\def\H#1{H_{#1}}
\def\V#1{V_{#1}}
\def\Hs#1{{H^{s}_{#1}}}
\def\sH#1{{H^{s}_{#1}}}
\def\Vs#1{{V^{s}_{#1}}}

\def\phm{\phantom{-}}

\def\half{{\textstyle{1\over 2}}}
\def\third{{\textstyle {1\over3}}}
\def\fourth{{\textstyle {1\over4}}}

\def\sixth{{\textstyle {1\over6}}}
\def\tenth{{\textstyle {1\over10}}}

\def\twothird{{\textstyle {2\over3}}}

\def\threefourth{{\textstyle {3\over4}}}

\def\inbar{\,\vrule height1.5ex width.4pt depth0pt}
\def\IT{\relax\hbox{$\inbar\kern-.3em{\rm T}$}}
\def\IS{\relax\hbox{$\inbar\kern-.3em{\rm S}$}}
\def\IC{\relax\hbox{$\inbar\kern-.3em{\rm C}$}}
\def\IQ{\relax\hbox{$\inbar\kern-.3em{\rm Q}$}}
\def\IR{\relax{\rm I\kern-.18em R}}
 \font\cmss=cmss10 \font\cmsss=cmss10 at 7pt

\def\IZ{\relax\ifmmode\mathchoice
 {\hbox{\cmss Z\kern-.4em Z}}{\hbox{\cmss Z\kern-.4em Z}}
 {\lower.9pt\hbox{\cmsss Z\kern-.4em Z}}
 {\lower1.2pt\hbox{\cmsss Z\kern-.4em Z}}\else{\cmss Z\kern-.4em Z}\fi}

\def\Io{\relax\ifmmode\mathchoice
 {\hbox{\cmss 1\kern-.4em 1}}{\hbox{\cmss 1\kern-.4em 1}}
 {\lower.9pt\hbox{\cmsss 1\kern-.4em 1}}
{\lower1.2pt\hbox{\cmsss 1\kern-.4em 1}}\else{\cmss 1\kern-.4em 1}\fi}

\def\ra{\rightarrow}

\def\al{\alpha}

\def\be{\begin{equation}}
\def\ee{\end{equation}}
\def\bea{\begin{eqnarray}}
\def\eea{\end{eqnarray}}
\begin{document}
\begin{titlepage}
\setcounter{page}{1}
\rightline{BU-HEPP-06-11}
\rightline{CASPER-06-05}
\rightline{\tt hep-ph/0703027}
\rightline{March 2007}

\vspace{.06in}
\begin{center}
{\Large \bf In Search of\\ the (Minimal Supersymmetric) Standard Model String}
\vspace{.12in}

{\large Gerald B. Cleaver,$^{1}$\footnote{gerald{\underline{\phantom{a}}}cleaver@baylor.edu}}
\\
\vspace{.12in}
{\it $^{1}$ Center for Astrophysics, Space Physics \& Engineering Research\\
            Department of Physics, Baylor University,
            Waco, TX 76798-7316\\}
\vspace{.06in}
\end{center}

\begin{abstract}
This paper summarizes several developments in string-derived (Minimal Supersymmetric) Standard Models.
Part one reviews the first string model containing solely the three generations of the Minimal Supersymmetric Standard Model and a single pair of Higgs as the matter in the observable sector of 
the low energy effective field theory. This model 
was constructed by Cleaver, Faraggi, and Nanopoulos in the $\IZ_2\otimes \IZ_2$ free fermionic formulation of weak coupled heterotic strings. Part two examines a representative collection of string/brane-derived MSSMs that followed.
These additional models were obtained from various construction methods, including 
weak coupled $\IZ_6$ heterotic orbifolds, 
strong coupled heterotic on elliptically fibered Calabi-Yau's,
Type IIB orientifolds with magnetic charged branes, and
Type IIA orientifolds with intersecting branes (duals of the Type IIB).
Phenomenology of the models is compared.\\
\\
{\it To appear in String Theory Research Progress, Ferenc N. Balogh, editor.,
(ISBN 978-1-60456-075-6), Nova Science Publishers, Inc.}

\end{abstract}
\end{titlepage}
\setcounter{footnote}{0}

%%%%%%%%%%%%%%%%%%%%%%%%%%%%%%%%%%%%%%%%%%%%%%%%%%%%%%%%%%%%%%%%%%%%%%%%%%%%%%%%%%%%%%%%%%%%%%%%%%%%%%%%%%%%%%%%%%%%%
% intro at here 1
%%%%%%%%%%%%%%%%%%%%%%%%%%%%%%%%%%%%%%%%%%%%%%%%%%%%%%%%%%%%%%%%%%%%%%%%%%%%%%%%%%%%%%%%%%%%%%%%%%%%%%%%%%%%%%%%%%%%%
\section{Introduction: The Mission of String Phenomenology}

\no
Opponents of string theory \cite{woit:2006a,richter:2006a,Schroer:2006na} have suggested that research in string theory\footnote{The term {\it string theory} denotes in this chapter its manifestation as a whole, from the collection of 10-dimensional string theories pre-second string revolution to its current all encompassing, but still incompletely understood, 11-dimensional M-theory.} may be moving beyond the bounds of science, especially with the rise of the theorized string landscape containing an estimated $10^{100\, {\rm to}\, 500}$ vacua, among which exactly one vacua corresponds to our visible universe. The claim is that ``what passes for the most advanced theory in particle science these days is not really science" \cite{richter:2006a}. Instead, the opposition argues that string theory has ``gone off into a metaphysical wonderland" producing ``models with no testable consequences," rather than models ``with testable and falsifiable consequences." 
Richter, for example, maintains that real progress in physics comes when ``why" questions turn into ``how" questions; he views string theory as trespassing into metaphysical speculation rather than answering the how questions. For Richter, ``why" answers only provide vague, unscientific answers, whereas ``how" answers reveal a clear, physical process \cite{richter:2006a}. 
 
Alternately, many within the string community contend that the role of string theory is, indeed, to answer the ``why" questions. What's at odds are the meanings attached to ``why" and ``how". Munoz \cite{Munoz:2003au}, for example, maintains that while the standard model of particle physics ``adequately describes how the elementary particles behave," it is insufficient ``to explain {\it why} a particular set of elementary particles, each with its own particular properties, exists."  In this context, answering the ``why" is ``precisely one of the purposes of string theory." Munoz ``why" is Richter's ``how," which shows the importance of precisions of terms.

Before string theory can truly provide the ``why" to the standard model (SM), two prerequisites must be satisfied: First, string theory must be shown to contain (at least) one, or a set of, vacua within the landscape that exactly reproduce all known features of the standard model. Second, string theory must then explain why that particular vacuum, or one from among the set, are realized in the observable universe. The first of these requirements may be viewed as the (primary) mission of {\it string phenomenology} \cite{Munoz:2003au}. 

In this chapter we review and summarize developments in the search for the SM, and for its supersymmetric extension, the minimal supersymmetric standard model (MSSM). While the claim that ``no [string theory] solution that looks like our universe has been found" \cite{richter:2006a,Munoz:2003au} is correct 
(and acknowledged by both critics and proponents of string theory \cite{Munoz:2003au}), much progress has been made in this pursuit since the first string revolution and especially since the second string revolution. In this chapter we examine several representative advancements in the search for the SM and MSSM from string theory and some predictions such models might have.

We begin in section 2 with a review of the very first string model for which the observable sector is composed of exactly 
the MSSM gauge particles, the three matter generations of the Minimal Supersymmetric Standard Model, and a pair of Higgs, without {\it any} unwanted MSSM-charged exotics. It is a weak coupling free fermionic heterotic $\IT^6/(\IZ_2\otimes \IZ_2)$ model constructed by Cleaver, Faraggi, and Nanopoulos \cite{Cleaver:1998sa,Cleaver:1999cj,Cleaver:1999mw,Cleaver:2000aa,Cleaver:2001ab}. 

While the first string-based MSSM model, it has been by no means been the last. Several others have appeared since 
\cite{Cleaver:1998sa}. If the density of MSSM-like models within the landscape is truly around $1/10^9$ as asserted in \cite{Gmeiner:2005vz}, then there should indeed be {\it many, many} more yet to come!
Section three examines a representative collection of string- or brane-derived MSSMs that have followed the 
heterotic $\IT^6/(\IZ_2\otimes \IZ_2)$ free fermionic model. 
The various methods of construction of these additional MSSM models, 
include
$\IZ_6$ heterotic orbifolds, 
strong coupled heterotic with elliptically fibered Calabi-Yau,
Type IIB with magnetic charged branes, and
Type IIA with intersecting branes (dual of the Type IIB).
Phenomenological aspects of models are compared. 
Section four concludes with some discussion of the goal of string phenomenology and some common features of the MSSM string models.

%\section{Defining the (Minimal Supersymmetric) Standard Model}

Before beginning a review of string-based (MS)SM models, we should first specify 
what is truly meant by (MS)SM for the sake of clarity. 
The definition of an MSSM model is critical to any claims of constructing one from string theory
and was discussed in \cite{Munoz:2003au} by Mu\~ noz: A true string derived (MS)SM must possess many more phenomenological properties than simply having the (MS)SM matter content of 
12 quarks and 4 leptons\footnote{We assume the existence of the left-handed neutrino singlet, and as such each generation forms a $\mathbf{16}$ representation (rep) of $SO(10)$}
and a (pair of) Higgs doublet(s) (and corresponding supersymmetric partners), without any MSSM-charged exotics, in its low energy effective field theory (LEEFT). Obviously particles identified with the (MS)SM must have 
the correct mass hierarchy, a viable CKM quark mixing matrix, a realistic neutrino mass and mixing matrix, 
and strongly suppressed proton decay. The correct gauge coupling strengths must also result from it.
Further, potential dangers associated with string theory must also be prevented, such as hidden sector (kinetic) mixing via extra $U(1)$ charges. For MSSM candidates, viable non-perturbative supersymmetry breaking methods must ultimately be found and supersymmetric particle mass predictions must correspond to (hoped-for) LHC measurements. Ultimately the stringy characteristics of the model must be understood sufficiently well to also explain the value of the cosmological constant.

No model has yet been produced that contains all of the necessary properties. Which, then, of these characteristics have been found in a model? How deep does the phenomenology go beyond the initial matter requirement?
Following the first string revolution of Schwarz and Green in 1984, numerous string models with a {\it net} number of three generations were constructed from Calabi-Yau compactification of the heterotic string. 
The vast majority of these models contain numerous extra vector-like pairs of generations. 
Further, even those with exactly three generations (and no extra vector-like generation pairs) contain many dozens of MSSM exotics and numerous unconstrained moduli.
Eventually several models were constructed using free fermionic and orbifold compactifications that contained exactly three net MSSM generations \cite{Faraggi:1989ka,Faraggi:1991be,Faraggi:1991jr,Faraggi:1992fa}. Nevertheless, dozens of MSSM-charged exotics are endemic to these also \cite{Faraggi:1990af}.

Proof of the existence of at least one string model containing in its observable sector solely the MSSM matter, and absent any MSSM-charge exotics, waited many years \cite{Cleaver:1998sa}. The first model 
with this property was found by systematic analysis of the $D$ and $F$-flat directions of a free fermionic model constructed much earlier \cite{Faraggi:1989ka}. The model of \cite{Faraggi:1989ka} contains several dozen MSSM-charged exotics that where shown in \cite{Cleaver:1998sa} to gain string-scale mass along a highly constrained set of flat directions \cite{Cleaver:1999cj,Cleaver:1999mw,Cleaver:2000aa,Cleaver:2001ab}.
Since construction of this first string model with exactly the MSSM matter content in its observable sector, several 
similar models have been constructed from alternate compactifications. 
In the next section we start of our reviews with the very first string model with the MSSM content. 

%%%%%%%%%%%%%%%%%%%%%%%%%%%%%%%%%%%%%%%%%%%%%%%%%%%%%%%%%%%%%%%%%%%%%%%%%%%%%%%%%%%%%%%%%%%%%%%%%%%%%%%%%%%%%%%%%%%%%
% heterotic free fermionic at here 2
%%%%%%%%%%%%%%%%%%%%%%%%%%%%%%%%%%%%%%%%%%%%%%%%%%%%%%%%%%%%%%%%%%%%%%%%%%%%%%%%%%%%%%%%%%%%%%%%%%%%%%%%%%%%%%%%%%%%%
\section{First MSSM Spectrum}

\subsection{Heterotic Free Fermionic Models}
\no
The first model to contain exactly the MSSM gauge group, three generations of MSSM matter models, a pair of Higgs doublets, and no exotic MSSM-charged states in the observable sector is a heterotic model constructed in the free
fermionic formulation \cite{Antoniadis:1986rn,Kawai:1986va, Kawai:1987ew,Antoniadis:1987wp} corresponding to $\IZ_2\times \IZ_2$ orbifold models with nontrivial
Wilson lines and background fields. In the free fermionic formulation, each of the six compactified directions is replaced by two real worldsheet fermions. The related left moving bosonic modes $X_{i}$,
for $i=1\, ...,\, 6$ are replaced by real fermion modes $(y_i,w_i)$ and the right moving bosonic modes 
$\overline{X}_{i}$ are replaced by corresponding $(\overline{y}_i,\overline{w}_i)$. 
For the supersymmetric left-moving sector, each real worldsheet boson $X_{i}$ 
is accompanied by a real worldsheet fermion $x_{i}$. Thus, the supersymmetric left-moving compactified modes are represented by a total of 18 real worldsheet fermions, arranged in six sets of three, $(x_i,y_i,w_i)$.
(Consistency with the supersymmetry sector requires the boundary conditions of $x_{2I-1}$ and $x_{2I}$ match, pairing these up to form a complex fermion $\chi_{I} = x_{2I-1} + i x_{2I}$, $I = 1,\, ...,\, 3$.)

In light cone gauge the uncompactified left-moving degrees of freedom are two spacetime bosons $X_{\mu}$ and the associated real fermions $\psi_{\mu}$, for $\mu = 1,\, 2$. 
For the bosonic right moving sector, the six real worldsheet bosonic modes $\overline{X}_{i}$ are unpartnered. Thus in the free fermionic formulation there are just the six pairs of real fermions, $(\overline{y}_i,\overline{w}_i)$ from the compactified space. The  $\overline{X}_{\mu}$ are likewise unpaired.

For heterotic $E_8\times E_8$ strings, the degrees of freedom for the observable sector $E_8$ are
represented by five complex fermions $\overline{\psi}_{j}$, $j=1,\, ...,\, 5$, and three complex fermions $\overline{\eta}_{k}$, $k=1,\, ...,\, 3$, (in anticipation of $E_8 \rightarrow SO(10)\times SO(6)$), 
while the 8 hidden sector bosonic degrees of freedom for the hidden sector $E_8$ 
are represented by 8 complex worldsheet fermions, $\overline{\phi}_{m}$, $m=1,\, ...,\, 8$.
Thus, in light-cone gauge heterotic free fermionic models contain 20 left-moving real fermionic degrees of freedom and 44 right-moving real fermionic degrees of freedom.

A free fermionic heterotic string model is specified by
two objects \cite{Antoniadis:1986rn,Kawai:1986va, Kawai:1987ew,Antoniadis:1987wp}. 
The first is a $p$-dimensional basis set 
of free fermionic boundary vectors $\{\mV_i$, $i= 1$, ... , $p \}$. 
Each vector $\mV_i$ has 64 components, $-1< V_i^{m}\leq 1$, $m=1$, ... , 64,
with the first 20 components specifying boundary conditions for the 20 real free fermions 
representing  worldsheet degrees of freedom for the left-moving supersymmetric string, and the
latter 44 components specifying boundary conditions for the 44 real free fermions representing 
worldsheet degrees of freedom for the right-moving bosonic string.
(Components of $\mV_i$ for complex fermions are double-counted.) 
Modular invariance dictates the basis vectors, $\mV_i$, span a finite
additive group $\Xi = \{\sum_{n_i=0}^{N_i -1} \sum_{i=1}^{p}{{n_i}{\mV_i}} \}$,
with $N_i$ the lowest positive integer such that $N_i \mV_i = \mzero \mod{2}$. 
$\mV_1 = \mone$ (a 64-component unit vector)  
must be present in all models.   
In a given sector 
{\boldmath$\alpha$\unboldmath}$\equiv a_i \mV_i \mod{2}\in \Xi$, 
with $a_i\in\{0,\, 1,\, \dots\, , N_i -1\}$, 
a generic worldsheet
fermion, denoted by $f_m$, transforms as $f_m\rightarrow -\exp\{ \pi \alpha_m \} f_m$ around 
non-contractible loops on the worldsheet. Boundary vector components for real fermions 
are thus limited to be either 0 or 1, 
whereas boundary vector components for complex fermions can be rational.

The second object necessary to define a free fermionic model 
(up to vacuum expectation values (VEVs) of fields in the LEEFT)
is a $p\times p$-dimensional matrix $\mk$ of rational numbers 
$-1<k_{i,j}\leq 1$, $i,j= 1$, ..., $p$, that determine the GSO operators 
for physical states. The $k_{i,j}$ are related to the phase 
weights $C{{\mV_i}\choose {\mV_j}}$  in the one-loop partition 
function $Z$: 
\beqn
C{{\mV_i}\choose {\mV_j}}
= (-1)^{s_i+s_j}
        {\rm exp}( \pi i k_{j,i}- \half \mV_i \cdot\mV_j), 
\label{kijpw}
\eeqn
where $s_i$ is the 4-dimensional spacetime component of $\mV_i$.
The inner product of boundary (or charge) vectors is lorentzian, 
taken as left-movers minus right movers. Contributions to inner products 
from real fermion boundary components are weighted by a factor of $\half$ compared to 
contributions from complex fermion boundary components. 

The phase weights $C{ { 
\mbox{\boldmath$\alpha$\unboldmath}
}
\choose {
\mbox{\boldmath$\beta$\unboldmath}
} 
}$ for general sectors 
\beqn
\mbox{\boldmath$\alpha$\unboldmath}
= \sum_{j=1}^{p} a_j \mV_j \in \Xi ,\,\,\,\,
\mbox{\boldmath$\beta$\unboldmath}
  = \sum_{i=1}^{p} b_i \mV_i \in \Xi 
\label{bab} 
\eeqn
can be expressed in terms of the components in the
$p\times p$-dimensional matrix $\mk$ for the basis vectors:
\beqn
C{{
\mbox{\boldmath$\alpha$\unboldmath}
}\choose {
\mbox{\boldmath$\beta$\unboldmath}
}}
= (-1)^{s_{\alpha}+s_{\beta}}
        {\rm exp}\{ \pi i \sum_{i,j} b_i( k_{i,j} - \half \mV_i \cdot\mV_j) a_j \}.  
\label{kab}
\eeqn

Modular invariance simultaneously imposes constraints on the basis vectors $\mV_i$ 
and on components of the GSO projection matrix $\mk$:
\beqn
k_{i,j} + k_{j,i} &=& \half\, \mV_i\cdot \mV_j\, \mod{2}
\label{kseta1}\\
N_j k_{i,j} &=& 0 \, \mod{2}   
\label{kseta2}\\
k_{i,i} + k_{i,1}&=& - s_i + \fourth\, {\mV_i}\cdot {\mV_i}\, \mod{2}.
\label{kseta3}
\eeqn

The dependence upon the $k_{i,j}$
can be removed from equations (\ref{kseta1}-\ref{kseta3}),
after appropriate integer multiplication, 
to yield three constraints on the $\mV_i$:
\beqn
&&N_{i,j} \mV_i\cdot \mV_j =  0 \, \mod{4} \label{ksetb1}\\
&&N_{i}   \mV_i\cdot \mV_i =  0 \, \mod{8} \label{ksetb2}\\
&&{\rm The~number~of~real~fermions~simultaneously~periodic}\nolabel\\
&&{\rm for~any~three~basis~vectors~is~even.}\label{ksetb3}
\eeqn
$N_{i,j}$ is the lowest common multiple of $N_i$ and $N_j$. 
(\ref{ksetb3}) still applies when two or more of the basis vectors are identical.
Thus, each basis vector must have an even number of real periodic fermions.

The physical massless states in the Hilbert space of a given sector
{\boldmath$\alpha$\unboldmath}$\in{\Xi}$, are obtained by acting on the vacuum with
bosonic and fermionic operators and by
applying generalized GSO projections. The $U(1)$
charges for the Cartan generators of the unbroken gauge group 
are in one to one correspondence with the $U(1)$
currents ${f^*_m}f_m$ for each complex fermion $f_m$, and are given by:
\beq
{Q^{
\mbox{\boldmath$\alpha$\unboldmath}
}_m = {1\over 2}\alpha_m + F^{
\mbox{\boldmath$\alpha$\unboldmath}
}_m},
\label{u1charges}
\end{equation}
where $\alpha_m$ is the boundary condition of the worldsheet fermion $f_m$
in the sector {\boldmath$\alpha$\unboldmath}, and
$F^{
\mbox{\boldmath$\alpha$\unboldmath}
}_m$ is a fermion number operator counting each mode of
$f_m$ once and of $f_m^{*}$ minus once. Pseudo-charges for non-chiral (i.e., with 
both left- and right-moving components) real Ising fermions $f_m$ can be similarly 
defined, with $F_m$ counting each real mode $f$ once.
 
For periodic fermions,
$\alpha_m =1$, the vacuum is a spinor representation of the Clifford
algebra of the corresponding zero modes.
For each periodic complex fermion $f_m$
there are two degenerate vacua ${\vert +\rangle},{\vert -\rangle}$ ,
annihilated by the zero modes $(f_m)_0$ and
$(f_m^{*})_0$ and with fermion numbers 
$F^{
\mbox{\boldmath$\alpha$\unboldmath}
}_m =0,-1$, respectively.

The boundary basis vectors $\mV_j$
generate the set of GSO projection operators for physical states from all sectors
{\boldmath$\alpha$\unboldmath}$ \in \Xi$.
In a given sector {\boldmath$\alpha$\unboldmath}, the surviving states are those that
satisfy the GSO equations imposed by all $\bV_j$ and determined by the 
$k_{j,i}$'s:
\beqn
 \mV_j\cdot {\mF}^{
\mbox{\boldmath$\alpha$\unboldmath}
} = \left(\sum_i k_{j,i} a_i\right) + s_j
    - \half\, \mV_j\cdot {
\mbox{\boldmath$\alpha$\unboldmath}
}\, \mod{2} ,
\label{gso1-a}
\eeqn
or, equivalently,
\beqn
\bV_j\cdot {\mQ}^{\mbox{\boldmath$\alpha$\unboldmath}
} 
= \left(\sum_i k_{j,i} a_i\right) + s_j\, \mod{2}.
\label{gso}
\eeqn

For a given set of basis vectors, the independent GSO matrix components
are $k_{1,1}$ and $k_{i,j}$, for $i>j$.    
This GSO projection constraint, when combined with equations (\ref{kseta1}-\ref{kseta3})
form the free fermionic re-expression of the even, self-dual  
modular invariance constraints for bosonic lattice models.
The masses, $m^2 = m^2_L + m^2_R$ (with $m^2_L = m^2_R$) of physical states can also be expressed as a simple function of the charge vector,
\beqn
\alpha^{'} m^2_L &=& -\half + \half {\mQ}^{\mbox{\boldmath$\alpha$\unboldmath}}_L\cdot 
                                    {\mQ}^{\mbox{\boldmath$\alpha$\unboldmath}}_L \label{m2l}\\
\alpha^{'} m^2_R &=& -1     + \half {\mQ}^{\mbox{\boldmath$\alpha$\unboldmath}}_R\cdot 
                                    {\mQ}^{\mbox{\boldmath$\alpha$\unboldmath}}_R \label{m2r}.
\eeqn

The superpotential for the physical states can be determined to all order.
Couplings are computable for any order in the free fermionic models, using 
conformal field theory vertex operators. The coupling constant can be expressed in terms of an $n$-point 
string amplitude $A_n$.
This amplitude is proportional to a world-sheet integral $I_{n-3}$ of the correlators of the 
$n$ vertex operators
$V_i$ for the fields in the superpotential terms \cite{Kalara:1990fb},
\beqn
A_n  &=&  \frac{g}{\sqrt{2}}(\sqrt{8/\pi})^{n-3}C_{n-3}I_{n-3}/(\MS)^{n-3}\,\, . 
\label{ainta}
\eeqn  

The integral has the form,
\beqn
I_{n-3}
     &=& \int d^2 z_3\cdots d^2z_{n-1}\,\,  \vev{V_1^f(\infty) V_2^f(1) V^b_3(z_3) \cdots V^b_{n-1}(z_{n-1}) V^b_{n}(0)}  
\label{intv}\\
     &=& \int d^2 z_3\cdots d^2z_{n-1}\,\, f_{n-3}(z_1=\infty,z_2=1,z_3,\cdots,z_{n-1},z_{n}=0),
\label{intf}
\eeqn
where $z_i$ is the worldsheet coordinate of the fermion (boson) vertex operator $V^f_i$ ($V^b_i$) of the $i^{\rm th}$ string state. 
$C_{n-3}$ is an ${\cal{O}}(1)$ coefficient that includes 
renormalization factors in the operator product expansion of the string vertex operators 
and target space gauge group Clebsch--Gordon coefficients.
$SL(2,C)$ invariance is used to fix the location of three of the
vertex operators at $z= z_\infty, 1, 0$. When $n_v$ of the fields 
$\prod_{i=1}^l\X_{i}$ take on VEVs, $\vev{\prod_{i=1}^{n_v}\X_{i}}$, then the
coupling constant for the effective $n_e= (n-n_v)$-th order term becomes
$A^{'}_{n_e}\equiv A_{n} \vev{\prod_{i=1}^{n_v}\X_{i}}$.

In order to simply determine whether couplings are non-zero, 
worldsheet selection rules (in addition to the demand of gauge invariance) 
predicting nonvanishing correlators were formulated
in \cite{Kalara:1990fb,Rizos:1991bm}. With $n$ designating the total
number of fields in a term, these rules are economically summarized as: \cite{Cleaver:1999cj}  
\begin{enumerate}
\item Ramond fields must be distributed equally, mod $2$, among all
categories, {\it and}
\item For $n=3$, either\, :
\begin{enumerate}
\item There is $1$ field from each $R$ category.
\item There is $1$ field from each $NS$ category.
\item There are $2 R$ and $1 NS$ in a single category.
\end{enumerate}
\item For $n>3$\, :
\begin{enumerate}
\item There must be at least $4 R$ fields.
\item All $R$ fields may not exist in a single category. 
\item If $R = 4$, then only permutations of $(2_R, 2_R, n-4_{NS})$ are allowed. 
\item If $R > 4$, then no $NS$ are allowed in the maximal $R$ category (if one exists).  
\end{enumerate}
\end{enumerate}

\subsection{$SO(10)$ $\IZ_2\times\IZ_2$ (NAHE based) Compactifications}
\no
In the NAHE class of realistic free fermionic models, the 
boundary condition basis is divided into two subsets.
The first is the NAHE set \cite{Faraggi:1990ac}, which contains 
five boundary condition basis vectors denoted 
$\{{\bf 1},\bS,\bb_1,\bb_2,\bb_3\}$.
With `$\bo$' indicating Neveu-Schwarz boundary conditions
and `$\bone$' indicating Ramond boundary conditions, 
these vectors are as follows:
\beqn
 &&\begin{tabular}{c|c|ccc|c|ccc|c}
 ~ & $\psi^\mu$ & $\chi^{12}$ & $\chi^{34}$ & $\chi^{56}$ &
        $\bar{\psi}^{1,...,5} $ &
        $\bar{\eta}^1 $&
        $\bar{\eta}^2 $&
        $\bar{\eta}^3 $&
        $\bar{\phi}^{1,...,8} $ \\
\hline
\hline
      $\bone$ &  1 & 1&1&1 & 1,...,1 & 1 & 1 & 1 & 1,...,1 \\
         $\S$ &  1 & 1&1&1 & 0,...,0 & 0 & 0 & 0 & 0,...,0 \\
\hline
  ${\bb}_1$ &  1 & 1&0&0 & 1,...,1 & 1 & 0 & 0 & 0,...,0 \\
  ${\bb}_2$ &  1 & 0&1&0 & 1,...,1 & 0 & 1 & 0 & 0,...,0 \\
  ${\bb}_3$ &  1 & 0&0&1 & 1,...,1 & 0 & 0 & 1 & 0,...,0 \\
\end{tabular}
   \nonumber\\
   ~  &&  ~ \nonumber\\
   ~  &&  ~ \nonumber\\
     &&\begin{tabular}{c|cc|cc|cc}
 ~&      $y^{3,...,6}$  &
        $\bar{y}^{3,...,6}$  &
        $y^{1,2},\omega^{5,6}$  &
        $\bar{y}^{1,2},\bar{\omega}^{5,6}$  &
        $\omega^{1,...,4}$  &
        $\bar{\omega}^{1,...,4}$   \\
\hline
\hline
  {$\bone$} & 1,...,1 & 1,...,1 & 1,...,1 & 1,...,1 & 1,...,1 & 1,...,1 \\
   $\bS$    & 0,...,0 & 0,...,0 & 0,...,0 & 0,...,0 & 0,...,0 & 0,...,0 \\
\hline
${\bb}_1$ & 1,...,1 & 1,...,1 & 0,...,0 & 0,...,0 & 0,...,0 & 0,...,0 \\
${\bb}_2$ & 0,...,0 & 0,...,0 & 1,...,1 & 1,...,1 & 0,...,0 & 0,...,0 \\
${\bb}_3$ & 0,...,0 & 0,...,0 & 0,...,0 & 0,...,0 & 1,...,1 & 1,...,1 \\
\end{tabular}
\label{nahe}
\eeqn
with the following
choice of phases which define how the generalized GSO projections are to
be performed in each sector of the theory:
\beq
      C\left( \matrix{\bb_i\cr \bb_j\cr}\right)~=~
      C\left( \matrix{\bb_i\cr \bS\cr}\right) ~=~
      C\left( \matrix{\bone \cr \bone \cr}\right) ~= ~ -1~.
\label{nahephases}
\eeq

The remaining projection phases can be determined from those above through
the self-consistency constraints.
The precise rules governing the choices of such vectors and phases, as well
as the procedures for generating the corresponding spacetime particle
spectrum, are given in refs.~\cite{Kawai:1987ew,Antoniadis:1987wp}.

Without the $\bb_i$ sectors, the NAHE sector corresponds to $\IT^6$
torus compactification at the self-dual radius. 
The addition of the sectors $\bb_1$, $\bb_2$ and $\bb_3$ produce an effective
$\IT^6/(\IZ_2\times \IZ_2)$ compactification. $\bb_1$, $\bb_2$ and $\bb_3$
correspond to the three
twisted sectors of the $\IZ_2\times \IZ_2$ orbifold model:
$\bb_1$ provides the $\IZ^a_2$ twisted sector,  
$\bb_2$ provides the $\IZ^b_2$ twisted sector, and 
$\bb_3$ provides the $\IZ^a_2\otimes \IZ^b_2$ twisted sector.

After imposing the NAHE set, the resulting model has gauge
group $SO(10)\times SO(6)^3\times E_8$ and $N=1$
spacetime supersymmetry. The model contains 48
multiplets in the $\mathbf{16}$ representation of $SO(10)$,
16 from each twisted sector $\bb_1$, $\bb_2$ and $\bb_3$.

In addition to the spin 2 multiplets and the spacetime
vector bosons, the untwisted sector produces 
six multiplets in the vectorial $\mathbf{10}$ representation
of $SO(10)$ and a number of $SO(10)\times E_8$ singlets. 
As can be seen from Table (\ref{nahe}), the model
at this stage possesses a cyclic permutation symmetry
among the basis vectors $\bb_1$, $\bb_2$ and $\bb_3$,
which is also respected by the massless spectrum.

\subsection{Minimal Heterotic Superstring Standard Model}
\no
The second stage in the construction of these NAHE-based
free fermionic models consists
of adding three additional basis vectors,
denoted $\lbrace \balpha,\bbeta,\bgamma\rbrace$ 
 to the above NAHE set.
These three additional basis vectors
correspond to ``Wilson lines'' in the orbifold construction.

The allowed fermion boundary conditions in these additional basis vectors are
also constrained by the string consistency constraints, 
and must preserve modular invariance and worldsheet supersymmetry.
The choice of these additional basis vectors
nevertheless distinguishes
between different models and determines their low-energy properties.
For example, three additional vectors are
needed to reduce the number of massless
generations to three, one from each sector $\bb_1$, $\bb_2$, and $\bb_3$,
and the choice of their boundary conditions for the internal fermions
${\{y,\omega\vert{\bar y},{\bar\omega}\}^{1,\cdots,6}}$
also determines the Higgs doublet-triplet splitting and
the Yukawa couplings. These
low-energy phenomenological requirements therefore impose strong
constraints \cite{Faraggi:1991be,Faraggi:1991jr,Faraggi:1992fa} 
on the possible assignment of boundary conditions to the
set of internal worldsheet fermions
${\{y,\omega\vert{\bar y},{\bar\omega}\}^{1,\cdots,6}}$.

The additional sectors corresponding to the first MSSM model are:
\beqn
 &\begin{tabular}{c|c|ccc|c|ccc|c}
 ~ & $\psi^\mu$ & $\chi^{12}$ & $\chi^{34}$ & $\chi^{56}$ &
        $\bar{\psi}^{1,...,5} $ &
        $\bar{\eta}^1 $&
        $\bar{\eta}^2 $&
        $\bar{\eta}^3 $&
        $\bar{\phi}^{1,...,8} $ \\
\hline
\hline
  ${\balpha}$     &  1 & 1&0&0 & 1~1~1~1~1 & 1 & 0 & 0 & 0~0~0~0~0~0~0~0 \\
  ${\bbeta}$   &  1 & 0&0&1 & 1~1~1~0~0 & 1 & 0 & 1 & 1~1~1~1~0~0~0~0 \\
  ${\bgamma}$  &  1 & 0&1&0 &
		${1\over2}$~${1\over2}$~${1\over2}$~${1\over2}$~${1\over2}$
	      & ${1\over2}$ & ${1\over2}$ & ${1\over2}$ &
                ${1\over2}$~0~1~1~${1\over2}$~${1\over2}$~${1\over2}$~1 \\
\end{tabular}
   \nonumber\\
   ~  &  ~ \nonumber\\
   ~  &  ~ \nonumber\\
     &\begin{tabular}{c|c|c|c}
 ~&   $y^3{y}^6$
      $y^4{\bar y}^4$
      $y^5{\bar y}^5$
      ${\bar y}^3{\bar y}^6$
  &   $y^1{\omega}^6$
      $y^2{\bar y}^2$
      $\omega^5{\bar\omega}^5$
      ${\bar y}^1{\bar\omega}^6$
  &   $\omega^1{\omega}^3$
      $\omega^2{\bar\omega}^2$
      $\omega^4{\bar\omega}^4$
      ${\bar\omega}^1{\bar\omega}^3$ \\
\hline
\hline
$\balpha$& 1 ~~~ 0 ~~~ 0 ~~~ 1  & 0 ~~~ 0 ~~~ 1 ~~~ 0  & 0 ~~~ 0 ~~~ 1 ~~~ 0 \\
$\bbeta$ & 0 ~~~ 0 ~~~ 0 ~~~ 1  & 0 ~~~ 1 ~~~ 0 ~~~ 1  & 1 ~~~ 0 ~~~ 1 ~~~ 0 \\
$\bgamma$& 0 ~~~ 0 ~~~ 1 ~~~ 1  & 1 ~~~ 0 ~~~ 0 ~~~ 1  & 0 ~~~ 1 ~~~ 0 ~~~ 0 \\
\end{tabular}
\label{fnymodel}
\eeqn
with corresponding generalized GSO coefficients:
\beqn
&&C\left(\matrix{\balpha\cr
                                    \bb_j,\bbeta\cr}\right)=
-C\left(\matrix{\balpha\cr
                                    {\bone}\cr}\right)=
-C\left(\matrix{\bbeta\cr
                                    {\bone}\cr}\right)=
C\left(\matrix{\bbeta\cr
                                    \bb_j\cr}\right)=\nonumber\\
&&-C\left(\matrix{\bbeta\cr
                                    \bgamma\cr}\right)=
C\left(\matrix{\bgamma\cr
                                    \bb_2\cr}\right)=
-C\left(\matrix{\bgamma\cr
                                    \bb_1,\bb_3,\balpha,\bgamma\cr}\right)=
-1\label{fnycs}
\eeqn
$(j=1,2,3),$
(The remaining GSO coefficients are specified by modular invariance and spacetime
supersymmetry.) 

The full massless spectrum of the model, together
with the quantum numbers under the right-moving
gauge group, are given in ref.~\cite{Faraggi:1989ka}.

\subsection{Initial Gauge Group}
\no
Prior to any scalar fields receiving non-zero VEVS, the observable gauge group
consists of the universal $SO(10)$ subgroup,
$SU(3)_C\times SU(2)_L\times U(1)_C\times U(1)_L$,
generated by the five complex worldsheet fermions
${\bar\psi}^{1,\cdots,5}$, and six observable horizontal,
flavor-dependent, Abelian symmetries $U(1)_{1,\cdots,6}$, generated by
$\{{\bar\eta}^1,{\bar\eta}^2,{\bar\eta}^3,{\bar y}^3{\bar y}^6,
{\bar y}^1{\bar\omega}^6,{\bar\omega}^1{\bar\omega}^3\}$,
respectively. The hidden sector gauge group is
the $E_8$ subgroup of 
$(SO(4)\sim SU(2)\times SU(2))\times SU(3)\times U(1)^4$, 
generated by ${\bar\phi}^{1,\cdots,8}$.

The weak hypercharge is given by 
\beq
U(1)_Y={1\over3}U(1)_C\pm{1\over2}U(1)_L,
\label{weakhyper}
\eeq
which has the standard effective level $k_1$ of $5/3$,
necessary for MSSM unification at $M_U$.\footnote{The sign ambiguity in eq.\  (\ref{weakhyper})
can be understood in terms of the two alternative embeddings of $SU(5)$ 
within $SO(10)$ that produce either the standard or flipped $SU(5)$.   
Switching signs in (\ref{weakhyper}) flips the representations,
$(e_L^c,u_L^c,h)\leftrightarrow (N_L^c,d_L^c,{\bar h})$.
In the case of $SU(5)$ string GUT models,
only the ``--'' (i.e., flipped version) is allowed, 
since there are no massless matter adjoint representations, which 
are needed to break the \NA gauge symmetry of the
unflipped $SU(5)$, but are not needed for the flipped version.
For MSSM-like strings, either choice of sign is allowed
since the GUT \NA symmetry is broken directly at the string level.} 
The ``+'' sign was chosen for the hypercharge definition in \cite{Faraggi:1989ka}.
\cite{Cleaver:1998sa} showed  that the choice of the sign
in eq.\  (\ref{weakhyper}) has interesting consequences
in terms of the decoupling of the exotic fractionally charged states.
The alternate sign combination of $U(1)_C$ and $U(1)_L$ is
orthogonal to $U(1)_Y$ and denoted by $U(1)_{Z^\prime}$.
Cancellation of the Fayet-Iliopoulis (FI) term by directions that are 
$D$-flat for all of the non-anomalous $U(1)$ requires that at least one
of the $U(1)_Y$ and $U(1)_{Z'}$ be broken. 
Therefore, viable phenomenology forced $SU(3)_C\times SU(2)_L\times U(1)_Y$
to be the unbroken $SO(10)$ subgroup below the string scale. 
This is an interesting example of how string dynamics may
force the $SO(10)$ subgroup below the string scale
to coincide with the Standard Model gauge group.

\subsection{Massless Matter}
\no
The full massless spectrum of the model, together
with the quantum numbers under the right-moving
gauge group, were first presented in ref.~\cite{Faraggi:1989ka}.
In the model, each of the three 
sectors $\bb_1$, $\bb_2$, and $\bb_3$ produce one generation in the 16 representation,
($\Q{i}$, $\uc{i}$, $\dc{i}$, $\L{i}$, $\ec{i}$, $\Nc{i}$),
of $SO(10)$ decomposed under $SU(3)_C\times SU(2)_L\times U(1)_C\times U(1)_L$,
with charges under the horizontal symmetries.

In addition to the gravity and gauge multiplets and several singlets,  
the untwisted Neveu-Schwarz (NS) sector produces three pairs of 
electroweak 
scalar doublets $\{h_1, h_2, h_3, {\bar h}_1, {\bar h}_2, {\bar h}_3\}$.
Each NS electroweak doublet set $(h_i,\bar{h}_i)$ may be 
viewed as a pair of Higgs with the potential to give renormalizable 
(near EW scale) mass to the corresponding $\bb_i$-generation of MSSM matter. 
Thus, to reproduce the MSSM states and generate a viable three 
generation mass 
hierarchy,
two out of three of these Higgs pairs must become massive near the string/FI 
scale.
The twisted sector provides some additional $SU(3)_C\times SU(2)_L$
exotics:
one $SU(3)_C$ triplet/antitriplet pair $\{\H{33},\,\, \H{40}\}$;  
one $SU(2)_L$ up-like doublet, $\H{34}$, and one down-like doublet, $\H{41}$;
and two pairs of vector-like $SU(2)_L$ doublets, 
$\{\V{45},\,\, \V{46}\}$ and $\{\V{51},\,\, \V{52}\}$,  
with fractional electric charges $Q_e= \pm\half$.
$\h{4}\equiv \H{41}$ and $\bh{4}\equiv \H{34}$ play the role of 
a fourth pair of MSSM Higgs. Hence, all exotics form vector-like pairs with regard to 
MSSM-charges, a generic requirement for their decoupling.

Besides the anti-electrons $\ec{i}$ and neutrino singlets $\Nc{i}$, 
the model contains another 57 non-Abelian singlets.
16 of these carry electric charge and 41 do not. 
The set of 16 are twisted sector states,\footnote{Vector-like 
representations of the hidden sector are denoted by a ``V'', 
while chiral representations are denoted by a ``H''. 
A superscript ``s'' indicates a \NA singlet.}
eight of which carry $Q_e= \half$,
$\{\Hs{3}, \Hs{5}, \Hs{7}, \Hs{9}, 
 \Vs{41}, \Vs{43}, \Vs{47}, \Vs{49}\}$,
and another eight of which carry $Q_e= -\half$,
$\{\Hs{4}, \Hs{6}, \Hs{8}, \Hs{10}, 
  \Vs{42}, \Vs{44}, \Vs{48}, \Vs{50})$.

Three of the 41 $Q_e= 0$ states, 
$\{\Phi_1,\Phi_2,\Phi_3 \}$,
are the completely uncharged moduli from the NS sector. 
Another fourteen of these singlets
form vector-like pairs, 
$(\p{12},\pb{12})$, $(\p{23},\pb{23})$,
$(\p{13},\pb{13})$, $(\p{56},\pb{56})$, 
$(\pp{56},\ppb{56})$, $(\p{4},\pb{4})$,
$(\pp{4},\ppb{4})$,
possessing charges of equal magnitude, but opposite sign, 
for all local Abelian symmetries. The remaining 24 $Q_e= 0$ singlets,
$\{\Hs{15},\, \Hs{16},\, \Hs{17},\, \Hs{18},\, \Hs{19},\, 
   \Hs{20},\, \Hs{21},\, \Hs{22},\, \Hs{29},\,
   \Hs{30},\, \Hs{31},\, \Hs{32},\, \Hs{36},\,
   \Hs{37},\, \Hs{38},\, \Hs{39},$ 
and $\{\Vs{1},\,  
   \Vs{2},\,  \Vs{11},\, \Vs{12},\, \Vs{21},\, 
   \Vs{22},\, \Vs{31},\, \Vs{32} \}$,
are twisted sector states carrying
both observable and hidden sector Abelian charges. 

The model contains 34 hidden sector \NA states, 
all of which also carry both observable and hidden $U(1)_i$ charges: 
Five of these are $SU(3)_H$ triplets,
$\{\H{42},\, \V{4},\, \V{14},\, \V{24},\, \V{34} \}$, 
while another five are antitriplets, 
$\{\H{35},\, \V{3},\, \V{13},\, \V{23},\, \V{33} \}$.
The remaining hidden sector states are  
12 $SU(2)_H$ doublets, 
$\{\H{1},\, \H{2},\,
\H{23},\, \H{26},\, \V{5},\, \V{7},\, \V{15},\, \V{17},\,  
\V{25},\, \V{27},\, \V{39},\, \V{40} \}$
and a corresponding 12 $SU(2)^{'}_H$ doublets,
$\{\H{11},\, \H{13},\,
\H{25},\, \H{28},\, \V{9},\, \V{10},\, \V{19},\, \V{20},\,  
              \V{29},\, \V{30},\, \V{35},\, \V{37} \}$.
The only hidden sector NA states with non-zero $Q_e$ (in half-integer units) 
are the four of the hidden sector doublets, $\H{1}$, $\H{2}$, $\H{11}$, and $\H{13}$.   

For a string derived MSSM to result from this model, 
the several exotic MSSM-charged states must be eliminated from the LEEFT. Along with  
three linearly independent combinations of the $\h{i}$, and of the
$\hb{i}$, for $i=1,\, ...,\, 4$, the 26 states,
$\{ \H{33},\,  \H{40},\,  \V{45},\, 
     \V{46},\,  \V{51},\,  \V{52},\,                
      \H{1},\,   \H{2},\,   \H{11},\,  \H{12}\}$,
$\{  \Hs{3},\,  \Hs{5},\,  \Hs{7},\,  \Hs{9},\,   
     \Vs{41},\, \Vs{43},\, \Vs{47},\, \Vs{49}\}$, and
$\{  \Hs{4},\,  \Hs{6},\,  \Hs{8},\,  \Hs{10},\,  
     \Vs{42},\, \Vs{44},\, \Vs{48},\, \Vs{50} \}$   
must be removed.

Examination of the MSSM-charged state superpotential
shows that three out of four of each of the $h_i$ and $\bar{h}_i$ Higgs, 
and {\it all} of the 26 states above can 
be decoupled from the LEEFT via the terms,
\beqn
&& \p{12} \h{1} \hb{2} + \p{23} \h{3} \hb{2} 
+ \Hs{31} \h{2} \H{34} + \Hs{38} \hb{3} \H{41}+\nolabel\\ 
&&  \p{4} [ \V{45} \V{46} + \H{1} \H{2} ] +  
\pb{4} [ \Hs{3} \Hs{4} + \Hs{5} \Hs{6} + \Vs{41} \Vs{42} + \Vs{43} \Vs{44} ]
                                           +   \label{minveva}\\
&& \pp{4}  [ \V{51} \V{52} + \Hs{7} \Hs{8} + \Hs{9} \Hs{10} ]+
\ppb{4} [ \Vs{47} \Vs{48} + \Vs{49} \Vs{50} +\H{11} \H{13}]+
\nolabel\\
&& \p{23} \Hs{31} \Hs{38} [ \H{33} \H{40} + \H{34} \H{41} ]\,\, .\nolabel
\eeqn

This occurs when all states in the set 
\beqn
\{\p{4},\,\,  \pb{4},\,\, \pp{4},\,\, \ppb{4},\,\,
 \p{12},\,\, \p{23},\,\, \Hs{31},\,\, \Hs{38} \},
\label{minvev}
\eeqn
take on near string scale VEVs through FI term anomaly cancellation. 
All but one of the dominant terms in (\ref{minveva}) are of third order 
and will result in unsuppressed FI scale masses, while the remaining term 
is of fifth order, giving a mass suppression of $\frac{1}{10}$ for $\H{33}$ and $\H{40}$. 

\subsection{Anomalous $U(1)$}
\no
All known quasi-realistic chiral three generation  
$SU(3)_C\times SU(2)_L \times U(1)_Y$
heterotic models, of lattice, orbifold, or free fermionic construction,
contain an anomalous local $U(1)_A$ \cite{Kobayashi:1996pb,Cleaver:1997rk}. 
Anomaly cancellation provides a means 
by which VEVs naturally appear. While non-perturbatively chosen, some perturbative possibilities may exist that provide the needed effective mass terms to eliminate all unwanted MSSM exotics from models generically containing them.

An anomalous $U(1)_A$ 
has non-zero trace of its charge over the massless states of the LEEFT, 
\beqn
         \Tr Q^{(A)} \neq 0 .
\label{audef}
\eeqn

String models often appear to have not just one, but several anomalous
Abelian symmetries $U(1)_{A,i}$ ($i= 1$ to $n$), 
each with $\Tr Q^{(A)}_i \neq 0$. 
However, there is always a rotation
that places the entire anomaly into a single $U(1)_{A}$,
uniquely defined by 
\beq
         U(1)_{\rm A} \equiv c_A\sum_{i=1}^n \{\Tr Q^{(A)}_{i}\}U(1)_{A,i},
\label{rotau}
\eeq
with $c_A$ a normalization coefficient. 
Then $n-1$ traceless $U(1)'_j$ are formed from linear combinations of
the $n$ $U(1)_{A,i}$ that are orthogonal to $\UA$.

Prior to rotating the anomaly into a single $\UA$, 
six of the FNY model's twelve $U(1)$ symmetries are anomalous:
Tr${\, U_1=-24}$, Tr${\, U_2=-30}$, Tr${\, U_3=18}$,
Tr${\, U_5=6}$, Tr${\, U_6=6}$ and  Tr${\, U_8=12}$.
Thus, the total anomaly can be rotated into a single 
$U(1)_{\rm A}$ defined by 
\beq
U_A\equiv -4U_1-5U_2+3U_3+U_5+U_6+2U_8.
\label{anomau1infny}
\eeq

Five mutually orthogonal $U^{'}_s$, for $s = 1,\, ...,\, 5$ are then formed that 
are all traceless and orthogonal to $U_A$.
A set of vacuum expectations values (VEVs) will automatically appear
in any string model with an anomalous $\UA$ as a result of the 
Green-Schwarz-Dine-Seiberg-Witten anomaly cancellation mechanism \cite{Dine:1987xk,Atick:1987gy}.
Following the anomaly rotation of \eq{rotau}, the universal Green-Schwarz 
(GS) relations,
\beqn
\frac{1}{k_m k_A^{1/2}}\mathop{\Tr}_{G_m}\, 
T(R)Q_A &=& \frac{1}{3k_A^{3/2}}\Tr Q_A^3
                      = \frac{1}{k_i k_A^{1/2}}\Tr Q_i^2 Q_A 
                      =\frac{1}{24k_A^{1/2}}\Tr Q_A 
\nolabel\\
&\equiv& 8\pi^2 \delta_{\rm GS} \, ,
\label{gsa}\\
\frac{1}{k_m k_i^{1/2}}\mathop{\Tr}_{G_m}\, 
T(R)Q_i &=& \frac{1}{3k_i^{3/2}}\Tr Q_i^3
                     = \frac{1}{k_A k_i^{1/2}}\Tr Q_A^2 Q_i 
                     = \frac{1}{(k_i k_j k_A)^{1/2}}\Tr Q_i Q_{j\ne i} Q_A 
\nolabel\\ 
        &=&\frac{1}{24k_i^{1/2}}\Tr Q_i = 0 \, ,
\label{gsna}
\eeqn
where $k_m$ is the level of the \NA gauge group $G_m$ and
$2 T(R)$ is the index of the representation $R$ of $G_m$, defined by
\beq
\Tr\, T^{(R)}_a T^{(R)}_b = T(R) \delta_{ab}\, ,
\label{tin}
\eeq
removes all Abelian triangle anomalies except those involving
either one or three $U_A$ gauge bosons.\footnote{The GS relations are a by-product of modular invariance constraints.}

The standard anomaly cancellation mechanism breaks $U_A$ and, 
in the process, generates a FI $D$-term, 
\beq
      \eps\equiv \frac{g^2_s M_P^2}{192\pi^2}\Tr Q^{(A)}\, ,
\label{fidt}
\eeq
where $g_{s}$ is the string coupling
and $M_P$ is the reduced Planck mass, 
$M_P\equiv M_{Planck}/\sqrt{8 \pi}\approx 2.4\times 10^{18}$ . 

\subsection{Flat Direction Constraints}
\no
Spacetime supersymmetry is broken in a model
when the expectation value of the scalar potential,
\beqn
 V(\varphi) = \half \sum_{\alpha} g_{\alpha}^2 D_a^{\alpha} D_a^{\alpha} +
                    \sum_i | F_{\varphi_i} |^2\,\, ,
\label{vdef}
\eeqn
becomes non-zero. 
The $D$-term contributions in (\ref{vdef}) have the form,   
\beqn
D_a^{\alpha}&\equiv& \sum_m \varphi_{m}^{\dagger} T^{\alpha}_a \varphi_m\,\, , 
\label{dtgen} 
\eeqn
with $T^{\alpha}_a$ a matrix generator of the gauge group $g_{\alpha}$ 
for the representation $\varphi_m$, 
while the $F$-term contributions are, 
\beqn
F_{\Phi_{m}} &\equiv& \frac{\partial W}{\partial \Phi_{m}} \label{ftgen}\,\, . 
\eeqn

The $\varphi_m$ are the scalar field superpartners     
of the chiral spin-$\half$ fermions $\psi_m$, which together  
form a superfield $\Phi_{m}$.
Since all of the $D$ and $F$ contributions to (\ref{vdef}) 
are positive semidefinite, each must have 
a zero expectation value for supersymmetry to remain unbroken.

For an Abelian gauge group, the $D$-term (\ref{dtgen}) simplifies to
\beqn
D^{i}&\equiv& \sum_m  Q^{(i)}_m | \varphi_m |^2 \label{dtab}\,\,  
\eeqn
where $Q^{(i)}_m$ is the $U(1)_i$ charge of $\varphi_m$.  
When an Abelian symmetry is anomalous, 
the associated $D$-term acquires the FI term
(\ref{fidt}),
\beqn
D^{(A)}&\equiv& \sum_m  Q^{(A)}_m | \varphi_m |^2 
+ \eps \, .
\label{dtaban}  
\eeqn  
$g_{s}$ is the string coupling and $M_P$ is the reduced Planck mass, 
$M_P\equiv M_{Planck}/\sqrt{8 \pi}\approx 2.4\times 10^{18}$ GeV. 

The FI term breaks supersymmetry near the string scale,
$V \sim g_{s}^{2} \eps^2$, 
unless its can be cancelled by a set of scalar VEVs, $\{\vev{\varphi_{m'}}\}$, 
carrying anomalous charges $Q^{(A)}_{m'}$,
\beq
\vev{D^{(A)}}= \sum_{m'} Q^{(A)}_{m'} |\vev{\varphi_{m'}}|^2 
+ \eps  = 0\,\, .
\label{daf}
\eeq

To maintain supersymmetry, a set of anomaly-cancelling VEVs must 
simultaneously be $D$-flat 
for all additional Abelian and the non-Abelian gauge groups, 
\beq
\vev{D^{i,\alpha}}= 0\,\, . 
\label{dana}
\eeq

A non-trivial superpotential $W$ also imposes numerous constraints on allowed
sets of anomaly-cancelling VEVs, through the $F$-terms in (\ref{vdef}).
$F$-flatness (and thereby supersymmetry) can be broken through an 
$n^{\rm th}$-order $W$ term containing $\Phi_{m}$ when all of the additional 
fields in the term acquire VEVs,
\beqn
\vev{F_{\Phi_m}}&\sim& \vev{{\frac{\partial W}{\partial \Phi_{m}}}} 
         \sim \lambda_n \vev{\varphi}^2 (\frac{\vev{\varphi}}{\MS})^{n-3}\,\, ,
\label{fwnb2}
\eeqn
where $\varphi$ denotes a generic scalar VEV.
If $\Phi_{m}$ additionally has a VEV, then
supersymmetry can be broken simply by $\vev{W} \ne 0$.

$F$-flatness must be retained up to an order
in the superpotential that is consistent with observable sector
supersymmetry being maintained down to near the electroweak (EW) scale. 
However, it may in fact be {\it desirable} to allow such a term to escape at some elevated order, since it is known that supersymmetry does {\it not} survive down to `everyday' energies.
Depending on the string coupling strength, $F$-flatness cannot be broken
by terms below eighteenth to twentieth
order. As coupling strength increases,
so does the required order of flatness.

Generically, there are many more $D$-flat directions 
that are simultaneously $F$-flat  
to a given order in the superpotential   
for the LEEFT of a string model
than for the field-theoretic counterpart.
In particular, there are usually several $D$-flat directions 
that are $F$-flat to all order in a string model,  
but only flat to some finite, often low, order in the 
corresponding field-theoretic model. 
This may be attributed to the
string worldsheet selection rules, which impose 
strong constraints on allowed superpotential terms beyond gauge invariance.

\subsection{MSSM Flat Directions}
\no
The existence of an all-order $D$- and $F$-flat direction containing VEVs for 
all of the fields in (\ref{minvev}) was proved in \cite{Cleaver:1998sa}.
This $U(1)_A$ anomaly-cancelling flat direction
provided the {\it first} known example of a superstring
derived model in which, of all the 
$SU(3)_C\times SU(2)_L \times U(1)_Y$-charged states,
only the MSSM spectrum remains light below the string scale. 

From the first stage of a systematic study of $D$- and $F$-flat directions,
a total of four all-order flat directions formed solely of non-Abelian singlet fields with zero hypercharge
were found \cite{Cleaver:1999cj}. The scale of the overall VEV for all of these directions was computed to be
$|<\alpha>| \sim 10^{17}$ GeV.
Other than these four, no other directions were found to
be $F$-flat past $12^{\rm th}$ order. 
The resulting phenomenology of the all-order flat directions was presented in 
\cite{Cleaver:1999mw}.
For these directions renormalizable mass terms appeared for one complete
set of up-, down-, and electron-like fields and their conjugates.
However, the apparent top and bottom quarks did not appear in the 
same $SU(2)_L$ doublet. Effectively, these flat directions gave the
strange quark a heavier mass than the bottom quark. This inverted 
mass effect was a result of the field $\p{12}$ receiving a VEV
in all of the above direction. 
A search for MSSM-producing singlet flat directions 
that did not contain $\vev{\p{12}}$ was then performed. None were found. 
This, in and of itself, suggested the need for 
non-Abelian VEVs in more phenomenologically
appealing flat directions. 
Too few first and second generation down and electron mass terms
implied similarly. 

Thus, in the second stage of the study, hidden sector non-Abelian fields (with zero hypercharge) 
were also allowed to take on VEVs.
This led to discovery of additional all-order flat directions containing non-Abelian VEVs 
\cite{Cleaver:2000aa} with a similar overall VEV scale $\sim 10^{17}$ GeV.
Their phenomenology was explored in \cite{Cleaver:2001ab}. Several phenomenological
aspects improved when non-Abelian VEVs appeared.

All MSSM flat directions were found via a computer search that generated
combinations of maximally orthogonal $D$-flat basis directions \cite{Cleaver:1997cr}. 
{\it Maximally orthogonal} means that each of the basis directions contained at least one VEV unique to itself. Thus, unless a basis direction's defining VEV
was vector-like, the basis vector could only appear in a physical flat direction multiplied by positive (real) weights. The physical $D$-flat directions generated where required to minimally contain VEVs for the set of states,
$\{ \p{4},\,  \pb{4},\, \pp{4},\, \ppb{4},\, \p{12},\, \p{23},\, \Hs{31},\, \Hs{38} \}$
necessary for decoupling of all 32 SM-charged MSSM exotics, comprised of the  
three extra pairs of Higgs doublets and the 28 MSSM-charged exotics. 
The all-order flat directions were found to eliminate no less than seven of the extra non-anomalous $U(1)'$,
thereby greatly reducing the horizontal symmetries.

Stringent $F$-flatness \cite{Cleaver:1999cj} was demanded of the all singlet flat directions. 
This requires that the expectation value of each component of a given $F$-term 
vanish, rather than allowing elimination of an $F$-term via cancellation between 
expectation values of two or more components. 
Such stringent demands are clearly not necessary for $F$-flatness.
Total absence of all individual non-zero VEV 
terms can be relaxed: collections of such terms 
appear without breaking $F$-flatness, so long as the terms 
separately cancel among themselves in each $\vev{F_{\Phi_m}}$ and in $\vev{W}$. 
However, even when supersymmetry is retained at a given order in the superpotential 
via cancellation between several terms in a specific $F_{\Phi_m}$,
supersymmetry could well be broken at a slightly higher order.
Thus, stringent flatness enables 
all-order flat directions to be found without demanding precise fine-tuning among the field VEVs.

Non-Abelian VEVs offer one solution to the stringent $F$-flatness issue.  
Because non-Abelian fields contain more than one field component, 
{\it self-cancellation} of a dangerous $F$-term can sometimes occur along 
\NA directions. That is, for some directions it may be possible to maintain
``stringent'' $F$-flatness even when dangerous 
$F$-breaking terms appear in the stringy superpotential.
Since Abelian $D$-flatness constraints limit only VEV magnitudes, 
the gauge freedom of each group remains (phase freedom, in particular, is
ubiquitous) with which to attempt a cancellation between terms (whilst
retaining consistency with non-Abelian $D$-flatness).
However, it can often be the 
case that only a single term from $W$ becomes an offender in a given
$F$-term. If a contraction of \NA fields (bearing
multiple field components) is present it may be possible to effect a
self-cancellation that is still, in some sense, stringently flat. 

Self-cancellation was first demonstrated in \cite{Cleaver:2000aa}.
In the model, eighth order terms containing \NA
fields posed a threat to $F$-flatness of two \NA directions. 
It was showed that a set of \NA VEVs exist that is consistent 
with $D$-flat constraints and by which self-cancellation 
of the respective eighth order terms can occur. 
That is, for each specific set of \NA VEVs imposed by 
$D$-flatness constraints, 
the expectation value of the dangerous $F$-term is zero. 
Hence, the ``dangerous'' superpotential terms posed no problem
and two directions became flat to all finite order. 
 
The \NA fields taking on VEVs were the doublets of the two hidden sector
$SU(2)$ gauge symmetries, with doublet fields of only one of the $SU(2)_H$ taking on VEVs in 
a given flat directions. Each flat direction with field VEVs charged under one of the $SU(2)_{H}$ was matched by 
a corresponding flat direction with isomorphic field VEVs charged under the other $SU(2)_H$. 
Example directions wherein self-cancellation was not possible were compared to examples where
self-cancellation occurred. Rules for $SU(2)$ self-cancellation of dangerous $F$-terms 
were developed. As examples, one direction involving 
just the $SU(2)_H$ doublet fields $\{\H{23},\, \H{26},\, \V{40}\}$ was compared to
another containing both $SU(2)_H$ and $SU(2)'_H$ doublets: $\{\H{23}, \V{40}, \H{28}, \V{37}\}$.

In generic \NA flat directions, the norms of the VEVs of all fields are fixed by
Abelian $D$-term cancellation, whereas the signs of the VEV components of a \NA field
are fixed by non-diagonal mixing of the VEVs in the corresponding \NA $D$-terms (\ref{dtgen}). 
For the first direction, Abelian $D$-term cancellation required 
the ratio of the norms-squared of the $\H{23}$, $\H{26}$, $\V{40}$ VEVs to be $1:1:2$, while the 
\NA $D$-term cancellation required 
\beqn
\vev{D^{SU(2)_H}}=  
\vev{\H{23}^{\dagger} T^{SU(2)} \H{23} 
   + \H{26}^{\dagger} T^{SU(2)} \H{26}  
   + \V{40}^{\dagger} T^{SU(2)} \V{40}}= 0\,\, ,
\label{dfsu2}
\eeqn
where
\beqn
T^{SU(2)}\equiv \sum^{3}_{a= 1} T^{SU(2)}_a = 
\left ( 
\begin{array}{c c }
%row 1 
1 & 1-i \\
%row 2
1+i & -1 
\end{array} \right )\,\, .
\label{TSU2}
\eeqn

The only solutions (up to a $\alpha \leftrightarrow -\alpha$ transformation)
to (\ref{dfsu2}) are
\beqn
&&\vev{\H{23}} =\left ( 
\begin{array}{c}
%row 1 
\phm\alpha \\
%row 2
-   \alpha
\end{array} \right )\, , \quad 
\vev{\H{26}} =\left ( 
\begin{array}{c}
%row 1 
\phm\alpha \\
%row 2
-   \alpha
\end{array} \right )\,\, \quad 
%{\rm and}\quad
\vev{\V{40}} =\left ( 
\begin{array}{c}
%row 1 
\phm\sqrt{2} \alpha \\
%row 2
\phm\sqrt{2} \alpha
\end{array} \right )\, , 
\label{fdna1a} 
\eeqn
and
\beqn
&&\vev{\H{23}} =\left ( 
\begin{array}{c}
%row 1 
\phm\alpha \\
%row 2
\phm\alpha
\end{array} \right )\, , \quad 
\vev{\H{26}} =\left ( 
\begin{array}{c}
%row 1 
\phm\alpha \\
%row 2
\phm\alpha
\end{array} \right )\,\, \quad 
%{\rm and}\quad
\vev{\V{40}} =\left ( 
\begin{array}{c}
%row 1 
\phm\sqrt{2} \alpha \\
%row 2
-\sqrt{2}\alpha
\end{array} \right )\, . 
\label{fdna1b}
\eeqn

A ninth-order superpotential term
jeopardizes flatness of this \NA $D$-flat direction via,  
\beqn
\vev{F_{\V{39}}} & \equiv& \vev{\frac{\partial W}{\partial \V{39}}}
\label{v35a}\\
                 & \propto & \vev{\p{23}\pb{56}\pp{4}\Hs{31}\Hs{38}}
                             \vev{\H{23}\cdot\H{26} \V{40} + 
                                  \H{23}\H{26}\cdot \V{40} +
                                  \H{26}\H{23}\cdot \V{40} }\, .\nolabel\\
\label{v35b}
\eeqn

Self-cancellation of this $F$-term could occur if the \NA VEVs resulted in 
\beqn
\vev{\H{23}\cdot\H{26} \V{40} +  \H{23}\H{26}\cdot \V{40} 
   + \H{26}\H{23}\cdot \V{40} }=0\, .
\label{scna1}
\eeqn
However, neither (\ref{fdna1a}) nor (\ref{fdna1b}) are solution to this.

The contrasting, self-cancelling flat direction contains the \NA VEVs 
of $\H{23}$, $\V{40}$, $\H{28}$, $\V{37}$ with matching magnitudes.
The $SU(2)_H$ $D$-term,  
\beqn
\vev{D^{SU(2)_H}}=  
\vev{\H{23}^{\dagger} T^{SU(2)} \H{23} 
   + \V{40}^{\dagger} T^{SU(2)} \V{40}}= 0
\label{dfsu2sol}
\eeqn
has the two solutions  
\beqn
&&\vev{\H{23}} =\left ( 
\begin{array}{c}
%row 1 
\phm\alpha \\
%row 2
-   \alpha
\end{array} \right )\, , \quad 
\vev{\V{40}} =\left ( 
\begin{array}{c}
%row 1 
\phm\alpha \\
%row 2
\phm\alpha
\end{array} \right ), 
\label{fdna2a}
\eeqn
\no and
\beqn
&&\vev{\H{23}} =\left ( 
\begin{array}{c}
%row 1 
\phm\alpha \\
%row 2
\phm\alpha
\end{array} \right )\, , \quad  
\vev{\V{40}} =\left ( 
\begin{array}{c}
%row 1 
\phm\alpha \\
%row 2
-\alpha
\end{array} \right )\, .
\label{fdna2b}
\eeqn
(The $SU(2)'_H$ $D$-term solutions for $\H{28}$ and $\V{37}$
have parallel form.)

Flatness of this direction was threatened by an eighth-order superpotential
term through 
\beqn
\vev{F_{\V{35}}} & \equiv& \vev{\frac{\partial W}{\partial \V{35}}}
\label{v35e}\\
                 & \propto & \vev{\p{23}\pb{56}\Hs{31}\Hs{38}}
                             \vev{\H{23}\cdot\V{40}}\vev{\H{28}}\label{v35f}.
\eeqn

Either set of $SU(2)_H$ VEVs (\ref{fdna2a}) or (\ref{fdna2b})
results in $\vev{\H{23}\cdot\V{40}}=0$, which made this \NA direction flat to all finite order!

\subsection{Higgs $\mu$ \& Generation Mass Terms} 
\no
The all-order flat directions giving mass to all 
MSSM charged exotics
were also shown to generate string scale mass for all but one linear combination 
of the electroweak Higgs doublets $h_i$ and $\bar{h}_i$, for $i=1,\, ...,\, 4$ \cite{Cleaver:2001ab}.
Effective Higgs mass ``$\mu$-terms'' take the form of one $h_i$ field and one $\bar{h}_i$
plus one or more factors of field VEVs.
Collectively, they may be expressed in matrix form as the scalar contraction
$\h{i} M_{ij} \bh{j}$
with eigenstates $h$ and $\bar h$ formed from linear combinations of the $\h{i}$
and $\bh{i}$ respectively,\footnote{The possibility of linear  
combinations of MSSM doublets forming the physical Higgs is a feature 
generic to realistic free fermionic models.}  
\beqn
  h = \frac{1}{n_h}    \sum_{i=1}^{4} c_i h_i;
\quad\quad
\lh = \frac{1}{n_{\lh}} \sum_{i=1}^{4} \bc_i \bh{i}\, ,
\label{bhdef}
\eeqn
with normalization factors $n_h = {\sqrt{\sum_i (c_i)^2}}$,
and  $n_{\lh} = {\sqrt{\sum_i (\bc_i)^2}}$.  These combinations
will then in turn establish the quark and lepton mass matrices.

All MSSM-producing flat directions were shown to  
necessarily contain $\p{23}$, $\sH{31}$, and $\sH{38}$ VEVs. 
Together these three VEVs produce four (linearly dependent) terms in the Higgs
mass matrix: $\h{3}\bh{2} \vev{\p{23}}$, $\h{2}\bh{4} \vev{\H{31}}$,  
$\h{4}\bh{3} \vev{\H{38}}$, and $\h{4}\bh{4} \vev{\H{31}}$.
When these are the only non-zero terms in the matrix, 
the massless Higgs eigenstates
simply correspond to $c_1 = \bc_1 = 1$ and  $c_j = \bc_j = 0 $ for $j=2,3,4$.  
In this case all possible quark and lepton mass terms containing 
$\h{j}$ for $j\in \{2,3,4\}$, decouple 
from the low energy MSSM effective field theory. However, when one
or more of the $c_j$ or $\bc_j$ are non-zero, then some of these terms
are not excluded and provide addition quark and lepton mass terms.
In such terms, the Higgs components can be replaced by their corresponding
Higgs eigenstates along with a weight factor,
\beqn
h_i   \rightarrow \frac{c_i}{n_h} h;
\quad\quad 
\lh_i \rightarrow \frac{\bc_i}{n_{\lh}} \lh\, . \label{bhir}
\eeqn 

\setcounter{footnote}{0}
Thus, in string models such as this, two effects can contribute
to inter-generational (and intra-generational) mass hierarchies:
generic suppression factors of $\frac{\vev{\phi}}{\MP}$ in 
non-renormalizable effective mass terms and 
$\frac{c_i}{n_h}$ or $\frac{\bc_i}{n_{\lh}}$ 
suppression factors. This means a hierarchy of values among the 
${c_i}$
and/or among the ${\bc_i}$ holds the possibility of producing
viable inter-generational $m_t:m_c:m_u \sim 1: 7\times 10^{-3}: 3\times 10^{-5}$ mass ratios even when 
all of the quark and lepton mass terms are of renormalizable or very 
low non-renormalizable order. More than one generation
of such low order terms necessitates a hierarchy among the 
$c_i$ and $\bc_i$.\footnote{Generational hierarchy via suppression factor in Higgs components was first
used in free fermionic models of the flipped $SU(5)$ class \cite{Lopez:1989fb}.}

For any MSSM all-order flat direction, at least one generation must receive mass from renormalizable terms.
The up- and down-like renormalizable terms are 
\beqn
\bh{1} \Q{1 }  \uc{1}\, ,\quad
\h{2}  \Q{2 }  \dc{2}\, ,\quad  
\h{3}  \Q{3 }  \dc{3}\, .
\label{udrenorm}
\eeqn
Since $\bh{1}$ is either the only component or a primary component in
$\lh$, the top quark is necessarily contained in $\Q{1}$ and $\uc{1}$
is the primary component of the left-handed anti-top mass eigenstate.
Thus, the bottom quark must be the second component of $\Q{1}$. 
Since there are no renormalizable 
$\h{i} \Q{1} \dc{m}$ terms in eq.~(\ref{udrenorm}),
a bottom quark mass that is  hierarchically
larger than the strange and down quark masses requires that 
\beqn
\frac{|c_{j=2,3}|}{n_h}\ll 1\, .
\label{c23h}
\eeqn
Non-zero $c_{2,3}$ satisfying eq.\ (\ref{c23h}) could,
perhaps, yield viable strange or down mass terms.  

The first possible bottom mass term appears at fourth order, 
$\h{4}\Q{1 }\dc{3}\sH{21}$.
Realization of the bottom mass via this term requires
$h$ to contain a component of $\h{4}$ and for
$\sH{21}$ to acquire a VEV. Of all flat directions, only two, here denoted 
{\it FDNA1} and {\it FDNA2} give $\sH{21}$ a VEV. Of these two, only
{\it FDNA2} embeds part of $\h{4}$ in $h$.  

The physical mass ratio of the top and bottom quark is
of order $\sim 3\times 10^{-2}$. 
In free fermionic models there is  
no significant suppression to the  
effective third order superpotential coupling constant,
$\lambda_3^{\rm eff} = \lambda_4 \vev{\phi}$,
originating from a fourth order term \cite{Cvetic:1998gv}.
Hence, a reasonable top to bottom mass ratio would imply 
\beqn
\frac{|c_2|}{n_h},\frac{|c_3|}{n_h} \ll 
\frac{|c_4|}{n_h}\sim 10^{-2\,\,\, {\rm to}\,\,\, -3}
\label{c4ran}
\eeqn
when $\frac{|\bc_1|}{n_{\lh}}\sim 1$ and $\vev{h}\sim \vev{\bar{h}}$.
However, \cite{Cleaver:2001ab} showed that $\frac{|c_4|}{n_h}\gsim 10^{-3}$ value cannot be realized 
along any of the flat directions explored. 

\setcounter{footnote}{0}
The next possible higher order bottom mass terms do not occur 
until sixth order, all of which contain $h_j$, where  $j\in \{2,3,4\}$.
Beyond fourth order a suppression factor of 
$\frac{1}{10}$ per order is generally assumed \cite{Faraggi:1996pa}.
Thus, a sixth order down mass term would imply
$\frac{|c_j|}{n_h} \sim 1$, where  $j\in \{2,3,4\}$ as
appropriate, when $\frac{|\bc_1|}{n_{\lh}}\sim 1$. 
However, none of the flat directions 
transform any of such sixth order terms into mass terms \cite{Cleaver:2001ab}.

If not sixth order, then seventh order is the
highest order that could provide a sufficiently large bottom mass. 
There are no such seventh order terms containing $\h{1}$. However, 
$\h{2}$ is in 15 of these terms, one of which 
becomes a bottom mass term for {\it FDNA2}.
No seventh order terms containing $\h{3}$ or $\h{4}$ become $\mu$ terms for any of the flat directions.
Therefore, the only possible bottom quark mass terms 
resulting from the flat directions explored in \cite{Cleaver:2001ab} 
are the fourth order $\h{4}$ term and the seventh order $\h{2}$ terms.
However, no flat directions contain the VEVs required to transform any of the seventh or eighth
order terms into effective mass terms. 

While several flat directions can generate a particular ninth order 
Higgs $\mu$ term,
$\h{1}\bh{3}\vev{\Nc{3} \pp{4}\sH{15}\sH{30}\sH{31}\H{28}\cdot\V{37}}$,
{\it FDNA2} is the only one that 
simultaneously generates the fourth order bottom mass term. Thus, {\it FDNA2} was
singled out for detailed analysis.  

\cite{Cleaver:1999mw,Cleaver:2000aa} showed that a small $\pb{12}\ll$ FI-scale VEV 
produces superior quark and lepton mass matrix phenomenology. However, since $\pb{12}$ does not acquire a
VEV in any flat direction found, $\vev{\pb{12}}\ne 0$ was only allowed a second order effect at or below the 
MSSM unification scale. Thus, in the Higgs mass matrix, $\h{2}\hb{1}\vev{\pb{12}}$ was allowed to provide 
significantly suppressed mass term. When $\vev{\pb{12}}\sim 10^{-4}$ was assumed,
the numeric form of the {\it FDNA2} Higgs doublet mass matrix becomes (in string-scale mass units)
\beqn
M_{h_i,\hb{j}}
\sim{\left ( 
\begin{array}{cccc}
0 & 0 & 10^{-5} &  0 \\
10^{-4} & 0 & 0  & 1 \\ 
0 & 2 & 0  &  0 \\
0 & 0 & 1  & 10^{-1} 
\end{array} \right ).}
\label{mijb}
\eeqn

The massless Higgs eigenstates 
produced (by diagonalizing $M^{\dagger}M$)
are of order\footnote{In the 
limit of $\vev{\pb{12}}=0$, the $h$ eigenstate reduces to $\bh{1}$.} 
\beqn
 h  &=& \h{1}  +  10^{-7} \h{2}  - 10^{-5} \h{4},\label{hes}\\
\lh &=& \bh{1} +  10^{-6} \bh{3} - 10^{-4} \bh{4}\, .\label{bhes}
\eeqn

These Higgs eigenstates provide examples of how 
several orders of magnitude mass suppression factors can appear 
in the low order terms of a Minimal Standard Heterotic String Model (MSHSM).
Here, specifically, the $\h{4}$ coefficient in (\ref{hes}) can provide 
$10^{-5}$ mass suppression for one down-quark generation and one electron generation.
 When rewritten in terms
of the Higgs mass eigenstates, 
$\h{4} (\Q{i} \dc{j} + \L{i} \ec{j})$
contains a factor of $10^{-5} h (\Q{i} \dc{j} + \L{i} \ec{j})$. 
Similarly,             
the $\h{2}$ coefficient can provide $10^{-7}$ suppression. 
Further, the $\bh{4}$ coefficient in (\ref{bhes}) can provide 
$10^{-4}$ up like-quark mass suppression and            
the $\bh{3}$ coefficient a corresponding $10^{-6}$ suppression.

Under the assumption of the above Higgs (near) massless eigenstates, 
the quark and lepton mass matrices were calculated up to ninth order, the level at which
suppression in the coupling (assumed here to be $\sim {10}^{-5}$) is comparable
to that coming out of eqs.~(\ref{hes}, \ref{bhes}).  
The up-quark mass matrix contains only a single term
(corresponding to the top mass) when $\vev{\pb{12}}=0$, but develops an
interesting texture when $\vev{\pb{12}} \sim 10^{-4}$ FI-scale.
To leading order, the general form (in top quark mass units) is  
\beqn
M_{\Q{},\uc{}} 
&=&\left ( 
\begin{array}{ccc}
\hb{1} & .1 \hb{3}+ 10^{-3} \hb{4} & 0 \\
10^{-2}  \hb{3}+ 10^{-4} \hb{4} & 0 & 0 \\ 
0 & 0 & \hb{4} \\
\end{array} \right )
\sim \left ( 
\begin{array}{ccc}
1 & 10^{-6} & 0 \\
10^{-7} & 0 & 0 \\ 
0 & 0 & 10^{-4} \\
\end{array} \right ).\nolabel\\
&&\label{muijn}
\eeqn

The up-like mass eigenvalues are $1$, $10^{-4}$, and $10^{-13}$. 
A more realistic structure would appear for a $\hb{4}$ suppression 
factor of $10^{-2}$ rather than $10^{-4}$.   

The corresponding down-quark mass matrix has the form
\beqn
M_{\Q{},\dc{}}
={\left ( 
\begin{array}{ccc}
0 & 10^{-3}\h{2}+ 10^{-5} \h{4} & \h{4}\\
10^{-5} \h{2} & \h{2}+ .1 \h{4} & 10^{-5} \h{2} \\ 
0 & 10^{-4} \h{2}  & 0\\
\end{array} \right )
\sim \left ( 
\begin{array}{ccc}
0 & 10^{-9} & 10^{-5}  \\
10^{-11} & 10^{-6}  & 10^{-11} \\ 
0 & 10^{-10}  & 0 \\
\end{array} \right ).}\nolabel\\
&&\label{mdijn}
\eeqn

The resulting down-quark mass eigenvalues   
are $10^{-5}$, $10^{-6}$, and $10^{-15}$ 
(in top quark mass units).
This provides quasi-realistic
down and strange masses, but lacks a bottom
mass. Unfortunately, the down-like quark eigenstate corresponding
to a $10^{-5}$ mass is in $\Q{1}$, making it the bottom quark. 
The second and third generation masses would be more viable if
the $\h{4}$ suppression factor were $10^{-2}$ instead.    

The electron mass matrix takes the form
\beqn
M_{\L{},\ec{}}
=\left ( 
\begin{array}{ccc}
0 & 10^{-4}\h{2} & 0\\
10^{-4} \h{2} & \h{2}+ .1 \h{4} & 10^{-4} \h{2} \\ 
.1 \h{4} & 10^{-5} \h{2}  & 0\\
\end{array} \right )
\sim \left ( 
\begin{array}{ccc}
0 & 10^{-10} & 0  \\
10^{-10} & 10^{-6}  & 10^{-10} \\ 
10^{-5} & 10^{-11}  & 0 \\
\end{array} \right ).
\label{meijn}
\eeqn

The three corresponding electron-like 
mass eigenvalues $10^{-5}$, $10^{-6}$,
and $10^{-14}$.  
As with the down-like quark masses, more viable 
second and third generation electron masses would appear
if the $\h{4}$ suppression factor was only $10^{-2}$.    

\subsection{Hidden Sector Condensation}
\no
The free fermionic MSHSM model proved that non-Abelian VEVs can yield a three generation
MSSM model with no exotics while maintaining supersymmetry at the string/FI scale.
Supersymmetry must ultimately be broken slightly above the
electroweak scale, though. Along some of the flat directions (both Abelian and non-Abelian), this model
showed qualitatively how supersymmetry may be broken dynamically by 
hidden sector field condensation.
Two of the \NA flat directions break 
both of the hidden sector $SU(2)_H$ and $SU(2)^{'}_H$ gauge symmetries, 
but leave untouched the hidden sector $SU(3)_H$. 
Thus, condensates of $SU(3)_H$ fields can initiate supersymmetry breaking
\cite{Lopez:1995cs}.

The set of nontrivial $SU(3)_H$ fields is composed of five triplets, 
$\{\H{42},\, \V{4},\, \V{14},\, \V{24},\, \V{34}\}$,
and five corresponding anti-triplets,
$\{\H{35},\, \V{3},\, \V{13},\, \V{23},\,  \V{24}\}$.
Along these two \NA flat directions, singlet VEVs give unsuppressed FI-scale
mass to two triplet/antitriplet pairs 
via trilinear superpotential terms,\footnote{These mass terms even occur for the simplest Abelian flat directions.}
\beqn
\vev{\p{12}} \V{23} \V{24} + \vev{\p{23}} \V{33} \V{34}
\label{trilt}
\eeqn
and slightly suppressed mass to another triplet/anti-triplet pair
via a fifth order term, 
\beqn
\vev{\pb{56}\Hs{31}\Hs{38}} \H{42} \H{35}\,  .
\label{trilt2}
\eeqn
and a significantly suppressed mass to a fourth pair
via a tenth order term, 
\beqn
\vev{\p{23}\pb{56}\Hs{31}\Hs{38}\H{23}\V{40}\H{28}\V{37}} \V{4} \V{3}\,  .
\label{trilt3}
\eeqn

Before supersymmetry breaking, the last triplet/antitriplet pair, 
$\V{14}/\V{13}$, remain massless to all finite order. 

Consider a generic $SU(N_c)$ gauge group containing $N_f$ flavors 
of matter states in vector-like pairings 
$T_i \bar{T}_i$, $i= 1,\, \dots\, N_f$.
When $N_f < N_c$, the gauge coupling $g_s$, 
though weak at the string scale $\MS$, becomes strong
at a condensation scale defined by 
\beqn
\Lambda = \MP {\rm e}^{8 \pi^2/\beta g_s^2}\, ,
\label{consca}
\eeqn
where the $\beta$-function is given by,
\beqn
\beta = - 3 N_c + N_f\, .
\label{befn}
\eeqn

The $N_f$ flavors counted are only those that ultimately receive 
masses $m\ll \Lambda$.
Thus, in this model $N_c= 3$ and $N_f= 1$ 
(counting only the vector-pair, $\V{14}$ and $\V{13}$), which corresponds to  
$\beta = -8$ and results in an $SU(3)_H$ 
condensation scale
\beqn
\Lambda = {\rm e}^{-20}\MP 10^{10} \,\, {\rm GeV}. 
\label{consca2}
\eeqn

At this condensation scale $\Lambda$, the matter degrees of freedom are best
described in terms of the composite ``meson'' fields, $T_i \bar{T}_i$.
(Here the meson field is $\V{14}\V{13}$.)
Minimizing the scalar potential of the meson field induces
a VEV of magnitude,
\beqn
\vev{\V{14}\V{13}} = 
\Lambda^3 \left(\frac{m}{\Lambda}\right)^{N_f/N_c}\frac{1}{m}\, .
\label{ttv}
\eeqn

This results in an expectation value of
\beqn
\vev{W} = 
N_c \Lambda^3 \left(\frac{m}{\Lambda}\right)^{N_f/N_c}\, 
\label{wv}
\eeqn
for the non-perturbative superpotential.

Supergravity models are defined in terms of two functions,
the K\" ahler function, $G= K + {\rm ln}\, |W|^2$, where $K$ is the K\" ahler 
potential and $W$ the superpotential, and the gauge kinetic function $f$. 
These functions determine the supergravity interactions and the 
soft-supersymmetry
breaking parameters that arise after spontaneous breaking of supergravity, which is
parameterized by the gravitino mass $m_{3/2}$. The gravitino 
mass appears as a function of $K$ and $W$,
\beqn
m_{3/2} = \vev{{\rm e}^{K/2} W} .
\label{mkw}
\eeqn

Thus,
\beqn
m_{3/2} \sim \vev{{\rm e}^{K/2}} \vev{W} 
        \sim \vev{{\rm e}^{K/2}} N_c \Lambda^3 \left(\frac{m}{\Lambda}\right)^{N_f/N_c}\, . 
\label{mgeqa}
\eeqn

Restoring proper mass units explicitly gives,
\beqn
m_{3/2} \sim \vev{{\rm e}^{K/2}} N_c (\frac{\Lambda}{M_P})^3 
\left(\frac{m}{\Lambda}\right)^{N_f/N_c} M_P\, . 
\label{mgeqb}
\eeqn

Hence the meson field $\V{14}\V{13}$ 
acquires a mass of at least the supersymmetry breaking
scale. It was assumed in \cite{Cleaver:2001ab} that $m_{\V{14}\V{13}}\approx 1 $ TeV.
The resulting gravitino mass is 
\beqn
m_{3/2} &\sim& \vev{{\rm e}^{K/2}} 
\left(\frac{7\times 10^{9}\, {\rm GeV}}{2.4\times 10^{18}\,\, {\rm GeV}}
\right)^3 
\left(\frac{1000\, {\rm GeV}}{7\times 10^{9}\, {\rm GeV}}\right)^{1/3}
{2.4\times 10^{18}{\rm GeV}}
\nolabel\\  
&\approx& \vev{{\rm e}^{K/2}}\,  0.3\,\, {\rm eV} \, .
\label{mgeqd}
\eeqn

In standard supergravity scenarios, one generally obtains 
soft supergravity breaking parameters, such as scalar and gaugino masses and 
scalar interaction, that are comparable to the gravitino mass:
$m_o$, $m_{1/2}$, $A_o \sim m_{3/2}$.  
A gravitino mass of the order of the supersymmetry breaking scale 
would require $\vev{{\rm e}^{K/2}} \sim 10^{12}$ or $\vev{K}\sim 55$.
On the other hand, for a viable model, 
$\vev{{\rm e}^{K/2}}\sim {\cal{O}}(1)$ would 
necessitate a decoupling of local supersymmetry breaking (parametrized by
$m_{3/2}$) from global supersymmetry breaking (parametrized by
$m_{o}$, $m_{1/2}$). This is possible in the context of no-scale
supergravity \cite{Lahanas:1986uc}, endemic to weak coupled string models.

In specific types of 
no-scale supergravity, the scalar mass $m_o$ and the scalar coupling
$A_o$ have null values thanks to the associated form of the K\" ahler 
potential. Furthermore, the gaugino mass can go as a power of 
the gravitino mass,
$m_{1/2} \sim \left(\frac{m_{3/2}}{\MP}\right)^{1-\frac{2}{3}q} \MP$,
for the standard no-scale form of $G$ and a non-minimal gauge kinetic
function $f\sim {\rm e}^{-A z^q}$, where $z$ is a hidden sector moduli field 
\cite{Ellis:1984xe}.  
A gravitino mass in the range $10^{-5}$ eV $\lsim m_{3/2} \lsim 10^3$ eV
is consistent with the phenomenological requirement of $m_{1/2}\sim 100$ GeV 
for $\frac{3}{4}\gsim q \gsim \frac{1}{2}$.
%Note that decoupling between the local and global breaking of supersymmetry
%also appears to be realized in strongly coupled heterotic strings \cite{hora}.

\subsection{First MSHSM Model Summary and Inspired Research} 
\no
The NAHE-based free fermionic heterotic model presented initially in 
\cite{Cleaver:1998sa} was the first to succeed at removing {\it all} MSSM-charged exotic states
from the LEEFT. This model has since  
received the designation of Minimal Standard Heterotic String Model. 
The existence of the first Minimal Standard Heterotic String Models, which contain
solely the three generations of MSSM quarks and leptons and
a pair of Higgs doublets as the massless SM-charged states   
in the LEEFT, was a significant discovery.
The MSHSM offered the first potential realizations of possible equivalence, in the strong coupling limit, 
between the string scale and the minimal supersymmetric standard model
unification scale $M_U\approx 2.5\times 10^{16}$ GeV. 
This requires that the observable gauge group 
just below the string scale should be 
$SU(3)_C\times SU(2)_L\times U(1)_Y$ with charged spectrum 
consisting solely of the three MSSM generations and a pair of Higgs doublets.

In the MSHSM, masses of the exotic states were 
driven to be above the MSSM unification at around $\frac{1}{10}$ FI scale
by scalars taking on $D$- and $F$-flat VEVs to cancel an anomalous $U(1)_A$ (which are 
endemic to realistic free fermionic heterotic MSSM-like models. The first class of flat directions 
explored for this model involved only non-Abelian singlet fields 
\cite{Cleaver:1998sa,Cleaver:1999cj,Cleaver:1999mw}.
The MSHSM flat direction search was expanded in \cite{Cleaver:2000aa,Cleaver:2001ab}
to include hidden sector non-Abelian field VEVs. 
This provided for improved phenomenology, showing that 
quasi-realistic patterns to quark and charged-lepton mass matrices
can appear. 
This was a result of both the new non-Abelian VEVs
and the related structure of the physical Higgs doublets $h$ and $\bar{h}$.
In the more realistic free fermionic heterotic models, the physical 
Higgs can each contain up to four components with
weights vastly differing by several orders of magnitude. These components
generically have generation-dependent coupling strengths. Thus, mass suppression factors
for the first and second quark and lepton generations can appear
even at very low order as a result of the different weights of the
Higgs components.

In this model the top quark can receive a 
viable, unsuppressed mass (given realistic Higgs VEVs),
while masses for the bottom quark, most second generation and some 
first generation quarks and leptons were too small. This resulted from 
too small weight factors for one $h$ component and for one $\bar{h}$ component. 
Phenomenology would be significantly improved if
the $\h{4}$ and $\hb{4}$ weights in $h$ and $\bar{h}$
were larger by a factor of 100  
than their respective values of $10^{-5}$ and $10^{-4}$ in the 
best non-Abelian flat direction.

In this model the emergence of new techniques for the removal of
dangerous terms from $\vev{W}$ and from $\vev{F}$ was also observed.
Flatness of four \NA directions was lifted to all order by the vanishing of terms with
more than two \NA fields.  Non-Abelian self-cancellation within single terms
was developed in the MSHSM as a 
promising tool for extending the order to which a \NA direction is safe. 

The flat directions of this MSHSM present some interesting phenomenological
features such as multi-component physical Higgs that couple differently
to given quarks and leptons. 
Exploration of flat direction phenomenology 
for this model demonstrated that \NA VEVs are necessary (but perhaps not 
sufficient) for viable LEEFT MSSM phenomenology. This is in agreement with 
similar evidence presented suggesting this might be true 
as well for all MSHSM $\IZ_2\times \IZ_2$ models. 
Nonetheless, the stringent flat $F$- and $D$-flat directions
producing this MSHSM do not themselves lead to
viable quark and lepton mass matrices. 
This implies significant worth to exploring 
the generic properties of \NA flat directions in $\IZ_2\times \IZ_2$ models  
that contain exactly the MSSM three generations and two Higgs doublets
as the only MSSM-charged fields in the LEEFT. 

One direction suggested is non-stringent
MSHSM directions flat to a finite order due to 
cancellation between various components in an $F$-term. 
While the absence of any non-zero terms from within $\vev{F_{\Phi_m}}$ and 
$\vev{W}$ is clearly sufficient for $F$-flatness along a given $D$-flat direction, 
such stringent demands was not necessary.
Total absence of these terms can be relaxed, so long as they appear in
collections which cancel among themselves in 
each $\vev{F_{\Phi_m}}$ and in $\vev{W}$. 
It is desirable to examine the mechanisms of such cancellations
as they can allow additional flexibility
for the tailoring of phenomenologically viable particle properties while
leaving SUSY inviolate.

This insufficiency in phenomenology from stringent flat directions (Abelian or otherwise) for this model, 
and from non-Abelian singlet vacua for a range of other models inspired further analysis of non-Abelian flat direction technology and self-cancellation from a geometrical framework \cite{Cleaver:2005vv}.
The geometrical perspective facilitates manipulations of \NA VEVs, such as treating superpotential contractions with multiple pairings, and examining the new possibility of self-cancellation between elements of a single term.
When expressed in geometric language, the process of describing valid solutions, or compatibilities between the $F$ and $D$ conditions, can become more accessible and intuitive. This could provide further assistance in closing the gap between string model building and low energy experimental evidence \cite{Cleaver:2005vv}.

From the geometric pint of view, \cite{Cleaver:2005vv} showed that non-Abelian $D$-term flatness 
translates into the imperative that the adjoint space representation of all expectation
values form a closed vector sum. Furthermore, the possibility emerges that a single seemingly dangerous
$F$-term might experience a self-cancellation among its components.
In \cite{Cleaver:2005vv} it was examined whether this geometric language can provide an intuitive and
immediate recognition of when the $D$ and $F$ conditions are simultaneously
compatible, as well as a powerful tool for their comprehensive classification.
Some initial success to this process was found.
Geometric interpretation of $F$ and $D$ flat directions was made concrete by an examination of the specific 
cases of $SU(2)$ and $SO(2n)$ symmetries. As a rank $1$ group with a number of 
generators equal to its fundamental dimension $3$, $SU(2)$ represents the simplest specific case on which to initiate discussion.  On the other hand, $SO(2n)$ introduces new complications by way of higher rank groups and an adjoint space of dimension greater than the fundamental.\footnote{As an interesting aside, note that although the fields under consideration were all spacetime scalars, superpotential terms can inherit an 
induced symmetry property from the analytic rotationally invariant contraction form of the group under study.
$SU(2)$ will have a ``fermionic'' nature, with an antisymmetric contraction, while that of $SO(2n)$ will be symmetric, or ``bosonic''.} A geometric interpretation for simultaneity of $D$- and $F$-flatness for $SU(2)$-charged 
fields was introduced. 

Solutions to \NA $D$-flatness were found to
not necessarily be associated with gauge invariant superpotential terms. A counter example was presented in
\cite{Cleaver:2005vv} for which the VEVs within a single $\mathbf 6$ provide $SO(6)$ $D$-flatness. The constraint equations for non-trivial $D$-flatness were developed for any number of $\mathbf 6$'s.\footnote{Development of systematic methods for geometrical analysis of the landscape of $D$- and $F$-flat 
directions begun in \cite{Cleaver:2005vv} offers a possibility for further understanding of the geometry of brane-anti-brane systems. 
This is related to the connection between 4D supergravity $D$-terms
of a string and a $D_{3+q}-\bar{D}_{3+q}$ wrapped brane/antibrane system. In this association, 
the energy of a $D_{3+q}/\bar{D}_{3+q}$ system appears as an FI $D$-term. An open string tachyon connecting
brane and anti-brane is revealed as an FI-cancelling Higgs field, and a $D_{1+q}$-brane produced in an  
annihilation between a $D_{3+q}$-brane and a $\bar{D}_{3+q}$-anti-brane is construed to be a $D$-term string 
\cite{Dvali:2003zh}.}

The discovery of an MSHSM in the 
neighborhood of the string/M-theory parameter space allowing  
free-fermionic description strongly suggested a search for 
more phenomenologically realistic MSHSMs in the free fermion
region. The concrete results, obtained in the analysis of a specific
model, highlighted the underlying, phenomenologically successful,
structure generated by the NAHE set and promoted further investigation of 
the string vacuum in the vicinity of this
model. That is, it warranted further investigation of $\IZ_2\times \IZ_2$ models in the vicinity
of the self-dual radius in the Narain moduli space. 

One variation among MSHSM models in this neighborhood is the number of 
pairs of Higgs doublets. Investigation in this direction were recently conducted in
\cite{Faraggi:2006qa}, wherein the removal of some (all) of the three or four 
extra pairs of Higgs doublets through modifications to free fermion boundary conditions
was studied. A general mechanism was developed that achieves Higgs reduction through
asymmetric boundary conditions between the left-moving $(y^i,w^i)$ and 
right-moving $(\bar{y}^i,\bar{w}^i)$ internal fermions for the 6 compactified dimensions 
\cite{Faraggi:2006qa}.
By this, the number of pairs of Higgs doublets was reduced to one. 
However a correlation was found between the reduction in Higgs doublets and the size of the
flat direction moduli space. The change in boundary conditions substantially reduces
the number of non-hypercharged scalar field singlets. 
This vast reduction eliminated any flat directions that could simultaneously cancel the anomalous
$U(1)$ FI term while retaining the MSSM gauge group. Thus, the
only stable supersymmetric vacua in the model of \cite{Faraggi:2006qa} destroys 
the MSSM gauge group. 
If this pattern holds for all MSHSM models in the neighborhood, then
the physical Higgs must be a composite state with different coupling strengths to each generation.
This result could lead to some interesting phenomenological predictions for LHC physics.    

%%%%%%%%%%%%%%%%%%%%%%%%%%%%%%%%%%%%%%%%%%%%%%%%%%%%%%%%%%%%%%%%%%%%%%%%%%%%%%%%%%%%%%%%%%%%%%%%%%%%%%%%%%%%%%%%%%%%%
% survey of additional MSSM models
%%%%%%%%%%%%%%%%%%%%%%%%%%%%%%%%%%%%%%%%%%%%%%%%%%%%%%%%%%%%%%%%%%%%%%%%%%%%%%%%%%%%%%%%%%%%%%%%%%%%%%%%%%%%%%%%%%%%%
\section{Subsequent String-Derived MSSM Models ...}
\no
Following the construction of the first string-derived model containing exactly the MSSM states in the observable sector, several other string models with this feature have been generated from different heterotic compactifications and from Type IIA \& IIB theories. In this section a representative, but nonetheless incomplete, set of such models are reviewed. The discussions are arranged by model class and compactification method, rather than chronologically.
The methods of construction and phenomenological features of these models are summarized. For further details of these example models, the original papers should be consulted.

%%%%%%%%%%%%%%%%%%%%%%%%%%%%%%%%%%%%%%%%%%%%%%%%%%%%%%%%%%%%%%%%%%%%%%%%%%%%%%%%%%%%%%%%%%%%%%%%%%%%%%%%%%%%%%%%%%%%%
% heterotic orbifolds at here 3
%%%%%%%%%%%%%%%%%%%%%%%%%%%%%%%%%%%%%%%%%%%%%%%%%%%%%%%%%%%%%%%%%%%%%%%%%%%%%%%%%%%%%%%%%%%%%%%%%%%%%%%%%%%%%%%%%%%%%
\subsection{From Heterotic Orbifolds}
\no
An additional $E_8\times E_8$ heterotic model with solely the MSSM spectrum in the observable sector 
of the LEEFT was recently constructed using a $\IZ^{'}_6 = \IZ_3 \otimes \IZ_2$ orbifold alternative to 
$\IZ_2 \otimes \IZ_2$ \cite{Buchmuller:2005jr,Buchmuller:2006ik}. 
In the $\IZ^{'}_6$ model the quarks and leptons appear as three $\mathbf{16}$'s of $SO(10)$, two of which are localized at fixed points of the orbifold and one for which some of the quarks and leptons are distributed across the bulk space (untwisted sector) and some are localized in twisted sectors. Like the free fermionic model above, this model initially contains MSSM-charged exotics. This model specifically contains 
4 $SU(3)_C$ ${\mathbf 3}/\bar{\mathbf 3}$ pairs,
5 pairs of $SU(2)$ doublets with $\pm \half$ hypercharge, and 7 extra pairs of doublets with zero hypercharge. 
Flat directions, formed from a set of 69 \NA singlets without hypercharge, are argued to exist that may give near string-scale 
mass to all MSSM-charged exotics (depending on the severity of mass suppression for $7^{\rm th}$ and $8^{\rm th}$ \
order terms), thereby decoupling the exotics from the low energy effective field theory. 
  
\subsubsection{Orbifold Construction} 
\no
In the orbifold construction, all internal degrees of freedom are bosonized. For the left-moving supersymmetric sector, the 6 compactified bosonic modes $X^i_L$, $i=1$ to $6$, are complexified into 
$Z^j_L = \frac{1}{\sqrt{2}}(X_L^{2j-1} + i X_L^{2j})$, $j=1$ to $3$. 
The accompanying six fermionic worldsheet modes $x^i$ are likewise paired 
into three complex fermions $\chi^j = \frac{1}{\sqrt{2}}\{x^{2j-1} + i x^{2j}\}$, $j= 1,\, ...,\, 3\}$, which are then replaced by a real bosonic mode $\phi^j$ via the relation $\chi^j = \exp(-2i \phi^j)$. 
The single left-moving complex spacetime fermion $\psi^0$ in lightcone gauge 
is also replaced by a real bosonic mode $\phi^0$.
The corresponding right-moving modes are the compactified complex
$Z^j_R = \frac{1}{\sqrt{2}}(X_R^{2j-1} + i X_R^{2j})$
and the 16 real bosonic modes $\bar{X}^I$, $I= 1,\, ...,\, 16$, for which the momenta are on the $E_8\otimes E_8$ root lattice.

The compactification lattice factorizes as 
$\IT^6 = \IT^2\otimes \IT^2\otimes \IT^2$,
in parallel to the free fermionic models.  
However, in the orbifold model the torus is  
$\IT^6 = R^6/\Lambda_{G_2\otimes SU(3)\otimes SO(4)}$, where $\Lambda_{G_2\otimes SU(3)\otimes SO(4)}$ is 
the lattice for the $G_2\otimes SU(3)\otimes SO(4)$ algebra. Therefore, complex coordinates $z^i$ on the torus are
identified when they differ by a lattice vector, $\mz \sim \mz + 2\pi \ml$ (where $\mz$ is a vector with three complex components $z_j$) and $\ml = m_a \me_a$, with $\me_a$ the basis vectors of the three lattice planes and $m_a \in \IZ$. 
This lattice has a $\IZ^{'}_6 \equiv \IZ_3\otimes \IZ_2$ discrete symmetry, $\mz \rightarrow \mtheta \mz$, 
$\theta^i_j \equiv \exp(2 \pi i v^i_6) \delta^i_j$, for $i,j = 1,\, 2,\, 3$, with $\mtheta^6 = \mone$, and 
$6 v^i_6 = 0$ mod 1. 
$N=1$ supersymmetry is maintained if the $\IZ_6$ twist is a discrete symmetry within $SU(3)\in SO(6)$, which 
further requires that $\sum_i v^i_6 = 0$ mod 1.
The combination of lattice translations and twists of $\mz$ form the space group ${\IS}$ whose elements are 
$({\mtheta}^{k}, \ml)$, for $k = 0,\, ..., 5$. Points on the orbifolded torus $\IT^6/\IZ^{'}_6$ (alternately $\IR^6/{\IS}$) 
are then identified with $\mz \sim {\mtheta}^{k} \mz + 2\pi \ml$.

The $\IZ^{'}_6 = \IZ_3 \otimes \IZ_2$ twist acts on the $\mZ$ and $\mchi$ fields 
as $\mZ \equiv \mZ_L + \mZ_R \ra \mtheta \mZ$ and $\mchi \ra \mtheta \mchi$
and therefore on the $\mphi$ fields as $\mphi \rightarrow \mphi - \pi \mv_6$.
In travelling around non-contractible loops in the $\sigma$ direction, 
the fields transform as
\beqn
\mZ(\sigma + 2\pi) &=& \mZ(\sigma) + 2\pi m_a \me_a, \label{znclus}\\
\mchi (\sigma + 2\pi) &=& \pm \mchi(\sigma), \label{chinclus}
\eeqn
(where $+$ is for a Ramond fermion and $-$ is for a Neveu-Schwarz fermion)
in the untwisted sectors
and 
\beqn
\mZ(\sigma + 2\pi) &=& \mtheta^k \mZ(\sigma) + 2\pi m_a \me_a, \label{znclts}\\
\mchi (\sigma + 2\pi) &=& \pm \mtheta^k \mchi(\sigma), \label{chinclts}
\eeqn
in the $k^{\rm th}$ twisted sector.
Thus,
\beqn
\mphi (\sigma + 2\pi) = \mphi(\sigma) - \pi k \mv_6
\eeqn
in the $k^{\rm th}$ twisted sector.
Twisted strings are localized at the fixed points.

Orbifolding of the model is obtained by simultaneously modding the $E_8 \otimes E_8$ torus by a correlated $\IZ^{'}_6$ twist, as required for modular invariance. 
The bosonic fields on the lattice transform as $\bar{X}^I \rightarrow \bar{X}^I + \pi k V^I_6$. 
Discrete Wilson lines add shifts, $\bar{X}^I \rightarrow \bar{X}^I + \pi n_l W^I_l$, with integer $n_l$, to accompany torus lattice translations.\footnote{In a free fermion model, the corresponding complex fermion field is
$\bar{\lambda}^I = \pm \exp(2 \pi i \bar{X}^I)$ and transforms as
$\bar{\lambda}^I (\sigma + 2\pi) \ra \pm \Theta \bar{\lambda}^I(\sigma)$, 
where $\Theta^I = \exp(2\pi i k V^I_6)\delta^I_J$ without Wilson lines
and as 
$\bar{\lambda}^I(\sigma+2\pi) \ra \pm \Theta \bar{\lambda}^I(\sigma)$, where 
$\Theta^I = \exp(2\pi i \{k V^I_6 + m_a W^I_{la}\})\delta^I_J$
with Wilson lines.}
Thus, the twists and lattice shifts are embedded into the gauge degrees of freedom $\bar{X}^I$, as
$(\mtheta^k,m_a \me_a)\ra (\mone,k\mV_6 + m_a\mW_{la})$.

Modular invariance requires that both $6 \mV_6$ and $n \mW_l$, with $n\leq N$ the order of the Wilson line, must lie on the $E_8\otimes E_8$ lattice. In addition, the modular invariance rules, parallel to (\ref{kseta1}-\ref{kseta3}),
are 
\beqn
\half(\mV_6^2 - \mv_6^2) &=& 0\,\, {\rm mod}\, 1,\label{ovset1}\\
(\mV_6\cdot \mW_n) &=& 0\,\, {\rm mod}\, 1,\label{ovset2}\\
(\mW_n\cdot \mW_m) &=& 0\,\, {\rm mod}\, 1,\,\, (W_n\neq W_m)\label{ovset3}\\
\half \mW_n^2       &=& 0\,\, {\rm mod}\, 1,\label{ovset4}
\eeqn
which can be nicely unified \cite{Buchmuller:2005jr,Buchmuller:2006ik} as 
\beqn
\half\left[ (r \mV_6 + m_a \mW_{na})^2 
 - r \mv_6^2 \right] = 0\,\, {\rm mod}\, 1.\label{ovset5} 
\eeqn

The orbifold expression for the generalized GSO projections 
specifying the physical states in the untwisted sector
is
\beqn
\mv_6\cdot\mq - \mV_6\cdot \mbp = 0\,\, {\rm mod}\, 1;\,\,
\mW_6\cdot\mbp                    = 0 \,\, {\rm mod}\, 1,\label{outgso1}
\eeqn
where the components of vectors $({\mv}_{6},\mV_6,\mW_n)$ are the $(v^j_6,V^I_6,W^I_n)$.
This generalizes for the twisted sectors as
\beqn
\bar{k}\mv_6\cdot\left(\bar{\mN}_f - \bar{\mN}^{\ast}_f   \right)
&-&\bar{k}\mv_6\cdot \left(\mq + k\mv_6  \right)\nolabel\\
&+&\left(\bar{k}\mV_6 + \bar{m}_a \mW_{na}\right)\cdot
\left(\mbp + \bar{k}\mV_6 + \bar{m}_a \mW_{na}\right)= 0\,\, {\rm mod}\, 1,\label{otsgso1}
\eeqn
for all $\mW_n$ and $(\bar{k},\bar{m}_a)$ depending on conjugacy class (with modifications for non-prime orbifolds).
The components of $\mq$ are $q^j$, the momenta of the left-moving bosons $\phi^j$ on the $\IT^6$ torus, 
the components of $\mbp$ are $p^I$, the momenta of the right-moving $\bar{X}^I$ bosons on the $E_8\otimes E_8$ lattice,
and $\mV^I_f \equiv k \mV^I_6$.

In the untwisted sector the mass of the physical states are
\beqn
\alpha^{'} m^2_L &=& -\half + \half {\mq}^2 + N + N^{\ast}
                                     \label{orbutm2l}\\
\alpha^{'} m^2_R &=& -1     + \half {\mbp}^2 +\bar{N} + \bar{N}^{\ast}
                                    \label{orbutm2r},
\eeqn
for which $N^{(\ast)} = \sum_j N^{(\ast)j}$, where $N^{(\ast)j}$ are the sums of the eigenvalues of the number operators for the complex left-moving $Z^{(\ast)j}_L$ modes, respectfully, and
$\bar{N}^{(\ast})$ are the corresponding for the complex right-moving modes.
The generalization for the twisted sectors is,
\beqn
\alpha^{'} m^2_L &=& -\half + \half (\mq + k\mv_6)^2\nolabel\\ 
                 & & + \half \left[k \mv_{6}\cdot( \mone - k \mv_{6})
                                    + k \mv_{6}\cdot \mN_f -k \mv_{6}\cdot \mN_f^{\ast} \right]
                                     \label{orbtm2l}\\
\alpha^{'} m^2_R &=& -1     + \half (\mbp + \mV_f)^2\nolabel\\
                 & & + \half \left [k \mv_{6}\cdot( \mone - k v_{6j})
                                  + k \mv_{6}\cdot \bar{\mN}_f -k \mv_{6}\cdot \bar{\mN}_f^{\ast} \right].
                                     \label{orbtm2r}
\eeqn

At each fixed point on the orbifold, determined by the combination
$\mV_f$, $m_a \mW_{na}$, and $k \mv_n$,
$E_8\otimes E_8$ is broken to a subgroup. For the 
$G_2\otimes SU(3) \otimes SO(4)$ torus, the twist vector is chosen to be 
$\mv_6 = (\frac{1}{6},\frac{1}{3},-\frac{1}{2})$. For the associated order six $E_8\otimes E_8$ twist, only
two choices guarantee complete $\mathbf{16}$'s of $SO(10)$ in the twisted sector:  
\beqn
\mV_6     &=& (\half,\half,\third,0,0,0,0,0) (\third,0,0,0,0,0,0,0), \label{v6t}\\
\mV^{'}_6 &=& (\third,\third,\third,0,0,0,0,0) (\sixth,\sixth,0,0,0,0,0,0), \label{v6pt}
\eeqn
with the first being a $\IZ^{'}_6 \sim \IZ_3\otimes \IZ_2$ twist and the second a non-factorizable $\IZ_6$ twist. 
The orbifold model uses $\mV_6$ \cite{Buchmuller:2005jr,Buchmuller:2006ik}.

The simplest choice of additional Wilson lines leads to three equivalent fixed points with local
$SO(10)$ symmetry, arranging for one generation per fixed point. This would be analogous to the free fermionic model. However, it was found in \cite{Buchmuller:2005jr,Buchmuller:2006ik} that three fixed points always produced chiral, rather than vector-like, exotic MSSM states. These chiral exotic could only get EW mass, and therefore not be decoupled from the low energy effective field theory. Thus, two Wilson lines, 
\beqn
\mW_2 &=& (\half,0,\half,\half,\half,0,0,0) (-\threefourth,\fourth,\fourth,-\fourth,\fourth,\fourth,\fourth,-\fourth)
\label{w2t}\\
\mW_3 &=& (\third,0,0,\third,\third,\third,\third,\third)(1,\third,\third,\third,0,0,0,0),
\label{w3t}
\eeqn
were chosen that produce two generations at fixed points and a third spread between in the bulk $U$ and additional fixed points in the twisted (fixed point) sectors of $T_2$ and $T_4$. For these Wilson lines, the additional exotic MSSM states are all vector-like \cite{Buchmuller:2005jr,Buchmuller:2006ik}.

Since the $\mV$ twist is of order 6, the untwisted sector, $U$, is accompanied by five twisted sectors, $T_k$, $k=1$ to 5. The combination of orbifold and Wilson lines generate 2 equivalent sets of twisted sector fixed points. The local non-Abelian observable gauge groups at one set of the 6 fixed points are $SO(10)\otimes SO(4)$, $SO(12)$, $SU(7)$, $SO(8)\otimes SO(6)$, $SO(8)'\otimes SO(6)'$ and $SO(8)''\otimes SO(6)''$, respectively. The MSSM gauge group is realized as the intersection of these groups.\footnote{Notice that, as expected, formation of the net orbifold gauge group parallels that of free fermionic language.
In the free fermion case, the twisted sectors (Wilson lines) each act to reduce the untwisted sector gauge group and the net gauge group is the intersection of all twisted sector gauge reductions. However, free fermionic formulation 
does not yield to fixed point interpretation.}
The net gauge group, as common factor of all gauge groups at the 6 fixed points, is
\beqn 
[SU(3)_C \otimes SU(2)_L\otimes U(1)_Y \otimes \prod_{i=1}^{4} U(1)_i]\otimes [SU(4)_H \otimes SU(2)_H\otimes \prod_{i=5}^{8} U(1)_i]. 
\label{mssmorb}
\eeqn

The hypercharge has the standard $SO(10)$ embedding, 
\beqn
U(1)_Y \equiv U(1)_{1} = (0,0,0,\half,\half,-\third,-\third,-\third)(0,0,0,0,0,0,0,0)\label{hyper}
\eeqn
(following the notation of the $\mV$ and $\mW$), another aspect in common with the free fermionic model.
The remaining eight local Abelian symmetries are
\beqn
U(1)_2 &=& (1,0,0,0,0,0,0,0)(0,0,0,0,0,0,0,0)\label{qu1}\\
U(1)_3 &=& (0,1,0,0,0,0,0,0)(0,0,0,0,0,0,0,0)\label{qu2}\\
U(1)_4 &=& (0,0,1,0,0,0,0,0)(0,0,0,0,0,0,0,0)\label{qu3}\\
U(1)_5 &=& (0,0,0,1,1,1,1,1)(0,0,0,0,0,0,0,0)\label{qu4}\\
U(1)_6 &=& (0,0,0,0,0,0,0,0)(1,0,0,0,0,0,0,0)\label{qu5}\\
U(1)_7 &=& (0,0,0,0,0,0,0,0)(0,1,1,0,0,0,0,0)\label{qu6}\\
U(1)_8 &=& (0,0,0,0,0,0,0,0)(0,0,0,1,0,0,0,0)\label{qu7}\\
U(1)_9 &=& (0,0,0,0,0,0,0,0)(0,0,0,0,-1,-1,-1,1)\label{qu8}
\eeqn
$U(1)_2$ though $U(1)_8$ are all anomalous. As with all weak coupled 
heterotic models, the anomaly can be rotated into a single generator of the form
\beqn
U(1)_A &=& \sum_{i=2}^8 \tr\, Q_i\, U(1)_i\label{qua}
\eeqn
with $\tr\, Q_A =  88$.
A set of mutually orthogonal, anomaly-free $U(1)^{'}_k$, for $k=2$ to 7, are formed from linear combinations of 
$U(1)_i$, for $i=2$ to 8, orthogonal to $U(1)_A$. 
As with the free fermionic model, the anomaly is eliminated 
by the Green-Schwarz-Dine-Seiberg-Witten anomaly cancellation mechanism
\cite{Dine:1987xk,Atick:1987gy}. A non-perturbatively determined set of vacuum expectation values 
(VEVs) will turn on to cancel the resulting FI D-term $\sim \tr\, Q_A$ 
with a contribution of opposite sign. 

The cause of the anomaly, the massless matter, 
is found in the twisted sectors as well as the untwisted. 
Twisted matter, located at fixed points, appears in complete multiplet representations of the local gauge group, while untwisted bulk matter only appears in the gauge group formed from the intersection of all local gauge groups. 
Located at each of two equivalent fixed points on the $SO(4)$ plane in the twist sector $T_1$ is a complete MSSM generation. Each generation forms a complete $\mm 16$ rep of the untwisted sector $SO(10)$. Then, as a result of the Wilson lines, MSSM-charged matter from sectors $U$, $T_2$, and $T_4$ combine to form an additional complete $\mm 16$. The remaining matter (Higgs, MSSM exotics and hidden sector matter) appear in vector-like pairs. Since the Higgs originate in the untwisted sector, they naturally appear as MSSM reps, rather than as $SO(10)$ reps, providing a natural solution to the normal Higgs doublet/triplet splitting issue.

As in the free fermionic model, the string selection rules play an important role in matter Yukawa couplings. 
The orbifold worldsheet selections rules allow, at third order, only
$UUU$, $T_1 T_2 T_3$, $T_1 T_1 T_4$, $U T_2 T_4$, and $U T_3 T_3$ non-zero sector couplings. 
Of these, the four twisted sector couplings are prohibited for MSSM
states either by gauge symmetry or lack of MSSM states in the $T_3$ sector. 
For the heaviest generation, the quark doublet and anti-up quark originate in the untwisted sector, whereas the anti-down quark is from a twisted sector \cite{Buchmuller:2005jr,Buchmuller:2006ik}. This provides an unsuppressed top quark mass and suppressed higher order masses for all other MSSM matter, offering a possible explanation for the top to bottom mass hierarchy. 
Generational mass hierarchy is also afforded. Both are re-occurring themes in MSSM string models.

The exotic down-like and lepton-like states yield an interesting feature for the model. These states mix with the MSSM states in mass matrices and the physical states contain both. The 4 $d$-like quarks pair with all
7 $\bar{d}$-like quarks in a potentially rank 4 mass matrix, given sufficient non-zero VEVs from among the 69 non-Abelian singlet, hypercharge-free $s^{o}_m$ ($m=1$ to $69$) fields. 
The mass terms are formed by factors of 1 to 6 $s^o_m$ VEVs. 
The 8 lepton doublet fields with $-\half$ hypercharge appear with the 5 antilepton doublet fields 
in a potentially rank 5 mass matrix. The lepton mass terms are similarly formed by products of 1 to 6 $s^o_m$ VEVs.  
The 8 remaining lepton doublets with zero hypercharge appear in their own potentially rank 4 mass matrix. The 16 vector-like pairs of non-Abelian singlet with $\pm \half$ hypercharge similarly appear in their own 
potentially rank 8 mass matrix.
The VEVs required for mass terms in all four matrices sufficient to remove 4 pairs of exotic quarks, 4 pairs of exotic
lepton/anti-lepton hypercharged doublets (one of the pairs must be a pair of EW Higgs), 4 pairs of hypercharge-free lepton/anti-lepton doublets, and 16 pairs of hypercharged \NA singlets, correspond to 
$D$-flat directions \cite{Buchmuller:2005jr,Buchmuller:2006ik}.
 
Suppression of $R$-violating terms, in particular $\bar{u}\bar{d}\bar{d}$, imposes constraints on the allowed exotic down quark content of the physical states, thereby restricting the phenomenologically viable flat directions. 
Also, for generic non-Abelian singlet flat directions, all five pairs of non-generational hypercharged $SU(2)_L$ doublets acquire near FI scale mass, prohibiting any EW Higgs. Hence, fine tuning must be called upon to keep some 
VEVs far below the FI scale to allow for one Higgs doublet pair.

Like the free fermionic MSSM, the orbifold model has a large perturbative vacuum degeneracy that maintains supersymmetry after cancellation of the $U(1)_A$, while reducing gauge symmetries and making various combinations of states massive. In this model, a parameter space of VEVs from the set $s^o_m$ was found that is
simultaneously (i) $D$- and $F$-flat to all order, 
(ii) supplies near string-scale mass to most (or possibly all as claimed in \cite{Buchmuller:2006ik}) 
of the MSSM-charged exotics, and
(iii) breaks all of the Abelian gauge factors other than hypercharge. 
In this subspace of flat directions the gauge group is reduced to
\beqn
[SU(3)_C \otimes SU(2)_L\otimes U(1)_Y ]\otimes [SU(4)_H \otimes SU(2)_H],
\label{mssmfda}
\eeqn 
completely separating the observable and hidden sectors. 

When only the $s^o_m$ uncharged under a non-anomalous extended $U(1)_{B-L}$, defined as 
\beqn
U(1)_1 &=& (0,1,1,0,0,-\twothird,-\twothird,-\twothird)(\half,\half,\half,-\half,0,0,0,0),\label{qubl}
\eeqn
and some $SU(2)_H$ doublets uncharged 
under $U(1)_{B-L}$ are allowed to receive VEVs, the gauge group is alternately reduced to
\beqn
[SU(3)_C \otimes SU(2)_L\otimes U(1)_Y \otimes U(1)_{B-L}]\otimes [SU(4)_H].
\label{mssmfdb}
\eeqn 
(The standard $U(1)_{B-L}\in SO(10)$ is anomalous and required hidden sector extension to remove the anomaly.)
The fields with VEVs were shown to form $D$-flat monomials, but $F$-flatness was not required \cite{Buchmuller:2006ik}. Rather, it was assumed $F$-flatness can be satisfied by specific linear combinations of $D$-flat directions up to a sufficiently high order. Thus, the phenomenology associated with these VEVS is not necessarily as realistic as that for the prior parameter space.
With the reduced set of singlet VEVs, decoupling of the MSSM exotic states require mass terms up to 11th order in the 
superpotential. These masses are likely highly suppressed, with a value far below the MSSM unification scale. Therefore the related exotics likely do not decouple. The advantage to keeping a gauged $B-L$ symmetry is the removal of renormalizable $R$-parity violating couplings that otherwise lead to strong proton decay. When this $U(1)_{B-L}$
is retained, the up quarks are heavier than the down quarks for each generation, which are also heavier then the electrons. The up quark generational mass hierarchy is on the order of $1$: $10^{-6}$: $10^{-6}$. 
The down-quarks and for the electron quarks masses for the lighter two generations are also of the same order.
 
All order flatness for the initial collection of non-Abelian singlet directions was proved using the orbifold
parallels to the NSR free fermionic selections rules. First, these rules prohibit
self-coupling among the non-Abelian singlets in sectors $U$, $T_2$, and $T_4$. 
Additionally, singlets from these sectors can only couple to two or more states in $T_{1,3}$. Thus, any $D$-flat 
directions involving only VEVs of $U$, $T_2$, and $T_4$ non-Abelian singlets is guaranteed to be 
$F$-flat to all order. $F$-flat directions were formed as 
39 independent linear combinations of the $D$-flat directions 
containing only $s^o_m$ from $U$, $T_2$, and $T_4$. 
This set of flat directions allowed decoupling of many, {\it but not all}, of the MSSM exotics
\cite{Buchmuller:2005jr,Buchmuller:2006ik}. An increase in parameter space of all order $D$- and $F$- flat directions space, via hidden sector non-Abelian fields, that might induce mass to all MSSM exotics is reportedly under investigation, as are finite order $D$- and $F$-flat directions consistent with EW breaking \cite{Buchmuller:2006ik}.

Critically, \cite{Buchmuller:2006ik} reminds us that flat directions are not always necessary for decoupling of exotics. Rather, isolated special points generically exist in the VEV parameter space that are not located along flat directions, but for which all $D$- and $F$-terms are nonetheless zero.
\cite{Buchmuller:2006ik} calls upon the proof by Wess and Bagger \cite{Wess:1992cp} that
$D$-terms do not actually increase (except by one for FI term cancellation) 
the number of constraints for supersymmetric flat directions beyond the $F$-term constraints.\footnote{FI term cancellation requires the existence of one monomial that is $D$-flat for all non-anomalous symmetries, but that carries the opposite sign to the FI-term in for the anomalous $U(1)_A$, imposing an additional constraint.} 
Thus, the system of $D$- and $F$-equations is not over constraining.
Once a solution to $F_m = 0$, for all fields $s^0_m$, is found (to a given order), complexified gauge transformations of the fields $s^0_m$, that continue to provide a $F_m=0$ solution, can be performed that 
simultaneously arrange for $D$-flatness.
Thus, since the $F_m = 0$ equations impose $m$ (non-linear) constraints on $m$ fields, 
there should be at least one non-trivial solution for any set of fields $s^o_m$. A parallel proof for non-Abelian field VEVs also exists.\footnote{Complications to these proofs do arise when different scalar fields possess the same gauge charges.}
This reduction in apparent total constraints is possible because the $F$-term equations constrain
gauge invariant polynomials, which also correspond to non-anomalous $D$-flat directions \cite{Buchmuller:2006ik}.
Hence \cite{Buchmuller:2006ik} argues that since supersymmetric field configurations in generic string models form low dimensional manifolds or points, all of the MSSM singlets should generally attain non-zero VEVs.
Systematic surveys of these low dimensional manifolds for free fermionic models indicates the overall FI VEV scale 
is around $\tenth$ of the string scale, and therefore is around $10^{17}$ GeV. This argument nevertheless does not prove that {\it all} of the exotics will necessarily obtain mass. At even the special points some VEVs required for 
certain masses may be prohibited by supersymmetry.

The exotic states are given mass through superpotential terms (consistent with orbifold string selection rules \cite{Hamidi:1986vh,Font:1988mm}) involving up to six singlet VEVs (i.e., up to $8^{\rm th}$ order in the superpotential.\setcounter{footnote}{0}\footnote{In \cite{Buchmuller:2005jr,Buchmuller:2006ik} it is argued that the high order of some of these mass terms does not necessarily imply mass suppression, citing combinatorial effects in the coupling. However, for free fermionic models, the mass suppression factor is estimated to be
$\sim \tenth$ per order \cite{Faraggi:1996pa} beginning at $5^{\rm th}$ order \cite{Cvetic:1998gv}
Thus, if free fermionic patterns in mass suppression continue for orbifolds, then it should be expected that orbifold masses involving more than four VEVs should also fall below the MSSM unification scale and, therefore, do not actually decouple from the model.} The set of required $s^o_m$ singlet VEVs was shown to be provided by $D$-flat directions.
As discussed above, simultaneous $F$-flatness of a linear combination of the $D$-flat directions is then required. 

On par with the MSHSM free fermionic model, this orbifold model admits spontaneous supersymmetry breaking 
via hidden sector gaugino condensation $<\lambda \lambda>$ from the $SU(4)$. When the $\mm 6$'s and the ${\mm 4}/\bar{\mm 4}$'s of the hidden $SU(4)$
receive flat direction mass through allowed couplings, the condensation scale for 
$SU(4)$ gauginos is in the range of $10^{11}$ to $10^{13}$ GeV.  
Supersymmetry breaking then results after dilaton stabilization.
As with the free fermionic model, stabilization must be assumed to result from
non-perturbative corrections to the K\" ahler potential, since the multiple gaugino 
racetrack solution is not possible with the hidden $SU(2)$. The latter either does not condense or does so at too low of an energy.
\cite{Buchmuller:2006ik} assumes a dilaton non-perturbative correction of the form
\beqn
K = -\ln (S + S^{\ast}) + \Delta K_{np},
\label{knp}
\eeqn 
where Re$\, <S> \sim 2$. Supersymmetry is spontaneously broken by the dilaton $F$-term,
\beqn
F_S \sim \frac{<\lambda \lambda>}{\MP}
\label{conden}
\eeqn

The resulting soft SUSY-breaking terms from dilaton stabilization are universal with the only
independent parameter being the gravitino mass $m_{3/2}$. 
The universal gaugino and scalar masses are then $\sqrt{3} m_{3/2}$ and $m_{3/2}$, respectfully.

A nice feature of the orbifold model is possible mitigation of the factor-of-20 difference between the MSSM unification scale $\sim 2.5\times 10^{16}$ GeV and the weak coupled heterotic string scale 
$\sim 5\times 10^{17}$ GeV. One or two large orbifolded radii at the MSSM unification length scale are consistent with 
perturbativity and could lower the string scale down toward the MSSM scale. This is one aspect not possible
from the free fermionic perspective, since the free fermionic formulation effectively fixes compactification at the
self-dual radius, i.e., at the string length scale.\footnote{Alternatively, weak coupled free fermionic heterotic strings may offer a resolution \cite{Cleaver:2002qc,Perkins:2003tb,Perkins:2005zh} to the factor-of-20 difference through ``optical unification'' \cite{Giedt:2002kb} effects, whereby
an apparent MSSM unification scale below the string scale 
is guaranteed by the existence of certain classes of intermediate scale MSSM exotics.} 

In their search for MSSM heterotic orbifolds, the authors of \cite{Buchmuller:2005jr,Buchmuller:2006ik} reported 
finding roughly 
$10^4$ $\IZ^{'}_6 = \IZ_3\otimes \IZ_2$ orbifold models with the SM gauge group. Of these, approximately 100 have exactly 
3 MSSM matter generations plus vector-like MSSM matter. The model reviewed above was the only one for which
flat directions exist whereby all MSSM exotics can be made massive and, thereby, decouple from the LEEFT.
Thus, within the context of weak coupled heterotic strings a total of two MSSM models with no exotics have been 
identified. A vast array of MSSM heterotic models without exotics likely remains yet undiscovered 
in the weak coupling domain.

%%%%%%%%%%%%%%%%%%%%%%%%%%%%%%%%%%%%%%%%%%%%%%%%%%%%%%%%%%%%%%%%%%%%%%%%%%%%%%%%%%%%%%%%%%%%%%%%%%%%%%%%%%%%%%%%%%%%%
% heterotic elliptically fibered C-Y at here 4
%%%%%%%%%%%%%%%%%%%%%%%%%%%%%%%%%%%%%%%%%%%%%%%%%%%%%%%%%%%%%%%%%%%%%%%%%%%%%%%%%%%%%%%%%%%%%%%%%%%%%%%%%%%%%%%%%%%%%
\subsection{Heterotic Elliptically Fibered Calabi-Yau}
\no
The free fermionic and orbifold models are not the only heterotic MSSM's that have been found. An equal number of strong coupled heterotic models have also been constructed. (Strong coupling is implied by the presence of $D$-branes that provide needed anomaly cancellation.) 
Another $E_8\otimes E_8$ heterotic model containing solely the MSSM spectrum in the observable sector was constructed by compactification on an elliptically fibered Calabi-Yau three-fold 
$X = \tilde{X}/(\IZ_3 \otimes \IZ_3)$ containing a $SU(4)$ gauge instanton and a $\IZ_3 \otimes \IZ_3$ Wilson line 
\cite{Braun:2005nv}.\setcounter{footnote}{0}\footnote{Interestingly, the MSSM-charged content of this model is claimed to not vary between weak and strong string coupling strengths.}
The observable sector $E_8$ is spontaneously broken by the $SU(4)$ gauge instanton into $S0(10)$\footnote{The reduction is actually to Spin$(10)$, the universal  covering group of $SO(10)$, but from hereon this distinction will not be made.} which is further broken to $SU(3)_C \otimes SU(2)_L \otimes U(1)_Y \otimes U(1)_{B-L}$ by the Wilson line.\setcounter{footnote}{0}\footnote{The $U(1)_{B-L}$, which prohibits $\Delta L = 1$ and $\Delta B = 1$ dimension four nucleon decay terms must be broken above the electroweak scale.}
The Calabi-Yau $X$, with $\IZ_3 \otimes \IZ_3$ fundamental group, was first investigated in \cite{Braun:2004xv}, while
the $SU(4)$ instanton was obtained in \cite{Braun:2005ux,Braun:2005bw,Braun:2005zv} as a connection on a holomorphic vector bundle.  
This model was produced as a slight variation from another in \cite{Braun:2005ux,Braun:2005bw,Braun:2005zv}, with the MSSM-charged content of the latter varying by only an extra Higgs pair. Both models contain an additional 19 uncharged moduli, formed of 3 complex structure moduli, 3 K\" ahler moduli, and 13 vector bundle moduli. 

The renormalizable MSSM Yukawa couplings were computed in \cite{Braun:2006me}. 
Each superfield in the model is associated with a $\bar{\partial}$-closed $(0,1)$ form $\Phi_i$ taking values in some bundle over $X$. The tree-level (classical) couplings $\lambda_{i,j,k}$ can be 
expressed in the large-volume limit as 
\beqn
\lambda_{i,j,k} \sim \int_X \Omega \wedge \Phi_i \wedge \Phi_j \wedge \Phi_k, 
\label{yca}
\eeqn
where $\Omega$ is the Calabi-Yau's holomorphic $(3,0)$-form. (\ref{yca}) provides the unique way of generating a complex number from 3 superfields.
Rather than computing the numeric value of each $\lambda$,
general rules were constructed for non-zero values by evaluating the corresponding triplet products of cohomology classes for two matter fields and a Higgs.
The resulting Calabi-Yau worldsheet selection rules provide a cubic texture with only non-zero couplings for interactions of the first generation with the second and third generations. Thus, two of the three generations 
receive electroweak scale mass for their quarks and leptons, 
while masslessness of one generation can receive correction from higher order, non-renormalizable terms.
Couplings for non-zero moduli-dependent $\mu$-terms for the Higgs were similarly computed in \cite{Braun:2005xp}.
Selection rules were found to severely limit the number of moduli that can couple to a Higgs pair.
For the two Higgs pair model of \cite{Braun:2005ux,Braun:2005bw,Braun:2005zv} only four of the nineteen moduli can thereby produce renormalizable $\mu$-terms when they acquire non-zero VEVS. No third order mu-terms are allowed for the single Higgs pair in \cite{Braun:2005nv}. Thus in the minimal model, the Higgs mass is naturally suppressed.  
The coupling coefficients for both matter and Higgs cubic terms are not expected to be constant over the moduli space. Instead,
they likely depend on the moduli \cite{Braun:2005xp}.

A highly non-trivial consistency requirement of this model is the slope-stability of the $SU(4)$ vector bundle $V$,
which is necessary for $N=1$ SUSY. In \cite{Braun:2006ae} the vector bundle $V$ was proven to be slope-stable for any 
K\" ahler class $\omega$ in a maximum dimensional (i.e., 3-dimensional) subcone of the full K\" ahler cone.
A prerequisite for slope stability is the ``Bogomoli inequality" for non-trivial $V$, 
\beqn
\int_X \omega\wedge c_2(V) > 0. \label{vpom}
\eeqn

The viable hidden sector content of the model is so far undetermined--the possible range of hidden sector holomorphic vector bundles $V'$ has not yet been completely identified \cite{Braun:2005nv}. 
Any $V'$ must also be slope-stable, and thereby also satisfy (\ref{vpom}) for the same K\" ahler classes for which $V$ is slope-stable. 
Further, the topology of $V'$ must satisfy the anomaly cancellation condition (relating it to the topology of the observable vector bundle $V$),
\beqn
c_2(V) + c_2(V') = c_2(TX)  + [{\cal{W}}] - [\overline{\cal{W}}], \label{vpan}
\eeqn
where $c_2$ is the second Chern class, $TX$ denotes the tangent bundle of the Calabi-Yau three-fold $X$, and $[\cal{W}]$ and $[\bar{\cal{W}}]$ are the Poincare dual of the curves on which the brane and anti-branes are wrapped (if either one or both are present). The percentage of the subcone of $V$ stability for which $V'$ is also stable and can provide anomaly cancellation is claimed to be large \cite{Braun:2005nv}. 

A trivial vector-bundle $V'$, (with $c_2(V')=0$), corresponding to an unbroken hidden sector $E_8$ gauge symmetry
is one slope-stable choice. In this case (\ref{vpan}) becomes 
\beqn
c_2(V) = c_2(TX)  + [{\cal{W}}]- [\overline{\cal{W}}], \label{vpan2}
\eeqn
For this, \cite{Braun:2006th} investigated the general form of the potential when the combined anomaly from 
a non-trivial slope-stable $V$ and $X$ can be cancelled by the curve of a single five-brane rather than a brane/anti-brane combination. Their findings indicated ``that for a natural range of parameters, the potential energy function of the moduli fields has a minimum which fixes the values of all 
moduli.''\setcounter{footnote}{0}\footnote{Simultaneous stability of the $SU(4)$ vector bundle and anomaly cancellation with a trivial hidden sector vector bundle has been challenged 
in \cite{Gomez:2005ii,Bouchard:2005ag}.} 
Further, the minimum of the potential energy is negative and typically of order $-10^{-16} M^4_{\rm Pl}$, giving the theory a large negative cosmological constant.  They also showed that at the potential minimum the K\" ahler covariant derivatives vanish for all moduli fields. Since the moduli are uncharged, $D$-flatness is also retained. Therefore, supersymmetry remains unbroken in the anti-deSitter vacuum. All of this remains true even in the case of hidden sector gaugino condensation. It was argued that the pattern found for the example case continued for non-trivial hidden sector vector bundles $V'$.

In \cite{Braun:2006th} the addition of an anti-brane was shown to shift the minimum of the potential by an amount
proportional to
\beqn
{\cal J} = c_2(V) - c_2(TX) + [{\cal{W}}]+ [\overline{\cal{W}}], \label{shift}
\eeqn 
which by anomaly cancellation is
\beqn
{\cal J} = 2 [\overline{\cal{W}}] > 0. \label{shift2}
\eeqn 

For a viable range of the moduli, the minimum of the potential can be shifted to be positive and on the scale of the cosmological constant. The up-lifted vacuum becomes meta-stable, with a very long lifetime.

The same effect was found when a hidden $SU(4)$ vector bundle $V'$ is chosen \cite{Braun:2006th}. Without an anti-brane the potential minimum is highly negative, but this can be uplifted by an anti-brane. 
Further, when $SU(4)$ was chosen as the hidden sector vector bundle for the MSSM model described above, 
a consistent configuration was claimed with the addition of a  
5-brane/anti-5-brane pair for anomaly cancellation. (Since $V'=V$, 
slope stability was assured for the hidden sector.) With this, the total gauge group is 
$[SU(3)_C\otimes SU(2)_L \otimes U(1)_Y \otimes U(1)_{B-L}] \otimes SO(10)_H$. 

In \cite{Braun:2006em,Braun:2006da} the aspects of dynamical SUSY breaking process for the $SO(N_c = 10)$ hidden symmetry was analyzed in further detail, based on recent work in \cite{Intriligator:2006dd} regarding four dimensional $N=1$ theories with both supersymmetric and non-supersymmetric vacua. For SUSY $SO(10)$ moduli space, both stable  supersymmetric vacua and meta-stable non-supersymmetric vacua exist for an even number $N_f$ of massive $\mathbf{10}$ matter reps in the free-magnetic range, 
\beqn
N_c - 4 = 6\leq N_f \leq \frac{3}{2} (N_c -2) = 12.
\label{fmr}
\eeqn

At certain points in the $SO(10)$ moduli space some or all of these massive reps can become massless. In the neighborhood of the moduli space of these points, the matter reps gain slight masses and supersymmetry is also broken. The meta-stable non-supersymmetric vacua are long lived in the limit 
\beqn
\sqrt{\frac{m}{\Lambda}}<< 1
\label{ms}
\eeqn
where $m$ is the typical mass scale of the $\mathbf{10}$ reps and $\Lambda$ is the strong-coupling scale.
\cite{Braun:2006em,Braun:2006da} show that $SU(4)$ hidden sector bundle moduli can be consistently chosen to 
provide for long-lived meta-stable vacua. 

The stability and/or anomaly cancellation of the observable sector $SU(4)$ vector bundle, with 
trivial hidden sector vector bundle, has been claimed as likely, but is nonetheless uncertain.
An alternate model, with slightly modified elliptically fibered Calabi-Yau three-fold $X'$ and alternate $SU(5)$ vector bundle, has proven to be definitely stable. The latter model provides anomaly cancellation with a trivial hidden vector bundle \cite{Bouchard:2005ag}. 
The initial Calabi-Yau three-fold is the same, $\tilde{X}$, but this time
$X'= \tilde{X}/\IZ_2$ has a fundamental group $\phi_1(X') = \IZ_2$, rather than $\phi_1(X) = \IZ_3 \otimes \IZ_3$.
On $X'$, slope stability for an observable 
$SU(5)$ vector bundle $V'$ could be proved absolutely. The $SU(5)$ vector bundle breaks the initial
$E_8$ to the commutant of $SU(5)$, which is also $SU(5)$. A $\IZ_2$ Wilson line, allowed by the fundamental group,
then breaks $SU(5)$ to $SU(3)_C\times SU(2)_L \times U(1)_L$ \cite{Bouchard:2005ag}.

The particle spectrum is again given by the decomposition of the adjoint rep of $E_8$ under this breaking pattern.
The result is exactly three generations of MSSM matter, 0, 1, or 2 pairs of Higgs doublets (depending on location in the moduli space), and no MSSM exotics (other than perhaps one extra Higgs pair). 
Additionally, the particle spectrum includes at least 87 uncharged moduli \cite{Bouchard:2005ag}.
If a trivial hidden sector vector bundle is chosen (thereby giving $E_8$ local hidden symmetry), 
anomaly cancellation proves possible with addition of a single $M5$ brane wrapping an effective curve $[W]$ 
specified by ${c}_{2}(TX') - {c}_{2}(V') = [W]$ \cite{Bouchard:2005ag}. The presence of the
 $M\textsl{}5$ brane places this model in the strong coupling region. 

Trilinear Yukawa couplings for this model were analyzed in \cite{Bouchard:2006dn}. Non-zero couplings were computed for all three generations of up-quarks. Depending on the location in the moduli space, these couplings may provide a realistic mass hierarchy. Parallel non-zero down-quark and electron couplings were not found.  
All $R$-parity violating terms, including $B$ and $L$ violating terms leading to proton decay, were found to vanish.
$\mu$ mass parameters for the Higgs and neutrino mass terms were provided by vector bundle moduli \cite{Bouchard:2005ag}.

An enlarged class of $E_8\otimes E_8$ heterotic strings on elliptically fibered Calabi-Yau manifolds $X$ with vector bundles was introduced in \cite{Blumenhagen:2006ux}. These models have structure group 
$U(N)\sim SU(N)\otimes U(1)$ and M5-branes. This construction gives rise to GUT models containing $U(1)$ factors like flipped $SU(5)$ or directly to the MSSM, even on simply connected Calabi-Yau manifolds. MSSM-like models were 
constructed in which the only chiral states are the MSSM states, but nevertheless contain vector-like MSSM 
exotics. In the example given, the vector exotics were undetermined, as were the Yukawa couplings for all states.
The MSSM-like models constructed continue to possess the most attractive features of flipped $SU(5)$ such as doublet-triple splitting and proton stability. 

In the prior two elliptically-fibered models, the observable sector vector bundles 
$V= SU(4)$ or $SU(5)$ break $E_8$ to the respective
commutants $SO(10)$ or $SU(5)$. Reduction to the MSSM then depends on Wilson lines, since adjoint or higher rep Higgs are not possible for the level-1 Ka\v c-Moody algebras from which most heterotic gauge groups are derived. The Wilson
lines require Calabi-Yau three-folds with non-zero first fundamental group $\pi(CY_3)$, provided in the two prior models by freely-acting $\IZ_3\otimes \IZ_3$ or $\IZ_2$ orbifoldings of a simply connected Calabi-Yau three-fold.
By using $U(N)$ vector bundles that break ${E}_{8}$ to either flipped $SU(5)$ (for $N=4$) or directly to the MSSM 
(for $N=5$) the need for a non-trivial first fundamental group is eliminated. This allows a larger number of geometric backgrounds.

A vector bundle in this new class of models takes the form $W= V\oplus L^{-1}$ with $V$ being either a $U(4)$ or $U(5)$ vector bundle in the observable $E_8$ and $L^{-1}$ a line bundle included such that $c_1(W) = 0$. $W$ is embedded into $SU(6)\in E_8$ such that the commutant is $SU(3)_C\otimes SU(2)_L\otimes U(1)_1$ ($U(1)_1\neq U(1)_Y$). The $U(1)$ bundle in $V$ is embedded in $SO(10)$ as $Q_1 = (1,1,1,1,-5)$. The surviving $U(1)_1$ does not remain massless unless it is also embedded in the hidden sector, which is performed by embedding the line bundle also into 
the hidden $E_8$ (thereby bringing part of it into the observable sector). The line bundle breaks the hidden $E_8$
into $E_7\otimes U(1)_2$ and the surviving $U(1)$ becomes 
\beqn
U(1)_Y = \third (U(1)_1 + 3 U(1)_2).
\label{hypvbx}
\eeqn
$U(1)_Y$ was shown to remain massless as long as
\beqn
\int_X c_1(L) c_2(V) = 0;\quad \int_{\Gamma_a} c_1(L) = 0.
\label{uymassless}
\eeqn
where $\Gamma_a$ is the internal two-cycle wrapped by $N_a$ five branes.

As a result of the non-standard $U(1)_Y$ embedding, the tree level relations among the MSSM gauge couplings at the unification scale are changed. Rather than 
$\alpha_{MSSM} = \frac{5}{3} \alpha_Y$, this embedding yields
$\alpha_{MSSM} = \frac{8}{3} \alpha_Y$.
However, it was reported that through threshold corrections, a co-dimension one hypersurface in the K\" ahler moduli space allows MSSM gauge coupling unification. 
Relatedly, the non-standard hypercharge embedding may lead to some falsifiable
predictions. Since some ``hidden'' matter now carries hypercharge, it becomes coupled to
MSSM matter. Sufficiently low mass hidden matter could therefore have detectable effects at LHC.

The resulting MSSM-charged chiral massless spectrum contains
\beqn
g = \half \int_X c_3(V)
\label{numgen}
\eeqn
generations of 15-plets that do not include the neutrino singlet.
A few $g=3$ generation models were obtained.
As a result of the hidden sector contribution to hypercharge,
the anti-electrons carry both observable and hidden sector $E_8$ charge
unless
\beqn
\int_X c_3(L)= 0.
\label{aec}
\eeqn

The MSSM-charged spectrum also contains an undetermined number of 
vector-like pairs of exotics. However, exact MSSM models of this class without exotics 
may very well exist. Exploration of the MSSM-like models 
from vector bundles on either non-simply or simply connected Calabi-Yau 
threefolds has just begun.

%%%%%%%%%%%%%%%%%%%%%%%%%%%%%%%%%%%%%%%%%%%%%%%%%%%%%%%%%%%%%%%%%%%%%%%%%%%%%%%%%%%%%%%%%%%%%%%%%%%%%%%%%%%%%%%%%%%%%
% Type IIB at here 5
%%%%%%%%%%%%%%%%%%%%%%%%%%%%%%%%%%%%%%%%%%%%%%%%%%%%%%%%%%%%%%%%%%%%%%%%%%%%%%%%%%%%%%%%%%%%%%%%%%%%%%%%%%%%%%%%%%%%%
\subsection{Type IIB Magnetic Charged Branes}
\no
Realization of gauge groups and matter reps 
from D-brane stacks in compactified Type IIB models (with odd dimension branes) or 
Type IIA models (with even dimension branes)
provides another bottom-up route to (MS)SM-like models from string theory.
A supersymmetric Type IIB model with magnetized $D3$, $D5$, $D7$, and $D9$ branes is $T$-dual to a Type IIA 
intersecting $D6$ brane model. Relatedly, the model building rules for Type IIB and IIA models are very similar \cite{Bailin:2006rx,Bailin:2006zf,Kokorelis:2004dc,Kokorelis:2004tb}. 

The starting point in these models is two stacks, $a$, with $N_a= 3$ D-branes of the same type 
and similarly $b$, with $N_b = 2$, which generate a $U(3)_a\otimes U(2)_b$ symmetry. 
Along with $U(1)_a\in U(3)_a$ and $U(1)_b\in U(3)_b$, additional single (unstacked) branes $U(1)_i$ 
provide for $U(1)_Y$ and charge cancellation requirements.
The $U(3)_a = SU(3)_C \times U(1)_a$ ($U(2)_b = SU(2)_L \times U(1)_b$) gauge generators are open strings 
with both ends bound on the $a$ ($b$) stack. 
Quark doublet MSSM matter appears in bi-fundamental representations $({\mbf N}_a,\bar{\mbf N}_b) = ({\mbf 3}, \bar{\mbf 2})$ as open strings at the intersections of $a$ and $b$ 
with one end bound on the $a$ stack and the 
other end on the $b$ stack. 
The ${\mbf 3}$ has charge $Q_a = +1$ of $U(1)_a$, 
while the $\bar{\mbf 2}$ has charge $Q_b = -1$ of $U(1)_b$. 
Exotic $(\bar{\mbf 3}, {\mbf 1})$ quarks, $({\mbf 3}, {\mbf 1})$ anti-quarks, and $({\mbf 1},{\mbf 2})$ leptons \& Higgs can appear when open strings have one end attached to the $a$ ($b$) stack and the other end attached to an additional single $U(1)$ brane. 
The multiplicity of given chiral bi-fundamental 
reps is specified by the intersection numbers between the two related stacks.

In orientifold compactifications, the $a$ and $b$ stacks become paired with image stacks $a^{'}$ and $b^{'}$. 
Additional quarks appear in $({\mbf N}_a,\bar{\mbf N}_b) = ({\mbf 3}, {\mbf 2})$ reps at the intersections of $a$ and $b^{'}$ as open strings with one end bound on the $a$ stack and the other end on the $b^{'}$ stack. 
The ${\mbf 2}$ has charge $Q_b = +1$ of $U(1)_b$. Open strings can also be bound between stacks and their mirrors, producing
exotic matter in symmetric (adjoint) reps, that is $\mbf 8$ for $SU(3)_C$ or $\mbf 3$ for $SU(2)_L$ and
antisymmetric representations,  i.e., anti-quarks ($\overline{\mbf 3}$ for $SU(3)_C$ or (anti)-lepton singlets $\overline{\mbf 1}$ for $SU(2)_L$. Thus, models with only the MSSM content cannot produce
symmetric reps from the $SU(3)_C$ or $SU(2)_C$ stacks and must not produce more than three anti-symmetric $SU(3)$ representations.

Exact (MS)SM spectra place constraints on the intersection numbers. If the number of intersections of $a$ with $b$
is $p$ and the intersections of $a$ with $b^{'}$ is $q$, then exactly 3 quark doublets requires $p+q = 3$. The six 
anti-quark singlets can arise from a combination of $r$ antisymmetric intersections of $a$ with its image $a^{'}$
and $6-r$ intersections of $a$ with $U(1)$ from single $D$-brane stacks $c$, $d$, etc. The three quark doublets 
carry $Q_a = 2(+1)$, the 
$r$ antisymmetric states carry $Q_a = 1+1$, and the $6-r$ antiquarks carry $Q_a = -1$. Assuming no exotics
requires, $6 +2r - (6-r) = 0$ for tadpole cancellation for $Q_a$. 
Hence $r= 0$ and all antiquarks must come from open strings connecting $a$ stack and
single branes. 

Similarly, for $t$ copies of $({\mbf 1},{\mbf 2})$ and $u$ copies of $({\mbf 1},{\mbf {\bar{2}}})$, with $t+u=3$ $(t,u\geq 0)$, and 6 leptons singlets (three of which carry hypercharge), $s$ of which are from 
antisymmetric intersections of $b$ and $b^{'}$ and $6-s$ from intersections between generic $c$ and $d$ singlet 
branes, tadpole cancellation requires 
$t - (3 - t) + 2s -3 p + 3 (3 - p) = 0$, or equivalently, $t +s -3p = -3$. 
Solutions exist for all values of $0\leq p\leq 3$, which is thus a requirement for the (MS)MS exact spectra 
from orientifolds \cite{Bailin:2006rx,Bailin:2006zf,Blumenhagen:2001te}. $s = 3$ antisymmetric singlets 
with $t=0$, $s = 4$ antisymmetric singlets with $t=2$, and $s = 6$ antisymmetric singlets with $t=0$ 
allows for $(p,q)=(3,0)$ or $(0,3)$. All other viable values of $s$ and $t$ require $(p,q)=(2,1)$ or $(1,2)$
(for fixed definition of chirality, otherwise $\pm$ sign changes allowed). However, the antisymmetric states 
of $SU(2)$ do not have the MSSM Yukawa couplings to the Higgs \cite{Bailin:2006rx,Bailin:2006zf}.

Type IIB and Type IIA have produced models with exactly the (MS)SM states, i.e., with no MSSM-charged exotics, that
also yield somewhat realistic MSSM Yukawa terms.
The most successful Type IIB models  have been compactified on $\IT^6/(\IZ_2 \times \IZ_2)$, with $\IT^6 = \IT^2\times \IT^2\times \IT^2$ \cite{Berkooz:1996dw,Cvetic:2001tj}. This is similar to 
the compactification for the free fermionic MSSM-spectrum model, except that orientifolding replaces orbifolding.
The $\IZ_2 \times \IZ_2$ generators are $\theta$ and $\omega$, where $\theta : (z_1,z_2,z_3) \ra (-z_1,-z_2,z_3)$ and 
$\omega : (z_1,z_2,z_3) \ra (z_1,-z_2,-z_3)$. The additional orientifold modding is performed by the operator 
product $\Omega R$ where $R: (z_1,z_2,z_3) \ra (-z_1,-z_2,-z_3)$ and $\Omega$ is worldsheet parity. After orientifolding,
these models contain 64 $O3$-planes and 4 $O7_i$ planes, with $i$ denoting the $i^{\rm th}$ $\IT^2_i$ on which the plane is localized at a $\IZ_2$ fixed point and wrapped around the other two $\IT^2_{j\ne i}$. 

Crosscap tadpoles are produced in these models via non-trivial contribution to the Klein bottle amplitude. These tadpoles can be cancelled by Type IIB $D(3+2n)$-branes, for $n=0,1,2,3$, that fill up $D=4$ Minkowski space and wrap $2n$ cycles on the compact space. If desired, $D7$ and $D9$ branes can contain non-trivial magnetic field strength $F = dA$ in the compactified volume. These non-trivial gauge bundles generally reduce the rank of the gauge group and lead to $D=4$ chiral fermions. 

Of vital importance is that magnetic flux induces $D$-brane charges of lower dimension. That is, magnetic flux on $D9$ branes produces $D7$, $D5$, and $D3$ brane charges. The standard convention for magnetized $D$-branes was defined in \cite{Cascales:2003wn}, with topological information specified by the six integers $(n^i_a,\, m^i_a)$ with $n^i_a$ the unit of magnetic flux in the given torus and $m^i_a$ the number of times that the $D$-branes wrap the $i^{\rm th}$ $\IT^2$ \cite{Marchesano:2004yq}. The magnetic field flux constraint for any brane is
\beqn
\frac{m^i_a}{2\pi} \int_{\IT^2_i} F^i_a = n^i_a.
\label{magfielddef}
\eeqn

The chiral spectrum produced by two stacks of $D$-branes, $a$ and $b$, (and their images $a'$ and $b'$) is a result of the intersection products \cite{Marchesano:2004yq},
\beqn
I_{ab}  &=&   \prod^{3}_{i=1} (n^i_a m^i_b - m^i_a n^i_b), \label{iab}\\
I_{ab'} &=&  -\prod^{3}_{i=1} (n^i_a m^i_b + m^i_a n^i_b), \label{iabp}\\
I_{aa'} &=& -8\prod^{3}_{i=1} (n^i_a m^i_a), \label{iaap}\\
I_{aO}  &=&  8(-m^1_a m^2_a m^3_a + m^1_a n^2_a n^3_a + 
                n^1_a m^2_a n^3_a + n^1_a n^2_a m^3_a), \label{ia0}
\eeqn
where subscript $O$ denotes the contributions from intersections with the $O3$ and $O7_i$ planes.

In models of this class, the initial $U(N_a)$ from $N_a$ stacked $D$-branes reduce to 
$U(N_a/2)$ under $\IZ_2 \times \IZ_2$. 
Further, $\Omega R$ invariance requires that to each set of topological numbers 
$(n^i_a,m^i_a)$ be added its $\Omega R$ image, $(n^i_a,-m^i_a)$, for each $D$-brane $a$. The total $D9$- and $D5$-brane charge will vanish \cite{Marchesano:2004yq}. The $D$-branes fixed by some elements of 
$\IZ_2 \otimes \IZ_2$ and $\Omega R$ will carry $USp(N_a)$ gauge group. The complete physical spectrum must be invariant under $\IZ_2 \otimes \IZ_2 \otimes \Omega R$. 

Cancellation of the R-R tadpole for this class of model imposes \cite{Marchesano:2004yq}
\beqn
\sum_{a} N_a n^1_a n^2_a n^3_a &=& 16
\label{rr3}\\
\sum_{a} N_a m^1_a m^2_a n^3_a &=& \sum_{a} N_a m^1_a n^2_a m^3_a =
\sum_{a} N_a n^1_a m^2_a m^3_a  = -16.   
\label{rr7}
\eeqn 

 These conditions correspond to cancellation of non-Abelian triangle diagrams and 
of $U(1)$ anomalies. The R-R tadpole constraint is satisfied by
\beqn
g^2 + N_f = 14,
\label{n1sol}
\eeqn
where $g$ is the number of quark and lepton generations and 
$8 N_f$ is the number of $D3$-branes. 
$N=1$ SUSY (and NS-NS tadpole cancellation) additionally requires
\beqn
\sum_i \tan^{-1}\left(\frac{m^i_a A_i}{n^i_a}\right) = 0,
\label{nn}
\eeqn
with 
$A_i$ the area (in $\al^{'}$ units) of the $i^{\rm th}$ $\IT^2$, which 
is satisfied by $A_2 = A_3 = A$ and
\beqn
\tan^{-1}(A/3) + \tan^{-1}(A/4) = \frac{\pi}{2} + \tan^{-1}(A_1/2).
\label{tansusy}
\eeqn

In \cite{Marchesano:2004yq} an additional consistency, $K$-charge anomaly cancellation, was also examined. $K$-charge anomaly first arose in \cite{Witten:1998cd}. In addition to homological R-R charges, $D$-branes can also carry $K$-theory $\IZ_2$ charges that are invisible to homology. $K$-charge anomaly had been ignored before
\cite{Marchesano:2004yq}
because of its automatic cancellation in simple models.
However, it was found to be more severe than uncancelled NS-NS tadpoles, because there is no analogue of the Fischler-Susskind mechanism for $K$ charge anomalies \cite{Marchesano:2004yq}.   
In Type I theory, non-BPS $D$-branes exist that carry non-trivial $K$-theory $\IZ_2$ charges. Elimination of the anomaly requires that these non-BPS branes must be paired \cite{Uranga:2000xp,Chen:2005mj}. 
Since a Type I non-BPS $D7$-brane is regarded as a $D7$-brane and its worldsheet parity image 
$\overline{D7}$-brane in Type IIB theory, even numbers of these brane pairs must be required in Type IIB (and its dual Type IIA). For a $\IZ_2\times\IZ_2$ orientifold, this expresses itself as global cancellation of $\IZ_2$ R-R charges carried by the $D5_i \overline{D5}_i$ and $D9_i \overline{D9}_i$ brane pairs \cite{Uranga:2000xp,Chen:2005mj}. Cancellation conditions for $K$-charge were found stringent enough in \cite{Marchesano:2004yq} to
render inconsistent 
seemingly consistent flux compactified Pati-Salem (PS)-like vacua presented elsewhere. 
In contrast, the \cite{Marchesano:2004xz,Marchesano:2004yq} models were found $K$ anomaly-free.

In \cite{Marchesano:2004yq} a model with $SU(4)\times SU(2)_L \times SU(2)_R \times U(1)^3 \times USP(8N_f)$ gauge group (using $USP(2)\sim SU(2)$) was constructed. A generalized Greene-Schwarz mechanism breaks some of the extra $U(1)$. Further, the $USP(8N_f)$ reduces to $U(1)^{2N_f}$ as the $8N_F$ branes are moved away from an orbifold singularity, breaking the gauge group to,
\beqn
SU(4)\times SU(2)_L \times SU(2)_R \times U(1)^{'} \times U(1)^{2N_f}.
\label{aMSSM}
\eeqn

In addition to three generations of PS matter and a single Higgs bi-doublet, the model contains numerous 
exotic
$({\mathbf 4},{\mathbf 1}, {\mathbf 1})$, $({\mathbf {\bar{4}}},{\mathbf 1}, {\mathbf 1})$, 
$({\mathbf 1},{\mathbf 2}, {\mathbf 1})$, and  $({\mathbf 1},{\mathbf 1}, {\mathbf 2})$ PS reps, along with 196 moduli uncharged under PS (many more than in generic three generation MSSM-like free-fermionic models,
because the latter are fixed at the self-dual radius). The 196 moduli form a vast parameter space of $N=1$ flat directions, which corresponds in brane language to (anti)D9 recombination. 
It is possible to leave a $\bar{D}$-brane with a gauge bundle that makes massive {\it all} of the extra $U(1)^{2N_f}$,
reducing the gauge group to
\beqn
SU(4)\times SU(2)_L \times SU(2)_R \times U(1)^{'},
\label{bMSSM}
\eeqn
while making massive all but four of the PS-charged exotics, 
That is, all but two $({\mathbf 1},{\mathbf 2}, {\mathbf 1})$, and two $({\mathbf 1},{\mathbf 1}, {\mathbf 2})$ reps become massive. 
The standard PS reps are unaffected by the $D9$ recombination because the PS sector is associated with the three sets of $D7$ branes instead. The MSSM spectrum can be produced from the standard PS spectrum by higgsing.\setcounter{footnote}{0}\footnote{This model was similarly produced in \cite{Cremades:2003qj} by intersecting $D$-brane construction.} 
The MSSM Yukawa couplings were computed in terms of theta functions, which showed that exactly one generation could receive renormalizable EW-scale mass, while the other two generations remain massless at renormalizable order \cite{Marchesano:2004yq}.

A slight variation of the $D7$ stacks in \cite{Marchesano:2004yq} allows an improved model with gauge group 
\beqn
SU(3)\times SU(2)_L \times SU(2)_R \times U(1)_{B-L}.
\label{cMSSM}
\eeqn 

The corresponding exotic MSSM-charged mass is three $({\mbf 1},{\mbf 1},{\mbf 2})_{-1}$ states. A recombination process of the $D$-brane stacks generates VEVs for these three exotics (a frequent occurrence also in free fermionic models) and breaks
\beqn
SU(2)_R\times U(1)_{B-L} \ra U(1)_Y.
\label{dMSSM}
\eeqn

Following this, the MSSM-charged matter corresponds to exactly three generations {\it without}
a neutrino singlet. 

\subsubsection{With Background Flux}
\no
Problems with the magnetic brane models above include (1) the large number of uncharged closed string moduli, producing unobserved massless fundamental scalars in the low energy effective theory, and (2) the trivial hidden sector prohibiting SUSY breaking from gaugino condensation in the strong coupling limit. These two issues can be (partially) resolved by the introduction of non-trivial R-R and NS-NS 3-form fluxes, $F_3$ and $H_3$ respectively. 
Attempts to stabilize string moduli though the addition of compactified background fluxes can also induce SUSY-breaking soft terms. In addition meta-stable vacua can be induced. Nevertheless, even the best models of this class have shortcomings. In particular, they generally either have exotics or lack other critical phenomenological features \cite{Grana:2002nq,Camara:2003ku,Grana:2003ek,Lust:2004fi}. 

A systematic approach to embedding the Standard Model in flux compactifications was first offered in \cite{Cascales:2003wn}. Compatibility between chirality and $N=1$ flux compactifications was first shown in \cite{Marchesano:2004yq}. Therein, example models were presented with chiral $D= 4$ flux vacua leading to  
MSSM-like spectrum for both $N=1$ and $N=0$ SUSY. Since both R-R and NS-NS tadpoles cancel in these models, the broken SUSY case was free of the usually associated stabilization problems. In the latter model SUSY breaking induces soft terms in the LEEFT. Introduction of the magnetized $D9$-branes with large negative $D3$-brane charges enabled the construction of three-family standard-like models with one, two, and three units of quantized flux 
(the last case being supersymmetric).

In models with flux, both R-R and NS-NS fluxes must obey the Bianchi identities,
\beqn
dF_3 = 0;\, dH_3 = 0,
\label{fh}
\eeqn
for the case of no discrete internal $B$-field. These 3-form fluxes generate a scalar potential for the dilaton and complex structure moduli, freezing them at particular values. The $D3$-brane R-R charge given by a combined
 $G_3 = F_3 - \tau H_3$ field, with $\tau = a + i/g_s$ axion-dilaton coupling, is
\beqn
N_{\rm flux} = \frac{1}{(4\pi^2 \al^{'})^2} \int_{M_6} H_3\wedge {\overline{F}_3} 
             = \frac{i}{(4\pi^2 \al^{'})^2} \int_{M_6} \frac{G_3\wedge {\overline{G}_3}}{2{\rm Im}\tau} \in 64\IZ,
\label{NQ}
\eeqn
which is a topological quantity. To satisfy Dirac quantization, $N_{\rm flux}$ must also be quantized in units of 64.
The D3-brane tension is given by
\beqn
T_{\rm flux} = \frac{-1}{(4\pi^2 \al^{'})^2} \int_{M_6} \frac{G_3\wedge \ast_6 {\overline{G}_3}}{2{\rm Im}\tau} \in 64\IZ
\geq |N_{\rm flux}|.
\label{TQ}
\eeqn

This isn't a topological quantity, but depends on the dilaton and complex structure. The quantity
\beqn
V_{\rm eff} \equiv T_{\rm flux}  - N_{\rm flux} 
\label{VQ}
\eeqn
produces an effective potential for these moduli \cite{Marchesano:2004yq}.

The minimum of the potential is at either
$\ast_6 {\overline{G}_3} = iG_3$ (ensuring the contribution of $G_3$ to the R-R charge is positive), 
corresponding to $|N_{\rm flux}|$ $D3$ branes, or  
$\ast_6 {\overline{G}_3} = -iG_3$, corresponding to $|N_{\rm flux}|$ $\overline{D3}$ branes. 
The $\overline{D3}$ fluxes will necessarily break SUSY, while   
the $D3$ fluxes may or may not break SUSY, depending on their class.
 
(\ref{NQ}) modifies the $D3$-brane tadpole constraint (\ref{n1sol}) to
\beqn
g^2 + N_f + \frac{1}{16} N_{\rm flux} = 14,\label{3qmod}
\eeqn
as a result of (\ref{rr3}) being altered to 
\beqn
\sum_{a} N_a n^1_a n^2_a n^3_a + \frac{1}{2} N_{\rm flux} &=& 16.
\label{grr3}
\eeqn

The $g = 1$ generation solution is $N_{\rm flux} = 3\cdot 64$ (three units of quantized flux)
and $N_{f} = 1$ for which $G_3$, formed solely of $N=1$ SUSY preserving $(2,1)$ flux, 
can be chosen to fix all untwisted moduli and the dilaton. 
On the other hand, the $g=3$ generation solution has $N_{\rm flux} = 64$ and $N_{f} = 1$, for which
$G_3$ contains both $N=1$ SUSY preserving $(2,1)$ and SUSY breaking $(0,3)$ components.
In the latter case, NS-NS tadpole cancellation remains, but a gravitino mass is generated 
\beqn
m^2_{3/2} \sim \frac{\int |G_3\wedge \Omega |^2}{{\rm Im}\tau\, {\rm Vol}({\cal{M}}_6)^2},
\eeqn
along with soft terms in the low energy effective lagrangian for the MSSM.
The cosmological constant is, nonetheless, kept to zero at first order due to the no-scale structure of
$V_{\rm eff}$ \cite{Marchesano:2004yq}. The addition of discrete torsion to these models was also investigated in \cite{Marchesano:2004yq}.

Models of $(\IT^2)^3/(\IZ_2\otimes\IZ_2)$
compactification class with varying fluxes were also studied intensively in \cite{Kumar:2005hf,Cvetic:2005bn}. Issues among these continued to be realization of viable masses, mixings, and moduli stabilization. A self dual flux of the form
\beqn
G_3 = \frac{8}{\sqrt{3}}\exp^{-\pi/6}\left(d\bar{z}_1 dz_2 dz_3 + dz_1 d\bar{z}_2  dz_3 + dz_1 dz_2 d\bar{z}_3 \right)
\label{g3f} 
\eeqn
was shown to stabilize the remaining complex structure torus moduli (setting all to 
$e^{2\pi i/3}$), following prior stabilization of the first two moduli by NS-NS tadpole cancellation (\ref{tansusy}). The difficulty with this model class is cancellation of three-form flux contributions to the D3 charge; D3 charge conservation constraints are hard to satisfy in semi-realistic models \cite{Cvetic:2005bn}.

Three generation MSSM-like models with one, two, and three units of flux were presented in \cite{Cvetic:2005bn}.
Several models with a single unit of flux are analyzed. For two of these models the MSSM Higgs have no renormalizable couplings to the three generations of quarks and leptons. Three of these models contain four or five pairs of Higgs which have renormalizable couplings with quarks and leptons. In addition to the extra Higgs, the models also possess varying numbers of chiral MSSM-charged exotics, only a few of which can obtain large masses via coupling to the Higgs. Another of these models was the first MSSM-like model (with several exotics) containing flux to possess strong gauge dynamics, resulting in gaugino and matter condensation on $D7$-branes of the hidden sector. A two flux model was also reported that contains several MSSM-charged exotics. One pair of chiral exotics couple to the MSSM Higgs and, thus, obtains an electroweak scale mass. This model also comes with five pairs of MSSM higgs doublets with correct charges to create renormalizable Yukawa mass terms for all quarks and leptons.
In the supersymmetric flux model, the MSSM Higgs doublets do not have Yukawa couplings to quarks or leptons of any generation, due to wrong charges under broken anomalous $U(1)_A$ gauge symmetries. In addition this model contains several MSSM-charged exotics.

A first-of-its-kind MSSM-like model was reported in \cite{Chen:2005mj}. While the model is again a $\IT^6/(\IZ_2 \times \IZ_2)$ orientifold, its magnetized $D$9-branes with large negative charge where introduced into the hidden sector, rather than into the observable sector. Models with this property had not been studied previously because of the difficulty of arranging supersymmetric $D$-brane configurations with more than three stacks of $U(n)$ branes.
These problems were overcome in \cite{Chen:2005mj}. 

The initial gauge group of this model is 
\beqn
U(4)_C \times U(2)_L \times U(2)_R.
\label{dvnm1}
\eeqn
The intersection numbers were chosen to satisfy
\beqn
I_{ab} = 3;\, I_{ac} = -3; \, I_{bc}\geq 1,
\label{dvnm2}
\eeqn
which produces a model with one unit of flux. No filler branes are needed; hence, there is no $USp$ groups. 
This gauge group (\ref{dvnm1}) can be broken down to 
\beqn
SU(3)_C \times SU(2)_L \times U(1)_{I_3R} \times U(1)_{B-L}.
\label{dvnm3}
\eeqn
by the generalized Green-Schwarz mechanism and the splittings of the $U(4)_C$ and $U(2)_R$ stacks of $D6$-branes.
A significant feature of this model is that the $U(1)_{I_{3R}} \times U(1)_{B-L}$ gauge symmetry can only be broken to 
$U(1)_Y$ at the TeV scale by giving VEVs to the neutrino singlet scalar field or the neutral component from 
a $(\bar{\mathbf{4}},\mathbf{1},\bar{\mathbf{2}})$. As is typical, only one family gets mass because of anomalous charge constraints.

%%%%%%%%%%%%%%%%%%%%%%%%%%%%%%%%%%%%%%%%%%%%%%%%%%%%%%%%%%%%%%%%%%%%%%%%%%%%%%%%%%%%%%%%%%%%%%%%%%%%%%%%%%%%%%%%%%%%%
% Type IIA at here 6
%%%%%%%%%%%%%%%%%%%%%%%%%%%%%%%%%%%%%%%%%%%%%%%%%%%%%%%%%%%%%%%%%%%%%%%%%%%%%%%%%%%%%%%%%%%%%%%%%%%%%%%%%%%%%%%%%%%%%
\subsection{Type IIA Intersecting Branes}
\no
Type IIA intersecting $D6$-brane models are the $T$-duals to Type IIB with magnetic charged $D$-branes. As reported in \cite{Chen:2005mj}, a large number of non-supersymmetric three family SM-like models of Type IIA origin have been constructed that satisfy the R-R tadpole cancellation conditions 
\cite{Blumenhagen:2000wh,Aldazabal:2000cn,Aldazabal:2000dg,Blumenhagen:2000ea,Angelantonj:2000hi,Cremades:2003qj}. 
However, generically the NS-NS tadpoles remain uncancelled in these models. 
A gauge hierarchy issue is also generic to this class \cite{Chen:2005mj}. 

Translation of rules for Type IIA intersecting $D6$-brane model building from Type IIB with magnetic charged branes was provided in \cite{Cascales:2003zp,Blumenhagen:2003vr}. \cite{Chen:2005mj} applied this translation for the $\IT^6/(\IZ_2 \times \IZ_2)$, with
$\IT^6 = \IT^2\times \IT^2\times \IT^2$, compactification in particular. Related consistent Type IIA $N=1$ supersymmetric near-MSSMs with intersecting $D6$-branes have been constructed \cite{Cvetic:2001tj,Cvetic:2001nr,Cvetic:2003yd}. 
Following this, a wide range of MSSM-like (but not exact), PS-like, $SU(5)$, and flipped-$SU(5)$ three generation models were discovered \cite{Cvetic:2003xs,Cvetic:2002pj,Cvetic:2004ui,Cvetic:2004nk,Chen:2005ab}. 
In the phenomenological investigations of these models \cite{Cvetic:2002qa,Cvetic:2002wh}, the re-occurring concern was moduli stabilization in open and closed strings \cite{Chen:2005mj}. Hidden sector gaugino condensation 
was determined to stabilize the dilaton and complex structure in the Type IIA frame, but further 
stabilization remains via fluxes or the equivalent.

The Type IIB $(\IZ_2 \times \IZ_2)$ orientifold notation applies to this dual process, except now
$R: (z_1,z_2,z_3) \ra (\bar{z}_1,\bar{z}_2,\bar{z}_3)$. In the $D6$ models, $(n^i,m^i)$ are 
the respective brane wrapping numbers along the canonical bases of homology one-cycles $[a_i]$ and $[b_i]$,
on the $i^{\rm th}$ two torus $\IT^2_i$ and the general homology class for one cycle on $\IT^2_i$ is 
$n^i [a_i] + m^i [b_i]$. Hence the homology classes $[\Pi_a]$ for the three cycles wrapped by a stack of 
$N_a$ $D6$ branes are
\beqn
[\Pi_a] = \prod^{3}_{i=1} (n^i [a_i] + m^i [b_i]);\, [\Pi_{a^{'}}] = \prod^{3}_{i=1} (n^i [a_i] - m^i [b_i]).
\label{trcyc}
\eeqn
There are also homology classes for the four $O6$-planes associated with the orientifold projections 
$\Omega R$, $\Omega R \omega$, $\Omega R \Theta \omega$, and $\Omega R \Theta$ defined by
$[\Pi^1] = [a_1][a_2][a_3]$, $[\Pi^2] = -[a_1][b_2][b_3]$, $[\Pi^3] = [b_1][a_2][b_3]$, and
$[\Pi^3] = [b_1][b_2][a_3]$ \cite{Chen:2005mj}.

The R-R tadpole cancellation conditions (\ref{rr6}) are parallel to (\ref{grr3},\ref{rr7}),
\beqn
-N^{(1)} - \sum_{a} N_a n^1_a n^2_a n^3_a &=& 
-N^{(2)} + \sum_{\alpha} N_a m^1_a m^2_a n^3_a = \nolabel\\
-N^{(3)} + \sum_{a} N_a m^1_a n^2_a m^3_a &=&
-N^{(4)} + \sum_{a} N_a n^1_a m^2_a m^3_a  = -16,   
\label{rr6}
\eeqn 
with $N^{(i)}$ denoting the number of filler branes on the top of the $i^{\rm th}$ $O6$-brane.
$N=1$ SUSY survives the orientation projection if the rotation angle of any $D6$ brane with respect to the orientifold-plane is an element of $SU(3)$. This again translates into two additional constraints on triplets of wrapping numbers \cite{Chen:2005mj}. Last, $K$-theory charges must cancel in like manner as in Type IIB models.

From Type IIA with $\IT^6/(\IZ_2 \times \IZ_2)$ compactification (without flux), \cite{Chen:2005mj} constructed a three family trinification model with gauge group $U(3)_C\times U(3)_L \times U(3)_R$, wherein all MSSM fermions and Higgs fields belong to bi-fundamental reps. While the three $SU(3)$ gauge groups contain very large R-R charges, a supersymmetric intersecting $D6$-brane trinification model was found which satisfies the R-R tadpole and $K$-theory cancellation conditions. The trinification gauge group was then broken down to 
$SU(3)_C\times SU(3)_L \times SU(3)_R$ by the generalized Green-Schwarz mechanism, which involves the untwisted R-R forms
$B^{i=1,...,4}_2$. These four fields couple to the $U(1)$ field strength $F_a$ for each stack $a$ via,
\beqn
N_a m^1_a n^2_a n^3_a \int_{M4} B^1_2 {\rm tr} F_a,\quad N_a n^1_a m^2_a n^3_a \int_{M4} B^2_2 {\rm tr} F_a,
\nolabel\\
N_a m^1_a n^2_a m^3_a \int_{M4} B^3_2 {\rm tr} F_a,\quad N_a m^1_a m^2_a m^3_a \int_{M4} B^4_2 {\rm tr} F_a.
\label{extgs}
\eeqn

The next stage, reduction to 
$SU(3)_C\times SU(2)_L \times U(1)_{Y_L} \times U(1)_{I_{3R}} \times U(1)_{Y_R}$, results from
$D6$ brane splittings. The SM is then obtained by a higgsing that necessarily breaks $D$- and $F$-flatness. 
Quark Yukawa couplings are allowed for just one EW massive generation because, while there is only one Higgs doublet pair, the ${Q}^{i}_{L}$ arise from the intersections
on the second two-torus and the ${Q}^{i}_{R}$ arise from the intersections on the third two-torus. 
Lepton Yukawa couplings are forbidden by anomalous $U(1)_A$ symmetries \cite{Chen:2005mj}.
A Type IIB dual to this model has not yet been found. While this trinification model has very large R-R charge to start with, its dual Type IIB model also has supergravity fluxes to contend with, which makes R-R tadpole cancellation even more difficult \cite{Chen:2005mj}.\footnote{Thus, as a practical aspect Type IIA intersecting $D6$ models and Type IIB with magnetically charged branes, though $T$ dual, may offer some different opportunities for (MS)SM searches.}

$\IZ^{'}_6 = \IZ_3\otimes \IZ_2$ orientifolding of Type IIA models has been investigated by Bailin and Love 
\cite{Bailin:2006rx,Bailin:2006zf}. They determined that unlike $\IZ_6$ orientifold models to date,
$\IZ^{'}_6$ can support $(a \cap b,a \cap b')= \pm(2,1)$ or $\pm(1,2)$ intersection numbers, which
may lead to the most phenomenologically viable MSSM models without exotics. The initial torus was chosen as
$\IT^6 = \IT^2\times \IT^2\times \IT^2$, where $\IT^2_{1,2} = \IR^2/\Lambda_{SU(3)}$.
After $\IZ^{'}_6$ orbifolding, the two basis 1-cycles on each $\IT^2$ can be arranged in one of two 
configurations, producing a total of eight different configurations for the $\IZ^{'}_6$ orientifold.
Allowed intersection numbers for the eight different models with $\IZ^{'}_6$ orientifoldings  
were analyzed. By construction, 
there is no matter in the symmetric representations in any of the models.  
The models contain up to two generations of antiquarks and antileptons in the antisymmetric rep of the 
$SU(2)$ brane and its mirror. Unfortunately, none of the models could be enlarged to contain
exactly the MSSM content and no more. This is prohibited by Abelian charge cancellations. 
Comparison of the possible intersection numbers for $\IZ_6$ and $\IZ^{'}_6$ orbifolds, and the variation in the number of singlets within the $\IZ^{'}_6$ orbifolds models, showed that different orbifold point groups produce different physics, as does the same point group realized on different lattices. 

%%%%%%%%%%%%%%%%%%%%%%%%%%%%%%%%%%%%%%%%%%%%%%%%%%%%%%%%%%%%%%%%%%%%%%%%%%%%%%%%%%%%%%%%%%%%%%%%%%%%%%%%%%%%%%%%%%%%%
% Conclusion at here 7
%%%%%%%%%%%%%%%%%%%%%%%%%%%%%%%%%%%%%%%%%%%%%%%%%%%%%%%%%%%%%%%%%%%%%%%%%%%%%%%%%%%%%%%%%%%%%%%%%%%%%%%%%%%%%%%%%%%%%
\section{Conclusion}
\no
The Standard Model (SM) is truly one of the most outstanding theoretical and experimental achievement of the 20$^{th}$ century. Nevertheless, the very nature of the Standard Model (SM) and the inconsistency of quantum gravity beg the question of what more unified, self-consistent theory hides behind both. 
The details of the Standard Model cannot be explained from within; nor can the infinities of quantum gravity be resolved from within. ``Why'' is the SM as it is? ``Why'' does quantum theory seem inconsistent with gravity. 
A deeper, underlying theory is necessary to answer these ``why'' questions.

String theory proposes viable answers to the questions. While the concept of a string landscape has
significantly diminished hopes of locating {\it the} unique string vacuum that 
either resolves {\it all} the ``Why's'' of the SM and gravity by postdicting, in it low energy 
effect field theory limit, 
all experimentally verified SM physics or is eliminated as a viable theory by experimental disconfirmation.
Instead, string theory in its present form responds to these questions not with a single unique answer, but with a collection of viable answers, all of a geometrical or topological nature.
Just as a given Lie algebra has an infinite set of representations, so too
the underlying physical laws within string theory 
may well have a vast array of different representations, of which our observable universe is but one.

Several hundred years ago the debate was over a geocentric universe verses a heliocentric universe. 
Eventually the heliocentric perception was supplanted by a  ``galacticentric'' view.
Then within the last century Hubble proved how limited the latter outlook was also. 
Perhaps string theory is taking us to the next step of realization, that our universe may be but one of many. 
That to understand the ``why'' of our universe and its laws, we must look to a multiverse beyond.
Some may argue that this suggestion passes beyond science into the realm of philosophy. 
However, if string theory describes the nature of reality, then is it not still science? 
Is that not the very role of science, to uncover physical reality?
String theory may well be teaching us that a ``univercentric'' view 
is as outmoded today as the geocentric view was realized to be long ago.

Whether string theory ultimately predicts but one possibility for a consistent universe 
(if the landscape picture eventually collapses) or a vast array, to be viable 
string theory must predict {\it at least} our observable universe as a possible outcome. 
To prove that string theory allows for the physics of the observable universe is 
the mission of string phenomenology. String phenomenologists are pursuing this investigation from many fronts.
(MS)SM-like model realization from 
weak coupled heterotic models via free fermionic and orbifold constructions, strong coupled heterotic models
on elliptically fibered Calabi-Yau's, Type IIB orientifolds with magnetic charged branes, and Type IIA 
orientifolds with intersecting branes was reviewed in this chapter. 

A string derived (MS)SM must possess far more realistic features
than just the correct gauge forces and matter states. 
Also required for the model are realistic gauge coupling strengths, 
a correct mass hierarchy, 
a viable CKM quark mixing matrix, 
a realistic neutrino mass and mixing matrix, 
and a severely suppressed proton decay. 
The hidden sector must be sufficiently hidden. 
MSSM candidates must also provide viable non-perturbative supersymmetry breaking that yields testable 
predictions for supersymmetric particle masses. 
Finally, the physical value of the cosmological constant must be produced.
No current string-derived (MS)SM-like models are a perfect match with the (MS)SM, 
but significant progress had been made in the last decade. 
 
The first step in proving that the MSSM can be realized in string theory was accomplished when
Cleaver, Faraggi, and Nanopoulos constructed a string model that yields
exactly the matter content of the MSSM 
in the observable sector, with no MSSM-charged exotics.
This was shown in the context of a weak coupled heterotic model constructed in the free fermionic 
formalism \cite{Cleaver:1998sa,Cleaver:1999cj,Cleaver:1999mw,Cleaver:2000aa,Cleaver:2001ab}.
Glimpses of other necessary phenomenology were also shown by this model.
Generational mass hierarchy 
appears as an ubiquitous effect of vacuum expectation values of 
scalar fields induced via Abelian anomaly cancellation, 
a feature endemic to all quasi-realistic heterotic models.
In this model the physical Higgs is a mixed state of pure Higgs 
that carry extra generational charges. 
The mixed states are induced by off-diagonal terms in a Higgs mass matrix that produce exactly 
one pair of Higgs that is massless at the string scale.
The relative weights of the generational Higgs components in the physical Higgs
can vary by several orders of magnitude,
thereby allowing for a large mass hierarchy even with only low order couplings.
The model also provides for supersymmetry breaking through hidden sector condensates.

Following the $\IZ_2\otimes \IZ_2$ NAHE-based free fermionic model \cite{Cleaver:1998sa,Cleaver:1999cj,Cleaver:1999mw,Cleaver:2000aa,Cleaver:2001ab}, several additional MSSM models, also 
without exotic MSSM-charged states, were constructed by alternate means. 
A representative sample of these were reviewed herein, including
the heterotic $\IZ^{'}_6 = \IZ_3 \otimes \IZ_2$ orbifold of \cite{Buchmuller:2005jr,Buchmuller:2006ik};
the heterotic elliptically fibered Calabi-Yau's of 
\cite{Braun:2005nv,Braun:2004xv,Braun:2005ux,Braun:2005bw,Braun:2005zv,
Braun:2006me,Braun:2005xp,Braun:2006ae,Braun:2006em,Braun:2006da},
\cite{Bouchard:2005ag,Bouchard:2006dn}, and
\cite{Blumenhagen:2006ux};
the Type IIB magnetic charged branes
of \cite{Marchesano:2004yq,Marchesano:2004xz} without flux,
and those of \cite{Cascales:2003zp}, \cite{Kumar:2005hf}, \cite{Cvetic:2005bn}, and
\cite{Chen:2005mj} with flux; and the Type IIA intersecting brane models of \cite{Chen:2005mj} and \cite{Cvetic:2001tj,Cvetic:2001nr}. 

These examples showed several (differing) areas of progress toward more realistic phenomenology. 
For instance, some models reconfirmed the importance of a local anomalous Abelian symmetry and
the related flat direction VEVs, resulting from the
Green-Schwarz-Dine-Seiberg-Witten anomaly cancellation mechanism \cite{Dine:1987xk,Atick:1987gy}.
These VEVs invoke many necessary features of an MSSM: observable sector
gauge breaking to $SU(3)_C\otimes SU(2)_L\otimes U(1)_Y$,
decoupling of MSSM exotic matter, production of effective Higgs mu-terms, and 
formation of intergenerational and intragenerational mass hierarchy.

The greatest advancement was in the role of branes and antibranes,
especially with regard to supersymmetry breaking and moduli stabilization.
Brane-based MSSM-like models with stable supersymmetric anti-deSitter vacua have been constructed. Further,  
uplifting from stable anti-deSitter vacua to metastable deSitter vacua with
cosmological constants at a viable scale has been crafted into MSSM-like models
by the addition of antibranes.
In the years ahead
further realistic features of string models containing the (MS)SM gauge group, 
solely the three generations of (MS)SM matter and a Higgs pair 
will likely be found, as string phenomenologists continue to analyze the content of the string landscape.

\subsection*{Acknowledgments}
\no
G.C.\ wishes to acknowledge the authors of all papers reviewed herein and
of all MSSM-string model papers as a whole. Based on length limitations, reviews herein were limited to a representative set from most construction methods. 
%\vfill
%\newpage

%========================================================================
%          MACROS FOR REFERENCES
%========================================================================
\def\AP#1#2#3{{\it Ann.\ Phys.}\/ {\bf#1} (#2) #3}
\def\NPB#1#2#3{{\it Nucl.\ Phys.}\/ {\bf B#1} (#2) #3}
\def\NPBPS#1#2#3{{\it Nucl.\ Phys.}\/ {{\bf B} (Proc. Suppl.) {\bf #1}} (#2)
 #3}
\def\PLB#1#2#3{{\it Phys.\ Lett.}\/ {\bf B#1} (#2) #3}
\def\PRD#1#2#3{{\it Phys.\ Rev.}\/ {\bf D#1} (#2) #3}
\def\PRL#1#2#3{{\it Phys.\ Rev.\ Lett.}\/ {\bf #1} (#2) #3}
\def\PRT#1#2#3{{\it Phys.\ Rep.}\/ {\bf#1} (#2) #3}
\def\PTP#1#2#3{{\it Prog.\ Theo.\ Phys.}\/ {\bf#1} (#2) #3}
\def\MODA#1#2#3{{\it Mod.\ Phys.\ Lett.}\/ {\bf A#1} (#2) #3}
\def\MPLA#1#2#3{{\it Mod.\ Phys.\ Lett.}\/ {\bf A#1} (#2) #3}
\def\IJMP#1#2#3{{\it Int.\ J.\ Mod.\ Phys.}\/ {\bf A#1} (#2) #3}
\def\IJMPA#1#2#3{{\it Int.\ J.\ Mod.\ Phys.}\/ {\bf A#1} (#2) #3}
\def\JHEP#1#2#3{{\it JHEP}\/ {\bf #1} (#2) #3}
\def\nuvc#1#2#3{{\it Nuovo Cimento}\/ {\bf #1A} (#2) #3}
\def\RPP#1#2#3{{\it Rept.\ Prog.\ Phys.}\/ {\bf #1} (#2) #3}
\def\etal{{\it et al\/}}
%=========================================================================
%atbib

%\section*{References}

\label{lastpage-01}
%\hfill\vfill
%\newpage
%************************************************************************************


\begin{thebibliography}{99}
%%%% intro %%%%%%%%%%%%%%%%%%%%%%%%%%%%%%%%%%%%%%%%%%%%%%%%%%%%%%%%%%%%%%%%%%%%%%%%%%%%%%%%%%%%%%%%%%%%%%
% at here 1
\bibitem{woit:2006a}
  {P.~Woit, {\it Not Even Wrong: The Failure of String Theory and the Search for Unity in Physical Law},
     (Basic Books, New York, 2006).}
  

\bibitem{richter:2006a}
  {B.~Richter, {\it Theory in Particle Physics: Theological Speculation Versus Practical Knowledge},
  Physics Today 59 $\#$10, 8 (2006).}

%\cite{Schroer:2006na}
\bibitem{Schroer:2006na}
  B.~Schroer,
  %``String theory deconstructed (a detailed critique of the content of ST from
  %an advanced QFT viewpoint),''
  arXiv:hep-th/0611132.
  %%CITATION = HEP-TH 0611132;%%

%\cite{Munoz:2003au}
\bibitem{Munoz:2003au}
  C.~Munoz,
  %``Desperately seeking the standard model,''
  arXiv:hep-ph/0312091.
  %%CITATION = HEP-PH 0312091;%% 
 
%%%% fff %%%%%%%%%%%%%%%%%%%%%%%%%%%%%%%%%%%%%%%%%%%%%%%%%%%%%%%%%%%%%%%%%%%%%%%%%%%%%%%%%%%%%%%%%%%%%%
% at here 2 

%\cite{Cleaver:1998sa}
\bibitem{Cleaver:1998sa}
  G.~B.~Cleaver, A.~E.~Faraggi and D.~V.~Nanopoulos,
  %``String derived MSSM and M-theory unification,''
  \textit{Phys.\ Lett.\ B} {\bf 455}, 135 (1999)
  [arXiv:hep-ph/9811427].
  %%CITATION = HEP-PH 9811427;%%

%\cite{Cleaver:1999cj}
\bibitem{Cleaver:1999cj}
  G.~B.~Cleaver, A.~E.~Faraggi and D.~V.~Nanopoulos,
  %``A minimal superstring standard model. I: Flat directions,''
 \textit{ Int.\ J.\ Mod.\ Phys.\ A} {\bf 16}, 425 (2001)
  [arXiv:hep-ph/9904301].
  %%CITATION = HEP-PH 9904301;%%

%\cite{Cleaver:1999mw}
\bibitem{Cleaver:1999mw}
  G.~B.~Cleaver, A.~E.~Faraggi, D.~V.~Nanopoulos and J.~W.~Walker,
  %``Phenomenological study of a minimal superstring standard model,''
  \textit{Nucl.\ Phys.\ B} {\bf 593}, 471 (2001)
  [arXiv:hep-ph/9910230].
  %%CITATION = HEP-PH 9910230;%%

%\cite{Cleaver:2000aa}
\bibitem{Cleaver:2000aa}
  G.~B.~Cleaver, A.~E.~Faraggi, D.~V.~Nanopoulos and J.~W.~Walker,
  %``Non-Abelian flat directions in a minimal superstring standard model,''
  \textit{Mod.\ Phys.\ Lett.\ A} {\bf 15}, 1191 (2000)
  [arXiv:hep-ph/0002060].
  %%CITATION = HEP-PH 0002060;%%

%\cite{Cleaver:2001ab}
\bibitem{Cleaver:2001ab}
  G.~B.~Cleaver, A.~E.~Faraggi, D.~V.~Nanopoulos and J.~W.~Walker,
  %``Phenomenology of non-Abelian flat directions in a minimal superstring
  %standard model,''
  \textit{Nucl.\ Phys.\ B} {\bf 620}, 259 (2002)
  [arXiv:hep-ph/0104091].
  %%CITATION = HEP-PH 0104091;%%

%\cite{Gmeiner:2005vz}
\bibitem{Gmeiner:2005vz}
  F.~Gmeiner, R.~Blumenhagen, G.~Honecker, D.~Lust and T.~Weigand,
  %``One in a billion: MSSM-like D-brane statistics,''
  \textit{JHEP} {\bf 0601}, 004 (2006)
  [arXiv:hep-th/0510170].
  %%CITATION = HEP-TH 0510170;%%

%\cite{Faraggi:1989ka}
\bibitem{Faraggi:1989ka}
  A.~E.~Faraggi, D.~V.~Nanopoulos and K.~J.~Yuan,
  %``A STANDARD LIKE MODEL IN THE 4-D FREE FERMIONIC STRING FORMULATION,''
  \textit{Nucl.\ Phys.\ B} {\bf 335}, 347 (1990).
  %%CITATION = NUPHA,B335,347;%%

%\cite{Faraggi:1991be}
\bibitem{Faraggi:1991be}
  A.~E.~Faraggi,
  %``Hierarchical Top - Bottom Mass Relation In A Superstring Derived Standard -
  %Like Model,''
  \textit{Phys.\ Lett.\ B} {\bf 274}, 47 (1992).
  %%CITATION = PHLTA,B274,47;%%

%\cite{Faraggi:1991jr}
\bibitem{Faraggi:1991jr}
  A.~E.~Faraggi,
  %``A New standard - like model in the four-dimensional free fermionic string
  %formulation,''
  \textit{Phys.\ Lett.\ B} {\bf 278}, 131 (1992).
  %%CITATION = PHLTA,B278,131;%%

%\cite{Faraggi:1992fa}
\bibitem{Faraggi:1992fa}
  A.~E.~Faraggi,
  %``Construction of realistic standard - like models in the free fermionic
  %superstring formulation,''
 \textit{ Nucl.\ Phys.\ B} {\bf 387}, 239 (1992)
  [arXiv:hep-th/9208024].
  %%CITATION = HEP-TH 9208024;%%

%\cite{Faraggi:1990af}
\bibitem{Faraggi:1990af}
  A.~E.~Faraggi,
  %``Fractional charges in a superstring derived standard like model,''
  \textit{Phys.\ Rev.\ D} {\bf 46}, 3204 (1992).
  %%CITATION = PHRVA,D46,3204;%%

%\cite{Antoniadis:1986rn}
\bibitem{Antoniadis:1986rn}
  I.~Antoniadis, C.~P.~Bachas and C.~Kounnas,
  %``FOUR-DIMENSIONAL SUPERSTRINGS,''
  \textit{Nucl.\ Phys.\ B} {\bf 289}, 87 (1987).
  %%CITATION = NUPHA,B289,87;%%
  
%\cite{Kawai:1986va}
\bibitem{Kawai:1986va}
  H.~Kawai, D.~C.~Lewellen and S.~H.~H.~Tye,
  %``CONSTRUCTION OF FOUR-DIMENSIONAL FERMIONIC STRING MODELS,''
  \textit{Phys.\ Rev.\ Lett.\ } {\bf 57}, 1832 (1986)
  [Erratum-ibid.\  {\bf 58}, 429 (1987)].
  %%CITATION = PRLTA,57,1832;%%

%\cite{Kawai:1987ew}
\bibitem{Kawai:1987ew}
  H.~Kawai, D.~C.~Lewellen, J.~A.~Schwartz and S.~H.~H.~Tye,
  %``THE SPIN STRUCTURE CONSTRUCTION OF STRING MODELS AND MULTILOOP MODULAR
  %INVARIANCE,''
  \textit{Nucl.\ Phys.\ B} {\bf 299}, 431 (1988).
  %%CITATION = NUPHA,B299,431;%% 

%\cite{Antoniadis:1987wp}
\bibitem{Antoniadis:1987wp}
  I.~Antoniadis and C.~Bachas,
  %``4-D FERMIONIC SUPERSTRINGS WITH ARBITRARY TWISTS,''
  \textit{Nucl.\ Phys.\ B} {\bf 298}, 586 (1988).
  %%CITATION = NUPHA,B298,586;%%

%\cite{Kalara:1990fb}
\bibitem{Kalara:1990fb}
  S.~Kalara, J.~L.~Lopez and D.~V.~Nanopoulos,
  %``NONRENORMALIZABLE TERMS IN THE FREE FERMIONIC FORMULATION OF 4-D STRINGS,''
  \textit{Phys.\ Lett.\ B} {\bf 245}, 421 (1990).
  %%CITATION = PHLTA,B245,421;%%

%\cite{Rizos:1991bm}
\bibitem{Rizos:1991bm}
  J.~Rizos and K.~Tamvakis,
  %``Some selection rules for nonrenormalizable chiral couplings in 4-D
  %fermionic superstring models,''
  \textit{Phys.\ Lett.\ B} {\bf 262}, 227 (1991).
  %%CITATION = PHLTA,B262,227;%%

%\cite{Faraggi:1990ac}
\bibitem{Faraggi:1990ac}
  A.~E.~Faraggi and D.~V.~Nanopoulos,
  %``Naturalness of three generations in free fermionic Z(2)-n x Z(4) string
  %models,''
  \textit{Phys.\ Rev.\ D} {\bf 48}, 3288 (1993).
  %%CITATION = PHRVA,D48,3288;%%

%\cite{Kobayashi:1996pb}
\bibitem{Kobayashi:1996pb}
  T.~Kobayashi and H.~Nakano,
  %``*Anomalous* U(1) symmetry in orbifold string models,''
  \textit{Nucl.\ Phys.\ B} {\bf 496}, 103 (1997)
  [arXiv:hep-th/9612066].
  %%CITATION = HEP-TH 9612066;%%

%\cite{Cleaver:1997rk}
\bibitem{Cleaver:1997rk}
  G.~B.~Cleaver and A.~E.~Faraggi,
  %``On the anomalous U(1) in free fermionic superstring models,''
  \textit{Int.\ J.\ Mod.\ Phys.\ A} {\bf 14}, 2335 (1999)
  [arXiv:hep-ph/9711339].
  %%CITATION = HEP-PH 9711339;%%

%\cite{Dine:1987xk}
\bibitem{Dine:1987xk}
  M.~Dine, N.~Seiberg and E.~Witten,
  %``FAYET-ILIOPOULOS TERMS IN STRING THEORY,''
  \textit{Nucl.\ Phys.\ B} {\bf 289}, 589 (1987).
  %%CITATION = NUPHA,B289,589;%%

%\cite{Atick:1987gy}
\bibitem{Atick:1987gy}
  J.~J.~Atick, L.~J.~Dixon and A.~Sen,
  %``String Calculation Of Fayet-Iliopoulos D Terms In Arbitrary Supersymmetric
  %Compactifications,''
  \textit{Nucl.\ Phys.\ B }{\bf 292}, 109 (1987).
  %%CITATION = NUPHA,B292,109;%%                

%\cite{Cleaver:1997cr}
\bibitem{Cleaver:1997cr}
  G.~B.~Cleaver, \textit{Proceedings of Orbis Scientiae on Physics of Mass}, Miami, Florida, 12-15 Dec 1997. 
  %``Mass hierarchy and flat directions in string models,''
%\href{http://www.slac.stanford.edu/spires/find/hep/www?irn=4257286}{SPIRES entry}
%{\it Prepared for Orbis Scientiae on Physics of Mass, Miami, Florida, 12-15 Dec 1997}

%\cite{Lopez:1989fb}
\bibitem{Lopez:1989fb}
  J.~L.~Lopez and D.~V.~Nanopoulos,
  %``HIERARCHICAL FERMION MASSES AND MIXING ANGLES FROM THE FLIPPED STRING,''
  \textit{Nucl.\ Phys.\ B} {\bf 338}, 73 (1990).
  %%CITATION = NUPHA,B338,73;%%

%\cite{Cvetic:1998gv}
\bibitem{Cvetic:1998gv}
  M.~Cvetic, L.~L.~Everett and J.~Wang,
  %``Units and numerical values of the effective couplings in perturbative
  %heterotic string vacua,''
  \textit{Phys.\ Rev.\ D }{\bf 59}, 107901 (1999)
  [arXiv:hep-ph/9808321].
  %%CITATION = HEP-PH 9808321;%%

%\cite{Faraggi:1996pa}
\bibitem{Faraggi:1996pa}
  A.~E.~Faraggi,
  %``Calculating fermion masses in superstring derived standard-like  models,''
  \textit{Nucl.\ Phys.\ B }{\bf 487}, 55 (1997)
  [arXiv:hep-ph/9601332].
  %%CITATION = HEP-PH 9601332;%%

%\cite{Lopez:1995cs}
\bibitem{Lopez:1995cs}
  J.~L.~Lopez and D.~V.~Nanopoulos,
  %``A new scenario for string unification,''
  \textit{Phys.\ Rev.\ Lett.\ } {\bf 76}, 1566 (1996)
  [arXiv:hep-ph/9511426].

%\cite{Lahanas:1986uc}
\bibitem{Lahanas:1986uc}
  A.~B.~Lahanas and D.~V.~Nanopoulos,
  %``THE ROAD TO NO SCALE SUPERGRAVITY,''
  \textit{Phys.\ Rept.\  }{\bf 145}, 1 (1987).
  %%CITATION = PRPLC,145,1;%%

%\cite{Ellis:1984xe}
\bibitem{Ellis:1984xe}
  J.~R.~Ellis, K.~Enqvist and D.~V.~Nanopoulos,
  %``Noncompact Supergravity Solves Problems,''
  \textit{Phys.\ Lett.\ B} {\bf 151}, 357 (1985).
  %%CITATION = PHLTA,B151,357;%%


%\cite{Cleaver:2005vv}
\bibitem{Cleaver:2005vv}
  G.~B.~Cleaver, D.~V.~Nanopoulos, J.~T.~Perkins and J.~W.~Walker,
  %``On geometrical interpretation of non-Abelian flat direction constraints,''
  arXiv:hep-th/0512020.
  %%CITATION = HEP-TH 0512020;%%  

%\cite{Dvali:2003zh}
\bibitem{Dvali:2003zh}
  G.~Dvali, R.~Kallosh and A.~Van Proeyen,
  %``D-term strings,''
  \textit{JHEP} {\bf 0401}, 035 (2004)
  [arXiv:hep-th/0312005].
  %%CITATION = HEP-TH 0312005;%%     

%\cite{Faraggi:2006qa}
\bibitem{Faraggi:2006qa}
  A.~E.~Faraggi, E.~Manno and C.~Timirgaziu,
  %``Minimal standard heterotic string models,''
  \textit{Eur.~Phys.~J.~C} {\bf 50}, 701 (2007)
  [arXiv:hep-th/0610118].
  %%CITATION = HEP-TH 0610118;%%


%%%%%%% heterotic orbifold MSSM %%%%%%%%%%%%%%%%%%%%%%%%%%%%%%%%%%%%%%%%%%%%%%%%%%%%%%%%%%%%%%%%%%%%%%%%%%%%%%%
% at here 3

%\cite{Buchmuller:2005jr}
\bibitem{Buchmuller:2005jr}
  W.~Buchmuller, K.~Hamaguchi, O.~Lebedev and M.~Ratz,
  %``The supersymmetric standard model from the heterotic string,''
  \textit{Phys.\ Rev.\ Lett.\ } {\bf 96}, 121602 (2006)
  [arXiv:hep-ph/0511035].
  %%CITATION = HEP-PH 0511035;%%

%\cite{Buchmuller:2006ik}
\bibitem{Buchmuller:2006ik}
  W.~Buchmuller, K.~Hamaguchi, O.~Lebedev and M.~Ratz,
  %``Supersymmetric standard model from the heterotic string. II,''
  \textit{Nucl.\ Phys.\ B } {\bf 785}, 149 (2007)
  [arXiv:hep-th/0606187].
  %%CITATION = HEP-TH 0606187;%%

%\cite{Wess:1992cp}
\bibitem{Wess:1992cp}
  J.~Wess and J.~Bagger,
  ``Supersymmetry and supergravity,''
   {http://www.slac.stanford.edu/spires/find/hep/www?irn=5426545}{SPIRES entry}

%\cite{Hamidi:1986vh}
\bibitem{Hamidi:1986vh}
  S.~Hamidi and C.~Vafa,
  %``INTERACTIONS ON ORBIFOLDS,''
  \textit{Nucl.\ Phys.\ B} {\bf 279}, 465 (1987).
  %%CITATION = NUPHA,B279,465;%%

%\cite{Font:1988mm}
\bibitem{Font:1988mm}
  A.~Font, L.~E.~Ibanez, H.~P.~Nilles and F.~Quevedo,
  %``Yukawa Couplings In Degenerate Orbifolds: Towards A Realistic SU(3) X SU(2)
  %X U(1) Superstring,''
  \textit{Phys.\ Lett.\ } {\bf 210B}, 101 (1988)
  [Erratum-ibid.\ B {\bf 213}, 564 (1988)].
  %%CITATION = PHLTA,210B,101;%%

%\cite{Cleaver:2002qc}
\bibitem{Cleaver:2002qc}
  G.~Cleaver, V.~Desai, H.~Hanson, J.~Perkins, D.~Robbins and S.~Shields,
  %``On the possibility of optical unification in heterotic strings,''
  \textit{Phys.\ Rev.\ D} {\bf 67}, 026009 (2003)
  [arXiv:hep-ph/0209050].
  %%CITATION = HEP-PH 0209050;%% 

%\cite{Perkins:2003tb}
\bibitem{Perkins:2003tb}
  J.~Perkins, B.~Dundee, R.~Obousy, E.~Kasper, M.~Robinson, K.~Stone and G.~Cleaver,
  %``Heterotic string optical unification,''
  arXiv:hep-ph/0310155.
  %%CITATION = HEP-PH 0310155;%%

%\cite{Perkins:2005zh}
\bibitem{Perkins:2005zh}
  J.~Perkins {\it et al.},
  %``Stringent phenomenological investigation into heterotic string optical
  %unification,''
  \textit{Phys.\ Rev D} {\bf 75}, 026007 (2007)
  [arXiv:hep-ph/0510141].
  %%CITATION = HEP-PH 0510141;%%

\bibitem{Giedt:2002kb}
  J.~Giedt,
  %``Optical unification,''
  \textit{Mod.\ Phys.\ Lett.\ A }{\bf 18}, 1625 (2003)
  [arXiv:hep-ph/0205224].
  %%CITATION = HEP-PH 0205224;%%

%%%%%%% heterotic Calabi-Yau %%%%%%%%%%%%%%%%%%%%%%%%%%%%%%%%%%%%%%%%%%%%%%%%%%%%%%%%%%%%%%%%%%%%%%%%%%%%%%%
% at here 4

%\cite{Braun:2005nv}
\bibitem{Braun:2005nv}
  V.~Braun, Y.~H.~He, B.~A.~Ovrut and T.~Pantev,
  %``The exact MSSM spectrum from string theory,''
  \textit{JHEP} {\bf 0605}, 043 (2006)
  [arXiv:hep-th/0512177].
  %%CITATION = HEP-TH 0512177;%%

%\cite{Braun:2004xv}
\bibitem{Braun:2004xv}
  V.~Braun, B.~A.~Ovrut, T.~Pantev and R.~Reinbacher,
  %``Elliptic Calabi-Yau threefolds with Z(3) x Z(3) Wilson lines,''
  \textit{JHEP} {\bf 0412}, 062 (2004)
  [arXiv:hep-th/0410055].
  %%CITATION = HEP-TH 0410055;%%

%\cite{Braun:2005ux}
\bibitem{Braun:2005ux}
  V.~Braun, Y.~H.~He, B.~A.~Ovrut and T.~Pantev,
  %``A heterotic standard model,''
  \textit{Phys.\ Lett.\ B} {\bf 618}, 252 (2005)
  [arXiv:hep-th/0501070].
  %%CITATION = HEP-TH 0501070;%%

%\cite{Braun:2005bw}
\bibitem{Braun:2005bw}
  V.~Braun, Y.~H.~He, B.~A.~Ovrut and T.~Pantev,
  %``A standard model from the E(8) x E(8) heterotic superstring,''
  \textit{JHEP} {\bf 0506}, 039 (2005)
  [arXiv:hep-th/0502155].
  %%CITATION = HEP-TH 0502155;%%

%\cite{Braun:2005zv}
\bibitem{Braun:2005zv}
  V.~Braun, Y.~H.~He, B.~A.~Ovrut and T.~Pantev,
  %``Vector bundle extensions, sheaf cohomology, and the heterotic standard
  %model,''
  \textit{Adv.\ Theor.\ Math.\ Phys.\ } {\bf 10}, 4 (2006)
  [arXiv:hep-th/0505041].
  %%CITATION = HEP-TH 0505041;%%

%\cite{Braun:2006me}
\bibitem{Braun:2006me}
  V.~Braun, Y.~H.~He and B.~A.~Ovrut,
  %``Yukawa couplings in heterotic standard models,''
  \textit{JHEP} {\bf 0604}, 019 (2006)
  [arXiv:hep-th/0601204].
  %%CITATION = HEP-TH 0601204;%%

%\cite{Braun:2005xp}
\bibitem{Braun:2005xp}
  V.~Braun, Y.~H.~He, B.~A.~Ovrut and T.~Pantev,
  %``Moduli dependent mu-terms in a heterotic standard model,''
  \textit{JHEP} {\bf 0603}, 006 (2006)
  [arXiv:hep-th/0510142].
  %%CITATION = HEP-TH 0510142;%%

%\cite{Braun:2006ae}
\bibitem{Braun:2006ae}
  V.~Braun, Y.~H.~He and B.~A.~Ovrut,
  %``Stability of the minimal heterotic standard model bundle,''
  \textit{JHEP} {\bf 0606}, 032 (2006)
  [arXiv:hep-th/0602073].
  %%CITATION = HEP-TH 0602073;%%

%\cite{Braun:2006th}
\bibitem{Braun:2006th}
  V.~Braun and B.~A.~Ovrut,
  %``Stabilizing moduli with a positive cosmological constant in heterotic
  %M-theory,''
  \textit{JHEP} {\bf 0607}, 035 (2006)
  [arXiv:hep-th/0603088].
  %%CITATION = HEP-TH 0603088;%%

%\cite{Gomez:2005ii}
\bibitem{Gomez:2005ii}
  T.~L.~Gomez, S.~Lukic and I.~Sols,
  %``Constraining the Kaehler moduli in the heterotic standard model,''
  \textit{Commun.\ Math.\ Phys.\ } {\bf 276}, 1 (2007)
  [arXiv:hep-th/0512205].
  %%CITATION = HEP-TH 0512205;%%

%\cite{Bouchard:2005ag}
\bibitem{Bouchard:2005ag}
  V.~Bouchard and R.~Donagi,
  %``An SU(5) heterotic standard model,''
  \textit{Phys.\ Lett.\ B} {\bf 633}, 783 (2006)
  [arXiv:hep-th/0512149].
  %%CITATION = HEP-TH 0512149;%%

%\cite{Braun:2006em}
\bibitem{Braun:2006em}
  V.~Braun, E.~I.~Buchbinder and B.~A.~Ovrut,
  %``Dynamical SUSY breaking in heterotic M-theory,''
  \textit{Phys.\ Lett.\ B }{\bf 639}, 566 (2006)
  [arXiv:hep-th/0606166].
  %%CITATION = HEP-TH 0606166;%%

%\cite{Braun:2006da}
\bibitem{Braun:2006da}
  V.~Braun, E.~I.~Buchbinder and B.~A.~Ovrut,
  %``Towards realizing dynamical SUSY breaking in heterotic model building,''
  \textit{JHEP} {\bf 0610}, 041 (2006)
  [arXiv:hep-th/0606241].
  %%CITATION = HEP-TH 0606241;%%

%\cite{Intriligator:2006dd}
\bibitem{Intriligator:2006dd}
  K.~Intriligator, N.~Seiberg and D.~Shih,
  %``Dynamical SUSY breaking in meta-stable vacua,''
  \textit{JHEP} {\bf 0604}, 021 (2006)
  [arXiv:hep-th/0602239].
  %%CITATION = HEP-TH 0602239;%%


%\cite{Bouchard:2006dn}
\bibitem{Bouchard:2006dn}
  V.~Bouchard, M.~Cvetic and R.~Donagi,
  %``Tri-linear couplings in an heterotic minimal supersymmetric standard
  %model,''
  \textit{Nucl.\ Phys.\ B} {\bf 745}, 62 (2006)
  [arXiv:hep-th/0602096].
  %%CITATION = HEP-TH 0602096;%%

%\cite{Blumenhagen:2006ux}
\bibitem{Blumenhagen:2006ux}
  R.~Blumenhagen, S.~Moster and T.~Weigand,
  %``Heterotic GUT and standard model vacua from simply connected Calabi-Yau
  %manifolds,''
 \textit{ Nucl.\ Phys.\ B} {\bf 751}, 186 (2006)
  [arXiv:hep-th/0603015].
  %%CITATION = HEP-TH 0603015;%%

%%%%%%% Type IIB %%%%%%%%%%%%%%%%%%%%%%%%%%%%%%%%%%%%%%%%%%%%%%%%%%%%%%%%%%%%%%%%%%%%%%%%%%%%%%%
% at here 5

%\cite{Bailin:2006rx}
\bibitem{Bailin:2006rx}
  D.~Bailin and A.~Love,
  %``The supersymmetric standard model from the Z'(6) orientifold?,''
  \textit{AIP Conf.\ Proc.\ } {\bf 881}, 1 (2007)
  [arXiv:hep-th/0607158].
  %%CITATION = HEP-TH 0607158;%%

%\cite{Bailin:2006zf}
\bibitem{Bailin:2006zf}
  D.~Bailin and A.~Love,
  %``Towards the supersymmetric standard model from intersecting D6-branes on
  %the Z'(6) orientifold,''
  \textit{Nucl.\ Phys.\ B} {\bf 755}, 79 (2006)
  [arXiv:hep-th/0603172].
  %%CITATION = HEP-TH 0603172;%%

\bibitem{Kokorelis:2004dc}
  C.~Kokorelis,
  %``N = 1 supersymmetric standard models with no massless exotics from
  %intersecting branes,''
  arXiv:hep-th/0406258.

\bibitem{Kokorelis:2004tb}
  C.~Kokorelis,
  %``Standard model building from intersecting D-branes,''
  arXiv:hep-th/0410134.
  %%CITATION = HEP-TH/0410134;%%

%\cite{Blumenhagen:2001te}
\bibitem{Blumenhagen:2001te}
  R.~Blumenhagen, B.~Kors, D.~Lust and T.~Ott,
  %``The standard model from stable intersecting brane world orbifolds,''
  \textit{Nucl.\ Phys.\ B} {\bf 616}, 3 (2001)
  [arXiv:hep-th/0107138].
  %%CITATION = HEP-TH 0107138;%%

%\cite{Berkooz:1996dw}
\bibitem{Berkooz:1996dw}
  M.~Berkooz and R.~G.~Leigh,
  %``A D = 4 N = 1 orbifold of type I strings,''
  \textit{Nucl.\ Phys.\ B} {\bf 483}, 187 (1997)
  [arXiv:hep-th/9605049].
  %%CITATION = HEP-TH 9605049;%%

%\cite{Cvetic:2001tj}
\bibitem{Cvetic:2001tj}
  M.~Cvetic, G.~Shiu and A.~M.~Uranga,
  %``Three-family supersymmetric standard like models from intersecting  brane
  %worlds,''
  \textit{Phys.\ Rev.\ Lett.\ } {\bf 87}, 201801 (2001)
  [arXiv:hep-th/0107143].
  %%CITATION = HEP-TH 0107143;%%

%\cite{Cascales:2003wn}
\bibitem{Cascales:2003wn}
  J.~F.~G.~Cascales, M.~P.~Garcia del Moral, F.~Quevedo and A.~M.~Uranga,
  %``Realistic D-brane models on warped throats: Fluxes, hierarchies and  moduli
  %stabilization,''
  \textit{JHEP} {\bf 0402}, 031 (2004)
  [arXiv:hep-th/0312051].
  %%CITATION = HEP-TH 0312051;%%
  
%\cite{Marchesano:2004yq}
\bibitem{Marchesano:2004yq}
  F.~Marchesano and G.~Shiu,
  %``MSSM vacua from flux compactifications,''
  \textit{Phys.\ Rev.\ D} {\bf 71}, 011701 (2005)
  [arXiv:hep-th/0408059].
  %%CITATION = HEP-TH 0408059;%%

%\cite{Witten:1998cd}
\bibitem{Witten:1998cd}
  E.~Witten,
  %``D-branes and K-theory,''
  \textit{JHEP} {\bf 9812}, 019 (1998)
  [arXiv:hep-th/9810188].
  %%CITATION = HEP-TH 9810188;%%

%\cite{Uranga:2000xp}
\bibitem{Uranga:2000xp}
  A.~M.~Uranga,
  %``D-brane probes, RR tadpole cancellation and K-theory charge,''
  \textit{Nucl.\ Phys.\ B} {\bf 598}, 225 (2001)
  [arXiv:hep-th/0011048].
  %%CITATION = HEP-TH 0011048;%%

%\cite{Chen:2005mj}
\bibitem{Chen:2005mj}
  C.~M.~Chen, T.~Li and D.~V.~Nanopoulos,
  %``Standard-like model building on type II orientifolds,''
  \textit{Nucl.\ Phys.\ B} {\bf 732}, 224 (2006)
  [arXiv:hep-th/0509059].
  %%CITATION = HEP-TH 0509059;%%

%\cite{Marchesano:2004xz}
\bibitem{Marchesano:2004xz}
  F.~Marchesano and G.~Shiu,
  %``Building MSSM flux vacua,''
  \textit{JHEP} {\bf 0411}, 041 (2004)
  [arXiv:hep-th/0409132].
  %%CITATION = HEP-TH 0409132;%%

%\cite{Cremades:2003qj}
\bibitem{Cremades:2003qj}
  D.~Cremades, L.~E.~Ibanez and F.~Marchesano,
  %``Yukawa couplings in intersecting D-brane models,''
  \textit{JHEP} {\bf 0307}, 038 (2003)
  [arXiv:hep-th/0302105].
  %%CITATION = HEP-TH 0302105;%%

%\cite{Grana:2002nq}
\bibitem{Grana:2002nq}
  M.~Grana,
  %``MSSM parameters from supergravity backgrounds,''
  \textit{Phys.\ Rev.\ D} {\bf 67}, 066006 (2003)
  [arXiv:hep-th/0209200].
  %%CITATION = HEP-TH 0209200;%%

%\cite{Camara:2003ku}
\bibitem{Camara:2003ku}
  P.~G.~Camara, L.~E.~Ibanez and A.~M.~Uranga,
  %``Flux-induced SUSY-breaking soft terms,''
  \textit{Nucl.\ Phys.\ B} {\bf 689}, 195 (2004)
  [arXiv:hep-th/0311241].
  %%CITATION = HEP-TH 0311241;%%

%\cite{Grana:2003ek}
\bibitem{Grana:2003ek}
  M.~Grana, T.~W.~Grimm, H.~Jockers and J.~Louis,
  %``Soft supersymmetry breaking in Calabi-Yau orientifolds with D-branes  and
  %fluxes,''
  \textit{Nucl.\ Phys.\ B} {\bf 690}, 21 (2004)
  [arXiv:hep-th/0312232].
  %%CITATION = HEP-TH 0312232;%%

%\cite{Lust:2004fi}
\bibitem{Lust:2004fi}
  D.~Lust, S.~Reffert and S.~Stieberger,
  %``Flux-induced soft supersymmetry breaking in chiral type IIb  orientifolds
  %with D3/D7-branes,''
  \textit{Nucl.\ Phys.\ B} {\bf 706}, 3 (2005)
  [arXiv:hep-th/0406092].
  %%CITATION = HEP-TH 0406092;%%

%\cite{Kumar:2005hf}
\bibitem{Kumar:2005hf}
  J.~Kumar and J.~D.~Wells,
  %``Surveying standard model flux vacua on T**6/Z(2) x Z(2),''
  \textit{JHEP} {\bf 0509}, 067 (2005)
  [arXiv:hep-th/0506252].
  %%CITATION = HEP-TH 0506252;%%

%\cite{Cvetic:2005bn}
\bibitem{Cvetic:2005bn}
  M.~Cvetic, T.~Li and T.~Liu,
  %``Standard-like models as type IIB flux vacua,''
  \textit{Phys.\ Rev.\ D} {\bf 71}, 106008 (2005)
  [arXiv:hep-th/0501041].
  %%CITATION = HEP-TH 0501041;%%
 
 
%%%%%%% Type IIA %%%%%%%%%%%%%%%%%%%%%%%%%%%%%%%%%%%%%%%%%%%%%%%%%%%%%%%%%%%%%%%%%%%%%%%%%%%%%%%
% at here 6

%\cite{Blumenhagen:2000wh}
\bibitem{Blumenhagen:2000wh}
  R.~Blumenhagen, L.~Goerlich, B.~Kors and D.~Lust,
  %``Noncommutative compactifications of type I strings on tori with  magnetic
  %background flux,''
  \textit{JHEP} {\bf 0010}, 006 (2000)
  [arXiv:hep-th/0007024].
  %%CITATION = HEP-TH 0007024;%%

%\cite{Aldazabal:2000cn}
\bibitem{Aldazabal:2000cn}
  G.~Aldazabal, S.~Franco, L.~E.~Ibanez, R.~Rabadan and A.~M.~Uranga,
  %``Intersecting brane worlds,''
  \textit{JHEP} {\bf 0102}, 047 (2001)
  [arXiv:hep-ph/0011132].
  %%CITATION = HEP-PH 0011132;%%

%\cite{Aldazabal:2000dg}
\bibitem{Aldazabal:2000dg}
  G.~Aldazabal, S.~Franco, L.~E.~Ibanez, R.~Rabadan and A.~M.~Uranga,
  %``D = 4 chiral string compactifications from intersecting branes,''
  \textit{J.\ Math.\ Phys.\ } {\bf 42}, 3103 (2001)
  [arXiv:hep-th/0011073].
  %%CITATION = HEP-TH 0011073;%%

%\cite{Blumenhagen:2000ea}
\bibitem{Blumenhagen:2000ea}
  R.~Blumenhagen, B.~Kors and D.~Lust,
  %``Type I strings with F- and B-flux,''
  \textit{JHEP} {\bf 0102}, 030 (2001)
  [arXiv:hep-th/0012156].
  %%CITATION = HEP-TH 0012156;%% 

%\cite{Angelantonj:2000hi}
\bibitem{Angelantonj:2000hi}
  C.~Angelantonj, I.~Antoniadis, E.~Dudas and A.~Sagnotti,
  %``Type-I strings on magnetised orbifolds and brane transmutation,''
  \textit{Phys.\ Lett.\ B} {\bf 489}, 223 (2000)
  [arXiv:hep-th/0007090].
  %%CITATION = HEP-TH 0007090;%%

%\cite{Cascales:2003zp}
\bibitem{Cascales:2003zp}
  J.~F.~G.~Cascales and A.~M.~Uranga,
  %``Chiral 4d N = 1 string vacua with D-branes and NSNS and RR fluxes,''
  \textit{JHEP} {\bf 0305}, 011 (2003)
  [arXiv:hep-th/0303024].
  %%CITATION = HEP-TH 0303024;%% 
  
%\cite{Blumenhagen:2003vr}
\bibitem{Blumenhagen:2003vr}
  R.~Blumenhagen, D.~Lust and T.~R.~Taylor,
  %``Moduli stabilization in chiral type IIB orientifold models with fluxes,''
  \textit{Nucl.\ Phys.\ B} {\bf 663}, 319 (2003)
  [arXiv:hep-th/0303016].
  %%CITATION = HEP-TH 0303016;%%
  
%\cite{Cvetic:2001nr}
\bibitem{Cvetic:2001nr}
  M.~Cvetic, G.~Shiu and A.~M.~Uranga,
  %``Chiral four-dimensional N = 1 supersymmetric type IIA orientifolds from
  %intersecting D6-branes,''
  \textit{Nucl.\ Phys.\ B} {\bf 615}, 3 (2001)
  [arXiv:hep-th/0107166].
  %%CITATION = HEP-TH 0107166;%%

%\cite{Cvetic:2003yd}
\bibitem{Cvetic:2003yd}
  M.~Cvetic, P.~Langacker and J.~Wang,
  %``Dynamical supersymmetry breaking in standard-like models with  intersecting
  %D6-branes,''
  \textit{Phys.\ Rev.\ D} {\bf 68}, 046002 (2003)
  [arXiv:hep-th/0303208].
  %%CITATION = HEP-TH 0303208;%%
  
%\cite{Cvetic:2003xs}
\bibitem{Cvetic:2003xs}
  M.~Cvetic and I.~Papadimitriou,
  %``More supersymmetric standard-like models from intersecting D6-branes on
  %type IIA orientifolds,''
  \textit{Phys.\ Rev.\ D} {\bf 67}, 126006 (2003)
  [arXiv:hep-th/0303197].
  %%CITATION = HEP-TH 0303197;%%

%\cite{Cvetic:2002pj}
\bibitem{Cvetic:2002pj}
  M.~Cvetic, I.~Papadimitriou and G.~Shiu,
  %``Supersymmetric three family SU(5) grand unified models from type IIA
  %orientifolds with intersecting D6-branes,''
  \textit{Nucl.\ Phys.\ B} {\bf 659}, 193 (2003)
  [Erratum-ibid.\ B {\bf 696}, 298 (2004)]
  [arXiv:hep-th/0212177].
  %%CITATION = HEP-TH 0212177;%% 

%\cite{Cvetic:2004ui}
\bibitem{Cvetic:2004ui}
  M.~Cvetic, T.~Li and T.~Liu,
  %``Supersymmetric Pati-Salam models from intersecting D6-branes: A road to
  %the standard model,''
  \textit{Nucl.\ Phys.\ B} {\bf 698}, 163 (2004)
  [arXiv:hep-th/0403061].
  %%CITATION = HEP-TH 0403061;%%

%\cite{Cvetic:2004nk}
\bibitem{Cvetic:2004nk}
  M.~Cvetic, P.~Langacker, T.~Li and T.~Liu,
  %``D6-brane splitting on type IIA orientifolds,''
  \textit{Nucl.\ Phys.\ B} {\bf 709}, 241 (2005)
  [arXiv:hep-th/0407178].
  %%CITATION = HEP-TH 0407178;%%
  
%\cite{Chen:2005ab}
\bibitem{Chen:2005ab}
  C.~M.~Chen, G.~V.~Kraniotis, V.~E.~Mayes, D.~V.~Nanopoulos and J.~W.~Walker,
  %``A supersymmetric flipped SU(5) intersecting brane world,''
  \textit{Phys.\ Lett.\ B} {\bf 611}, 156 (2005)
  [arXiv:hep-th/0501182].
  %%CITATION = HEP-TH 0501182;%%

%\cite{Cvetic:2002qa}
\bibitem{Cvetic:2002qa}
  M.~Cvetic, P.~Langacker and G.~Shiu,
  %``Phenomenology of a three-family standard-like string model,''
  \textit{Phys.\ Rev.\ D} {\bf 66}, 066004 (2002)
  [arXiv:hep-ph/0205252].
  %%CITATION = HEP-PH 0205252;%%

%\cite{Cvetic:2002wh}
\bibitem{Cvetic:2002wh}
  M.~Cvetic, P.~Langacker and G.~Shiu,
  %``A three-family standard-like orientifold model: Yukawa couplings and
  %hierarchy,''
  \textit{Nucl.\ Phys.\ B} {\bf 642}, 139 (2002)
  [arXiv:hep-th/0206115].
  %%CITATION = HEP-TH 0206115;%%

%%%%%%%%%%%%%%%%%%%%%%%%%%%%%%%%%%%%%%%%%%%%%%%%%%


\end{thebibliography}
\end{document}